\documentclass[numberedappendix,twocolumn,onecolappendix]{openjournal}

\usepackage{newtxtext,newtxmath}
\usepackage{algorithm}
\usepackage[beginComment={}]{algpseudocodex}
\usepackage{framed}

\usepackage[scr=esstix,cal=boondox]{mathalfa} 

\usepackage{xcolor}
\usepackage[T1]{fontenc}
\usepackage{comment}
\usepackage[percent]{overpic}
\DeclareRobustCommand{\VAN}[3]{#2}
\let\VANthebibliography\thebibliography
\def\thebibliography{\DeclareRobustCommand{\VAN}[3]{##3}\VANthebibliography}

\newcommand{\modified}[1]{#1}
\newcommand{\mmodified}[1]{#1}

\usepackage{graphicx}	
\usepackage{amsmath}	
\usepackage{tikz}
\usepackage{hyperref}
\usepackage{hypcap}

\hypersetup{
    colorlinks=true,
    linkcolor=blue,
    citecolor=blue,
    filecolor=magenta,      
    urlcolor=cyan,
    pdfpagemode=FullScreen,
    }

\newcommand{\mx}{\mathbf{x}}
\newcommand{\mv}{\mathbf{v}}

\begin{document}

\title{The Ray Tracing Sampler: Bayesian Sampling of Neural Networks for Everyone}

\author{Peter Behroozi$^{1}$}
\affiliation{$^{1}$Department of Astronomy and Steward Observatory, University of Arizona, Tucson, AZ 85721, USA}



\begin{abstract}
We derive a \modified{family of} Markov Chain Monte Carlo (\modified{MCMC}) sampl\modified{ing methods} based on following ray paths in a medium where the refractive index $n(\mx)$ is a function of the desired likelihood $\mathcal{L}(\mx)$\modified{, extending past work on isokinetic sampling}. The \modified{simplest ray tracing} method propagates rays at constant speed through parameter space, leading to orders of magnitude higher resilience to heating for stochastic gradients as compared to \modified{both} Hamiltonian Monte Carlo (HMC) \modified{and Stochastic Gradient HMC}, as well as the ability to cross any likelihood barrier, including holes in parameter space.  Using the \modified{simplest} ray tracing method, we sample the posterior distributions of neural network outputs for a variety of different architectures, \modified{including a preliminary exploration of} the 1.5 billion-parameter GPT-2 (Generative Pre-trained Transformer 2) architecture, all on a single consumer-level GPU.  We also show that prior samplers including traditional HMC, \modified{the original} microcanonical HMC, Metropolis, Gibbs, and even Monte Carlo integration are special cases within a generalized ray tracing framework, which can sample according to an arbitrary weighting function.  Public code and documentation for  C, \texttt{JAX}, and \texttt{PyTorch} are available at \url{https://bitbucket.org/pbehroozi/ray-tracing-sampler/src}.
\end{abstract}




\section{Introduction}

Sampling algorithms, including MCMC methods, aim to draw independent samples from a probability distribution $P(\mx)$, often only using an unnormalized likelihood function $\mathcal{L}(\mx)\propto P(\mx)$.  This problem arises in all scientific disciplines, as it is equivalent to finding the distribution of model parameters \modified{or hidden variables} $\mx$ consistent with a set of observations (\modified{e.g., $P(\mx|\mathbf{D},\mathbf{P})$ for observations $\mathbf{D}$ and priors $\mathbf{P}$}).  Early sampling algorithms, such as \cite{Metropolis53} and \cite{Hastings70}, are hence widely used across fields.

Since these early papers, there have been many advances.  Sampling algorithms differ in their transition functions, often denoted as \mbox{$\pi(\mx:\rightarrow\mathbf{y})$} or $T(\mathbf{y}|\mx)$, defined as the probability density of moving to point $\mathbf{y}$ when starting at point $\mx$.  Sampling efficiency depends on how well the transition function aligns with the target distribution $P(\mx)$, so modern samplers have transition functions that adapt to the target distribution.  We can broadly classify modern algorithms by whether they use local or global information to adapt the transition function.  Local samplers use information near the current location, such as likelihood gradients (e.g., Hamiltonian/Hybrid Monte Carlo, \citealt{Duane87,Neal12}; the Metropolis-Adjusted Langevin Algorithm, \citealt{Besag94,RT96}; and recently, Bouncy Particle and Zig-Zag Sampling, \citealt{Peters12,BC15,Bierkens16,Corbella22}).  Global samplers instead learn approximate fitting functions to the global target distribution (e.g., mode jumping, \citealt{Tjelmeland01}; nested sampling, \citealt{Skilling04,Feroz08,Feroz19,Speagle20,Buchner23}; KDE sampling, \citealt{Rosenblatt56,Parzen62}; Gaussian processes, \citealt{Benavoli21}; and normalizing flows, \citealt{Esteban10,Dinh14,Rezende15}).  Hybrid methods that combine global and local information (e.g., \citealt{Hoffman19,Karamanis22,Wong22}) also exist.

Yet, the complexity of scientific modeling has grown exponentially in the meantime.  It is now common to train neural networks with millions to trillions of parameters in every scientific field (e.g., computer science, physics, biology, chemistry, and geoscience;  \citealt{Ahmad22,Choudhary22,Chowdhery22,Puzyrev22,Sriram22,Lin23,OpenAI23}).  In astronomy, this includes problems like initial condition recovery (e.g., reconstructing the initial matter distribution that led to the current Milky Way galaxy and its local environment, for billions to trillions of particles; see \citealt{Vogelsberger20} for a review) and blended source recovery (e.g., reconstructing intrinsic colors and luminosities for billions of partially overlapping galaxies as imaged by the \textit{Rubin} telescope; see \citealt{Melchior21} for a review).  All of these applications would benefit from Bayesian sampling to obtain uncertainties on model outputs, but it has remained challenging with current methods to do so.

\modified{Currently, neural posterior estimation (NPE) techniques are often used as fast estimators of $P(\mathbf{x}|\mathbf{D},\mathbf{P})$, including simulation-based inference generally, and methods like normalizing flows and diffusion models specifically (see \citealt{Cranmer20} and \citealt{Deistler25} for reviews).  However, NPE methods in fact output a different function: $P(\mathbf{x}|\mathbf{D},\mathbf{P},\mathbf{M})$, where $\mathbf{M}$ is the tensor of model parameters, corresponding to the fact that different models will give different outputs for the same inputs.  Indeed, the output of any individual neural posterior model can be arbitrarily far from the truth for a given data vector $\mathbf{D}$ (see, e.g., \citealt{Hayati25,Alokda26}).  The NPE output $P(\mathbf{x}|\mathbf{D},\mathbf{P},\mathbf{M})$ is related to the desired posterior probability $P(\mathbf{x}|\mathbf{D},\mathbf{P})$ via the law of total probability:}
\begin{equation}
    \mmodified{P(\mathbf{x}|\mathbf{D},\mathbf{P}) = \sum_\mathbf{M} P(\mathbf{x}|\mathbf{D},\mathbf{P},\mathbf{M})P(\mathbf{M}|\mathbf{D},\mathbf{P}).} \label{e:tp}
\end{equation}
\modified{Bayesian sampling is hence still necessary to compute this sum across NPE models $\mathbf{M}$ to obtain $P(\mathbf{x}|\mathbf{D},\mathbf{P})$ as desired.}

At present, with $\mathcal{O}(D^{1/4})$ scaling in the number of dimensions $D$, Hamiltonian Monte Carlo (HMC) techniques remain the leading methods for high-dimensional exploration \citep{Neal12}.  HMC techniques follow the trajectory of a massive particle in a gravitational potential given by $U(\mx) = -\ln \mathcal{L}(\mx)$, typically with a specific kinetic energy of $T = \frac{1}{2}|\mv|^2$ (defining $\mv\equiv \dot{\mx}$), thus leading to an acceleration of $\dot{\mv} = -\nabla U$.  While HMC and related techniques can work well for \modified{smaller} neural networks (\modified{$<10^6$ parameters}) with exact likelihood gradients \modified{\citep{Izmailov21,Sommer25}}, using stochastic gradients results in path heating that must be counteracted by an additional cooling term \citep{Chen14}.  Because the correct cooling term can be a complex function of the likelihood, it can be difficult to implement stochastic \modified{gradient} HMC in practice for Bayesian sampling of neural networks.

In this paper, we derive a simple and physically motivated alternative, which is to integrate the paths of light rays through a space where the refractive index varies as $n(\mx) = \mathcal{L}(\mx)^{1/(D-1)}$, which we show to result in fair sampling.  We implement light ray propagation with a constant speed through parameter space, which leads to very high resilience to heating from stochastic gradients.  This offers enormous performance advantages compared to traditional HMC \modified{as well as stochastic gradient HMC}, allowing us to perform approximate Bayesian sampling of neural networks on consumer hardware.  

\modified{The simplest ray tracing sampler has identical dynamics as an independently-derived gradient-based method (isokinetic sampling) that predates HMC \citep{Evans83,Tuckerman01}, which has also been rediscovered independently as microcanonical HMC \citep{Robnik23}.  However, the connection we make to Hamiltonian optics in this paper means that the framework derived here is more general than past efforts.  For example,} the \modified{ray tracing} method can also cross arbitrary disallowed regions in parameter space, which is not possible with past \modified{path-based} methods.  \modified{We also show that the generalized ray tracing} algorithm results in exact \modified{Metropolis-adjusted} sampling \modified{for arbitrary weighting and velocity schemes.  This allows us to rewrite}  
traditional HMC, along with many other past approaches, in terms of \modified{generalized} ray tracing samplers, providing a unified theoretical framework for interpreting local sampling methods and their relative efficiency at sampling high-likelihood peaks vs.\ exploring low-likelihood tails. 

\modified{Last, the ray tracing framework provides a simple way to derive the differential Jacobian of the transition function for arbitrary path-based samplers.  This results in a straightforward mapping between sampler dynamics (i.e., how a sampler changes direction in response to the likelihood gradient), the oversampling rate (i.e., how much the sampler overweights similar paths compared to the likelihood function), and the necessary path speed (or weight) required for fair sampling.  We use this connection to show the equivalence of sampling dynamics for HMC and isokinetic sampling for most high-dimensional problems with non-stochastic likelihoods, and derive new families of constant-speed samplers with unique dynamics compared to past methods.}

In Section \ref{s:methods}, we \modified{derive} the ray tracing \modified{framework} and connect it to past approaches, show example applications to Gaussian distributions and deep neural networks in Section \ref{s:applications}, and provide discussion and conclusions in Sections \ref{s:discussion} and \ref{s:conclusions}, respectively.

\begin{figure*}
\vspace{-9ex}
\capstart
\centering\includegraphics[width=0.8\textwidth]{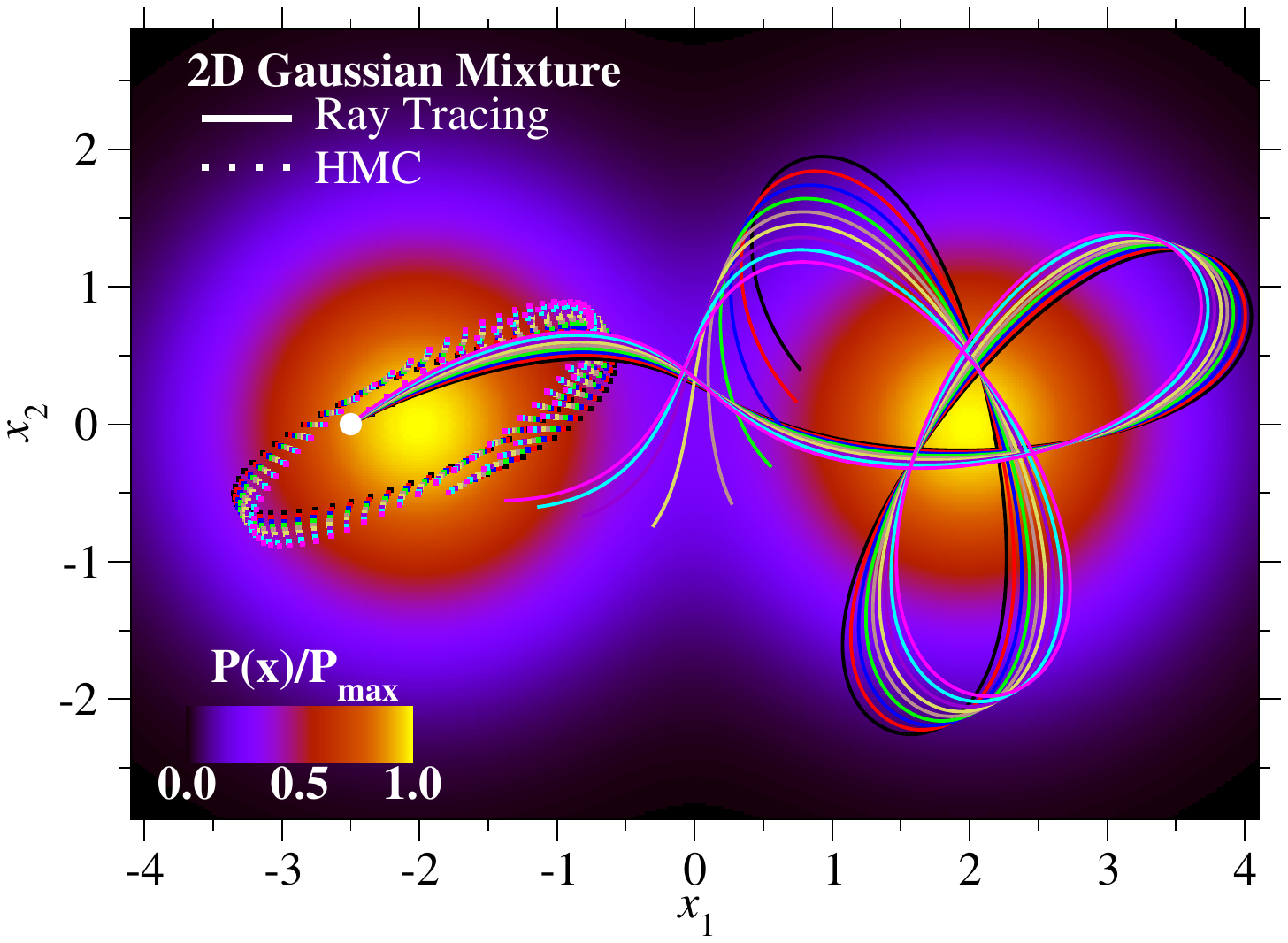}\\[-5ex]
    \caption{Example sampling dynamics for ray tracing and Hamiltonian Monte Carlo (HMC) for a 2D Gaussian mixture, with $\mathcal{L}(\mx)=[\exp(-0.5(x_1-2)^2)+\exp(-0.5(x_1+2)^2)]\exp(-0.5x_2^2)$.  For this case, as well as any other 2D distribution, fair sampling with ray tracing occurs when the spatially-varying refractive index is the same as the likelihood function (i.e., $n(\mx)=\mathcal{L}(\mx)$).  This figure illustrates the bending of light toward higher-probability regions due to Snell's law for a ray bundle originating from the white dot in the left of the figure.  Notably, the rays can explore a wide range of likelihood values.  In contrast, particle orbits starting from the same initial conditions using HMC are limited by the initial kinetic energy, chosen for this illustration to be the mean kinetic energy.  \modified{Eventually, HMC will sample a high enough initial kinetic energy to overcome the potential barrier between the two modes.} The background color scale corresponds to the underlying probability distribution function, scaled by its maximum value.}
    \label{f:overview}
    \vspace{1ex}
\end{figure*}

\section{Methods}

\label{s:methods}

\subsection{Overview}

\label{s:overview}

Here, we provide intuition for how ray tracing can sample fairly from likelihood distributions, with the formal mathematical derivation provided in Sections \ref{s:radiance}--\ref{s:generalized}.  

When light passes through the interface between two media, it is refracted according to Snell's law:
\begin{equation}
    n \sin(\theta) = \textit{constant},
\end{equation}
where $n$ is the refractive index and $\theta$ is the angle with respect to the interface normal vector.  This equation implies that light traveling from a lower-index medium to a higher-index medium will reorient to be closer to the interface normal vector.  Second, the mathematical form of Snell's law results in light being focused from a larger range of input solid angles in the lower-index medium to a smaller range of output solid angles in the higher-index medium, causing an overall radiance increase within the higher-index medium.

If we create a space in which the refractive index $n(\mx)$ is an increasing function of the likelihood $\mathcal{L}(\mx)$, Snell's law provides behavior useful for sampling.  For example, if we start from a low-likelihood (low-index) region, we would want our path to reorient closer to the likelihood (refractive index) gradient, to explore higher-likelihood (higher-index) regions.  As well, we would want to sample higher-likelihood (higher-index) regions with higher probability.  Both of these behaviors occur due to the refracting and focusing effects of Snell's law, as shown for example in Fig.\ \ref{f:overview}.

\begin{figure}
\capstart
    \centering
    \includegraphics[width=\columnwidth]{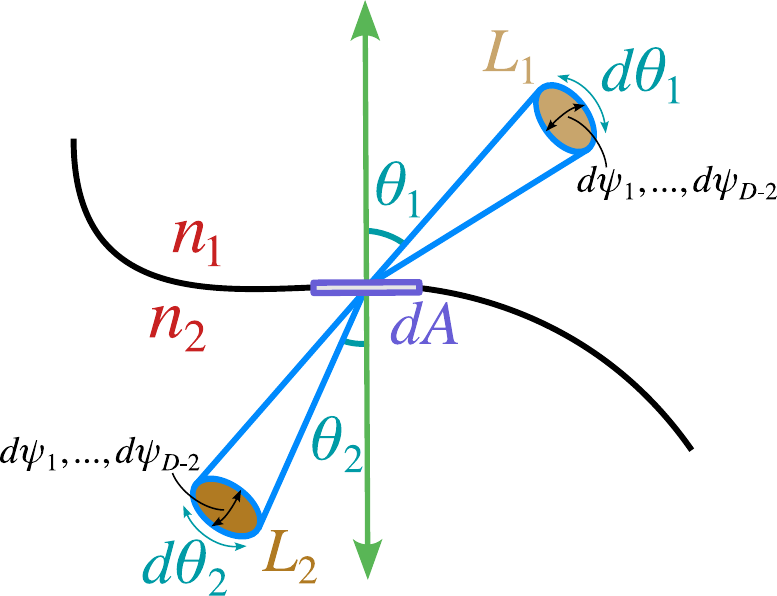}
    \caption{Refraction of light for an arbitrary dimensionality $D$.  This figure shows an incoming beam of light with radiance $L_1$ in a medium with refractive index $n_1$, crossing a differential area $dA$ at the interface with another medium with refractive index $n_2$, resulting in an outgoing beam of light with radiance $L_2$.  By Snell's law, $n_1 \sin \theta_1 = n_2 \sin \theta_2$, where $\theta_1$ and $\theta_2$ are the respective angles to the interface normal vector (\textit{in green}).  If $n_2$ is larger than $n_1$, the outgoing beam of light is compressed not only in the $\hat{\theta_2}$ direction (by Snell's law), but also along the directions perpendicular to $\hat{\theta}_2$, similar to how lines of longitude are compressed as one moves closer to the Earth's poles.  In addition, the outgoing beam becomes closer to the interface normal vector, increasing the projected area ($\cos\theta_2 dA$) from which it appears to originate. These two effects combine to result in an outgoing radiance $L_2$ that is a factor $(n_2/n_1)^{D-1}$ larger than the incoming radiance $L_1$ (Eq.\ \ref{e:radiance2}).}
    \label{f:lightbeam}
\end{figure}

\modified{One way of achieving fair sampling is for the sampling density to scale with the likelihood function (e.g., $|J|^{-1} \propto \mathcal{L}(\mx)$, with $J$ the Jacobian of the transition function) over a reversible path.  Ray tracing gives naturally reversible paths, and we show in Section \ref{s:sampler} that the sampling density (corresponding to the geometric density of ray paths) is the optical radiance $L$, defined as the} optical power per unit area per unit solid angle.  As we show in Section \ref{s:radiance}, Snell's law \modified{implies that radiance depends only on the index of refraction, not the path taken}:
\begin{equation}
    \mmodified{L(s) = L(\mx(s)) \propto n(\mx(s))^{D-1}},\label{e:radiance_evolution}
\end{equation}
where \modified{$s$ is the path length, so we can write $L(\mx)$ instead of $L(s)$ without loss of generality when Snell's law holds.   Using Eq.\ \ref{e:radiance_evolution}}, we can make an appropriate choice for the refractive index to enable fair sampling\modified{:}
\begin{equation}
    n(\mx) = \mathcal{L}(\mx)^{\frac{1}{D-1}}.
\end{equation}
With this choice, the radiance $L(\mx)$ \modified{(and hence the sampling density)} will automatically be proportional to the likelihood function $\mathcal{L}(\mx)$, so the ray paths will fairly sample the parameter-space posterior distribution.  \modified{Measuring the exactness of the equality in Eq.\ \ref{e:radiance_evolution} for an approximate integrator} also provides a simple test of the detailed balance of the system---similar to the energy conservation Metropolis test for Hamiltonian Monte Carlo systems---so that exact sampling can be \modified{achieved} even \modified{in this case}.

Finally, we briefly note that ergodicity is guaranteed as long as we introduce periodic scattering events that refresh the direction of travel, which results in ray diffusion throughout the parameter space.  Provided that parameter space boundaries are modeled as blackbody radiators, light paths can even reach parameter space islands, which is very difficult for other Hamiltonian-like techniques.  

In the next few sections, we derive the conservation of basic radiance and \'etendue for arbitrary dimensionality (Sections \ref{s:radiance} and \ref{s:etendue}), show how this leads to a fair sampling algorithm in Section \ref{s:sampler}, prove ergodicity in \ref{s:coverage}, discuss continuous momentum refreshment in  \ref{s:scattering}, generalize the sampler to arbitrary \modified{velocities and} sample weights in Section \ref{s:generalized}, provide pseudocode in Section \ref{s:pseudocode}, connect the sampler to prior work in Section \ref{s:prior}, \modified{discuss the connection between basic radiance and the oversampling rate in Section \ref{s:oversampling},} discuss adapting the sampling distribution in Section \ref{s:adaptive}, provide  specific recommendations for neural networks in Section \ref{s:practical}\modified{, and summarize the recipe for sampling posterior distributions in Section \ref{s:recipe}}.

\subsection{Conservation of basic radiance \modified{across interfaces} in \textit{D} dimensions}

\begin{figure}
\capstart
\centering
    \includegraphics[width=0.9\columnwidth]{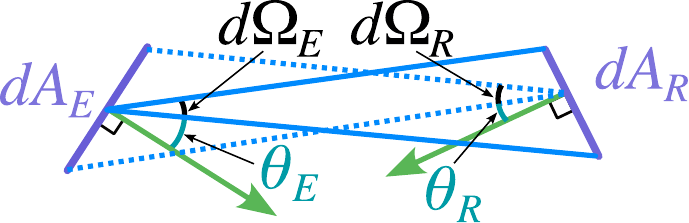}
    \caption{Parameters describing rays propagating from a differential emission area $dA_E$ to a differential receiving area $dA_R$.  The angles between the rays and the normal vectors of the emitter and receiver are $\theta_E$ and $\theta_R$, respectively.  The differential solid angle encompassed by the rays at the emitter is $d\Omega_E$, and that encompassed by the rays at the receiver is $d\Omega_R$.  As discussed in the text, the \'etendue at the emitter ($d^2G_E \equiv n_E^{D-1} dA_E \cos\theta_E d\Omega_E$) is equal to the \'etendue at the receiver ($d^2G_R\equiv n_R^{D-1} dA_R \cos\theta_R d\Omega_R$). }
    \label{f:etendue}
\end{figure}

\label{s:radiance}

In Hamiltonian optics, the canonical approach for dealing with light transport is to consider the evolution of a \textit{ray bundle}---that is, a group of light rays with differential cross section ($dA$) and differential solid angle ($d\Omega$).  We adopt this approach here, as it simplifies the derivation of the Jacobian (i.e., the compression or expansion of space) of the transition function and hence shortens the proof of stationarity.  In this section \modified{and the following one}, we show that \modified{Snell's law results in conservation of the basic radiance $R$, defined for arbitrary dimension $D$ as:}
\begin{equation}
\mmodified{R(s) \equiv n^{1-D}(\mx(s)) L(s)} \label{e:basic_radiance}.
\end{equation}

First, we consider a ray bundle (i.e., a collection of rays in a differential volume of position and direction phase space) crossing the interface between two transparent media with refractive indices $n_1$ and $n_2$.  This is shown in Fig.\ \ref{f:lightbeam} for $D>1$.  Within medium $i$, we define the angle with respect to the interface normal as $\theta_i$, and the opening angle of the bundle along the direction to the interface normal vector as $d\theta_i$.  We define $d\psi_{1},\ldots,d\psi_{D-2}$ to be the remaining opening angles in hyperspherical coordinates, which have no subscript $i$ as their values do not change across the interface \modified{by symmetry}.  The differential powers for the incoming/outgoing ray bundles are $d^2 P_i = L_i \cos(\theta_i)dA d\Omega_i$, where $L_i$ is the incoming or outgoing radiance, $\cos(\theta_i) dA$ is the differential projected hyper-area that the bundle hits, and $d\Omega_i = \sin^{D-2}(\theta_i)d\theta_i \sin^{D-3}(\psi_1)d\psi_1\sin^{D-4}(\psi_2)d\psi_2\ldots d\psi_{D-2}$ is the differential hyper-solid angle.

Because light (or, in our case, probability) cannot be created or destroyed at the interface, we find that $d^2 P_1 = d^2 P_2$.  Canceling like factors of $\sin^{D-3}(\psi_1)d\psi_{1},\ldots,d\psi_{D-2}$, we find:
\begin{equation}
L_1 \cos(\theta_1)d\theta_1 \sin^{D-2}(\theta_1) = L_2 \cos(\theta_2)d\theta_2 \sin^{D-2}(\theta_2). \label{e:radiance}
\end{equation}
By Snell's law, $n_1 \sin(\theta_1) = n_2 \sin(\theta_2)$, and differentiating with respect to $\theta_1$ or $\theta_2$ yields $n_1\cos(\theta_1)d\theta_1 = n_2\cos(\theta_2)d\theta_2$.  This gives the following useful substitutions:
\begin{equation}
    \frac{\sin(\theta_2)}{\sin(\theta_1)} = \frac{\cos(\theta_2)d\theta_2}{\cos(\theta_1)d\theta_1} = \frac{n_1}{n_2}.\label{e:differential_snell}
\end{equation}
As a result, Eq.\ \ref{e:radiance} reduces to:
\begin{equation}
    \frac{L_1}{L_2} = \left(\frac{n_1}{n_2}\right)^{D-1}, \label{e:radiance2}
\end{equation}
and we have proved the conservation of basic radiance \modified{across interfaces} for $D>1$.  For a one-dimensional space, there is no change in angle when moving from $n_1$ to $n_2$, so $L_1 = L_2$, and Eq.\ \ref{e:radiance2} also applies for $D=1$.  

In the derivation above, we neglected reflection.  This is necessary to consider for abrupt transitions of the refractive index, but in this paper, we are concerned with smoothly-varying refractive indices.  As one takes the limit of an increasing number $k$ of layers that linearly interpolate between an initial refractive index $n_i$ and final refractive index $n_f$, it is straightforward to show that the Fresnel reflection coefficients all asymptote to 0---i.e., there are no reflections \modified{between layers}, as with graded-index lenses in three dimensions.  Similarly, the radiance at the final layer will be
\begin{equation}
    L_f = L_i \prod_{j=1}^k \left(\frac{n_i + (j-1)(n_f-n_i)/k}{n_i + j(n_f-n_i)/k}\right)^{D-1},
\end{equation}
which reduces to Eq.\ \ref{e:radiance2} regardless of the number of layers.  As a result, Eq.\ \ref{e:radiance2} holds for all smoothly-varying refractive indices in arbitrary dimensions $D$---i.e., the quantity $\mmodified{R=\,} Ln^{1-D}$ is invariant.

\subsection{Conservation of \modified{basic radiance and of \'etendue in arbitrary media} in \textit{D} dimensions}

\label{s:etendue}

In Hamiltonian mechanics, Liouville's theorem guarantees invariance of phase-space volume, and hence a unit Jacobian for transitions.  In Hamiltonian optics, however, the Jacobian is \textit{not} unity, and instead there is another conserved quantity called \textit{\'etendue}, which is analogous to the phase-space volume of a ray bundle.

The derivation of Eq.\ \ref{e:radiance2} suggests that the following quantity will be conserved, which we will define as the \'etendue $\mmodified{d^2G}$ for $D$ dimensions:
\begin{equation}
    d^2G \equiv n\mmodified{(\mx)}^{D-1} dA \cos \theta d\Omega,\label{e:etendue}
\end{equation}
where $d^2G$ refers to the scaled phase-space volume of a ray bundle (as in Fig.\ \ref{f:lightbeam}), encompassing rays within a solid angle $d\Omega$ that are crossing a surface of area $dA$ at an angle $\theta$ to the surface normal, embedded in a medium with refractive index $n$.  Indeed, recognizing that basic radiance is connected to \'etendue by $R = \frac{d^2 P}{d^2 G}$, basic radiance will be conserved if \modified{and only if} \'etendue is conserved along \modified{nondissipative} ray path\modified{s}.

The standard conservation proof (in three dimensions) for the \'etendue $d^2 G$ generalizes very simply.  Consider the rays that start on a differential area $dA_E$ of an emitter and fall on a differential area $dA_R$ of a receiver (Fig.\ \ref{f:etendue}) in a medium of uniform refractive index $n$.  At the location of the emitter, the \'etendue will be $d^2G_E = n^{D-1} dA_E \cos \theta_E d\Omega_E$, where $\theta_E$ is the angle between the emitted rays and the surface normal, and $d\Omega_E$ is the differential solid angle of the rays that land on $dA_R$.  Since the refractive index is uniform, light will travel in straight lines, and so $d\Omega_E$ will correspond to the projected area of the receiver divided by the area of a hypersphere with radius equal to the distance $r$ between the source and the receiver:
\begin{equation}
    d^2G_E = n^{D-1} dA_E \cos \theta_E dA_R\cos\theta_R H_D^{-1} r^{1-D},
\end{equation}
where $\theta_R$ is the angle between the surface normal of $dA_R$ and the incoming rays, and $H_D r^{D-1}$ is the surface area of the hypersphere.  Likewise, the \'etendue at the receiver will be $d^2G_R = n^{D-1} dA_R \cos \theta_R d\Omega_R$, where $d\Omega_R$ is the differential solid angle of the incoming rays originating from $dA_E$.  As before, we relate $d\Omega_R$ to the projected area of the emitter as $d\Omega_R = dA_E\cos\theta_E H_D^{-1}r^{1-D}$, so:
\begin{equation}
    d^2G_R = n^{D-1} dA_R \cos \theta_R dA_E \cos \theta_E H_D^{-1} r^{1-D},
\end{equation}
and so $d^2G_E = d^2G_R$ by inspection.  Intuitively, as a ray bundle propagates, it will spread to a larger projected area ($dA_R\cos\theta_R$), but the apparent solid angle subtended by the emitting source ($d\Omega_R$) will become smaller and smaller, so that the product $d^2G$ remains constant.

Extending this to media with varying refractive indices is straightforward.  Considering a ray bundle passing through the boundary between two different media (Fig.\ \ref{f:lightbeam}), the incoming \'etendue is $d^2G_1 = n_1^{D-1}dA\cos\theta_1 d\Omega_1$, and the outgoing \'etendue is $d^2G_2 = n_2^{D-1}dA\cos\theta_2 d\Omega_2$.  We hence have:
\begin{eqnarray}
 \frac{d^2G_1}{d^2G_2} & = & \left(\frac{n_1}{n_2}\right)^{D-1} \frac{\cos(\theta_1)dAd\Omega_1}{\cos(\theta_2)dAd\Omega_2}\nonumber\\
 & = & \left(\frac{n_1}{n_2}\right)^{D-1} \frac{d^2P_1}{d^2P_2}\frac{L_2}{L_1}\nonumber\\
 & = & 1,
\end{eqnarray}
where $d^2P_1=d^2P_2$ are the incoming and outgoing differential powers as defined above.  Hence, \'etendue is conserved across media boundaries.  To extend this to media with continuously varying refractive indices, we can again approximate light paths using a discretized medium---i.e., wherein the refractive index is constant over a small volume.  Regardless of how finely we discretize, the \'etendue will not change, and \modified{thus} in the limit of infinitesimally small discretization, \'etendue \modified{as well as basic radiance are} conserved.

\subsection{The ray tracing sampler}

\label{s:sampler}

\modified{Conservation of \'etendue} implies that the phase-space volume of a ray bundle is compressed by a factor of $n\mmodified{(\mx)}^{D-1}$ depending on the refractive index.  We can use this to our advantage by linking the refractive index to the target likelihood distribution $\mathcal{L}(\mx)$, so that ray paths cluster in regions of higher likelihood.  The most natural way to do so is:
\begin{equation}
    n(\mx) = \mathcal{L}(\mx)^\frac{1}{D-1}, \label{e:index}
\end{equation}
so that the ray bundle phase-space density (i.e., the radiance $L$) is exactly proportional to $\mathcal{L}(\mx)$.  That is, the Jacobian of a ray tracing transition between two points under Eq.\ \ref{e:index} exactly cancels out the difference in probability between the two points.

\modified{We can show this more formally by leveraging} the results of the previous sections to \modified{prove} detailed balance.  Consider a differential volume $ds dA_\mx$ located at a position $\mx$, where $ds$ is the differential distance along the ray path and $dA_\mx$ is the differential cross-sectional area that is perpendicular to the direction of travel \modified{(so $\cos\theta_\mx = 1$)}.  Now suppose that we emit light rays from this volume in uniform directions with an intensity proportional to the likelihood $\mathcal{L}(\mx)$.  Formally, the radiant intensity density (power per volume per solid angle) emitted by this volume will be:
\begin{equation}
    \frac{d^3 P_\mx}{\mmodified{\cos \theta_\mx} d A_\mx ds d\Omega_\mx} = C \mathcal{L}(\mx), \label{e:untransformed}
\end{equation}
where $C$ is the constant of proportionality, which by definition does not depend on $\mx$ or the solid angle $\Omega_\mx$.

\modified{A ray bundle that travels a total path length $s$ will reach a parallelepiped} $\mmodified{ds \wedge dA_\mathbf{y}}$\modified{, which has total volume} $\mmodified{ds dA_\mathbf{y}\cos \theta_\mathbf{y}}$, \modified{where} $dA_\mathbf{y}$ \modified{is the wavefront area, and $dA_\mathbf{y}\cos \theta_\mathbf{y}$ is the cross-sectional area} perpendicular to the direction of travel.  This formulation explicitly preserves the extent along the direction of travel, so $ds$ is invariant.\footnote{To avoid potential confusion: as we are propagating the paths of ray bundles instead of photons, there is no explicit speed of light.  If we were tracking individual photons, they would travel slower in regions with higher refractive indices, but then the distances between photons would decrease correspondingly, canceling the effect of the photon speed on the luminosity.}  As the \'etendue $d^2G$ is preserved, we find:
\begin{equation}
    d^2G = n(\mx)^{D-1} \mmodified{\cos \theta_\mx}dA_\mx d\Omega_{\mx} = n(\mathbf{y})^{D-1} \mmodified{\cos \theta_\mathbf{y}}dA_\mathbf{y}d\Omega_{\mathbf{y}}. 
\end{equation}
Hence:
\begin{equation}
   \frac{d^3 P_\mx}{\mmodified{\cos \theta_\mathbf{y}}d A_\mathbf{y}ds d\Omega_\mathbf{y}} = C \frac{n(\mathbf{y})^{D-1}}{n(\mx)^{D-1}} \mathcal{L}(\mx) = C \mathcal{L}(\mathbf{y}).
\end{equation}
  That is, the differential power reaching the volume around $\mathbf{y}$ is \textit{non-uniformly} distributed in solid angle and cross-section in exactly the right way to balance the difference in likelihoods between the two points.  Because light rays propagate along reversible paths, the differential power emitted uniformly at $\mathbf{y}$ that reaches the differential volume $dsdA_\mx$ at $\mx$ will be non-uniformly distributed in the exact opposite way, ensuring detailed balance.

All practical ray tracing algorithms will of course introduce path errors, such that the product of the phase-space density and the likelihood will not be perfectly conserved.  Similar to HMC samplers, however, we can introduce an acceptance criterion based on the conservation of basic radiance.  As typically implemented, ray tracing algorithms update the angle of propagation of the ray path relative to the likelihood gradient by an amount $\mmodified{\Delta \theta_i}$ at each timestep $t_i$.  According to Eqs.\ \ref{e:radiance} and \ref{e:differential_snell}, the boost in radiance will then be:
\begin{equation}
    \frac{L_{i+1}\mmodified{(\mx_{i+1})}}{L_i\mmodified{(\mathbf{x}_i)}} = \left(\frac{\sin(\mmodified{\theta_i})}{\sin(\mmodified{\theta_i+\Delta \theta_i})}\right)^{D-1},\label{e:radiance_boost}
\end{equation}
so the net radiance boost $L(\mathbf{y})/L(\mx)$ between the starting and ending positions can be calculated as the product of the individual boosts.  This can then be compared to the actual change in likelihood, resulting in the following acceptance probability:
\begin{equation}
P(\mx\rightarrow\mathbf{y}) = \min\left(1, \frac{\mathcal{L}(\mathbf{y})L(\mx)}{\mathcal{L}(\mx)L(\mathbf{y})}\right).\label{e:acceptance}
\end{equation}

We note for completeness that ray tracing sampling is only meaningful in $D>1$ dimensions, as there are no degrees of freedom for the direction of propagation (and hence sampling density) in $D=1$, so Eq.\ \ref{e:index} becomes undefined.  In such cases, we recommend simply integrating the likelihood function over the single free parameter or using an alternate approach in Section \ref{s:prior}.

\subsection{Ergodicity}

\label{s:coverage}

Ergodicity requires that the direction of travel be refreshed \modified{from time to time}, for example, after a randomly chosen path length $s$.  This allows for full exploration of the direction space $\mv$ at any given location $\mx$.  Now, suppose that the reachable positions $\mx$ are a bounded set $B$.  By definition, we can choose $\mx$ to be a point at most an arbitrarily small distance $\epsilon$ away from the boundary of $B$.  Ray tracing involves a one-to-one mapping of initial directions to final directions because it is a reversible transformation (even for discrete integration, as in Eq.\ \ref{e:reversible_propagation} in Section \ref{s:pseudocode}).  Hence, an initial isotropic distribution for $\mv$ will be transformed into a non-isotropic distribution of outgoing $\mv$; however, all directions will be represented, including directions perpendicular to the boundary of $B$.  Hence, if the step size $\Delta s$ is equal to $\epsilon+\delta$ for some appropriately small $\delta$, then a single step of the ray tracing algorithm will reach a point outside the boundary $B$.  Provided that this point has a nonzero likelihood, then it will be reachable by the algorithm, and so the boundary for $B$ can only exist if the likelihood is identically zero at all points along the boundary---that is, if there are parameter space islands.

Parameter space islands are a special challenge for most Hamiltonian methods, since they cannot be reached by integrating the equations of motion.  However, for ray tracing methods, we can treat parameter space boundaries as black-body radiators that absorb ray bundles from arbitrary directions and then re-radiate them in a new direction according to black-body emissivity (i.e., where the emission probability is proportional to the product of $n^{D-1}$ and the cosine of the angle to the surface normal, also known as Lambertian emissivity).  Treating the non-allowed regions of parameter space as a medium with a uniform refractive index, this allows ray bundles to propagate between parameter space boundaries and be re-radiated back into allowed regions.  Noting that computing the full surface normal of a parameter space boundary may be prohibitive if the number of dimensions is large, Appendix \ref{a:lambertian} provides a simple algorithm for black-body emission that only requires \modified{measuring} the location of the parameter space boundary along two randomly chosen orthogonal directions.  

\subsection{Partial momentum refreshment, i.e., scattering}

\label{s:scattering}

Choosing a momentum refreshment rate involves a trade-off between exploring the parameter space along a fixed ray path (infrequent refreshes) vs.\ different ray paths (more frequent refreshes).  For problems where the typical sets (i.e., the regions with typical values of $\ln\mathcal{L}$; see also \citealt{Betancourt17}) are long, thin tubes (e.g., neural networks), it can be advantageous to have long timescales between momentum refreshes.  However, when momentum refreshes do occur, they can be disruptive to exploration rates and require short timesteps to avoid leaving the typical set.  Performing many partial momentum refreshes, analogous to an optical scattering process, provides an appealing alternative because step sizes can be kept high even as the momentum becomes uncorrelated over large time scales.

We can treat scattering/partial momentum refresh as a random rotation of the ray direction, which ensures that the Jacobian is unchanged.  However, some care is necessary for \textit{ when} the scattering is performed.  It is best to perform scattering at locations that would in principle be on the typical set, i.e., in between completed integration steps.  Performing scattering inside the integration step (e.g., drift--scatter--kick--scatter--drift), while allowed, may nonetheless lead to increased rejections, because the scattering may occur at locations outside of the typical set, and hence correspond physically to heating or cooling of the trajectories.  As discussed in \cite{Neal12}, if the direction of the momentum vector is not refreshed between Metropolis tests, it should be reversed upon Metropolis failure to maintain reversibility.

\subsection{Generalized ray tracing with sample weights}

\label{s:generalized}

The ray tracing method discussed so far has involved following paths at constant speed.  We can generalize the method by allowing the speed to change along the path.  Higher speeds $\mmodified{v}$ result in a lower sampling density per unit integration time, so the implicit sample weight $\mmodified{W_i}$ is related to the \modified{speed} as $W_i=v^{-1}$.  Alternately, we could retain a constant path speed and simply have an explicit weight $W$ applied to each sample.  Neither the sample weighting nor the speed affects the conservation of basic radiance (which is instead a statement about the geometric density of ray bundles) \modified{or the path that the sampler takes for a given $n(\mx)$}.  Instead, variable sample weighting and speed allow the sampler to be tuned so that it spends more time in lower- or higher-likelihood regions, depending on user needs.  As discussed in Section \ref{s:prior}, we can also use weighting to describe a large variety of past sampling methods within the context of ray tracing.

For an arbitrary explicit weight $W$ and path speed $v$, the sample density will scale as the product of the radiance $L$ and the weight $W$, divided by the speed $v$.  Letting $L(\mx)W(\mx)/v(\mx) = \mathcal{L}(\mx)$ to preserve fair sampling, this is equivalent to rescaling $\mathcal{L}(\mx)$ by $W(\mx)/v(\mx)$ and so we find that the refractive index is:
\begin{equation}
    n(\mx) = \left(\frac{\mathcal{L}(\mx)v(\mx)}{W(\mx)}\right)^{\frac{1}{D-1}}.\label{e:weighted_index}
\end{equation}
\vspace{1ex}
  The probability conservation test simply generalizes from Eq.\ \ref{e:basic_radiance}:
\begin{equation}
    P(\mx\rightarrow\mathbf{y}) = \min\left(1, \frac{n(\mx)^{1-D}L(\mx)}{n(\mathbf{y})^{1-D}L(\mathbf{y})}\right),\label{e:general_acceptance}
\end{equation}
which corresponds to the inverse of the excess basic radiance at $\mathbf{y}$ compared to the starting point $\mx$, 
and so exact sampling can be achieved for arbitrary weighting/velocity schemes.

\subsection{Algorithm pseudocode}

\label{s:pseudocode}

The standard form of Snell's law used for integrating ray paths is given in textbooks such as \cite{bornbook}:
\begin{equation}
    \frac{d}{ds} \left(n(\mx)\frac{d\mx}{ds}\right) = \nabla_\mx n(\mx),\label{e:ray_traditional}
\end{equation}
where $s$ is the path length and the gradient has a subscript $\mx$ to emphasize that it is taken with respect to the parameter space instead of the path.  We can rewrite to solve for the change in the direction of travel, $\frac{d\mx}{ds}$, as:
\begin{equation}
\frac{d}{ds}\left(\frac{d\mx}{ds}\right) = \nabla_\mx \ln(n(\mx)) - \left(\frac{d\mx}{ds}\cdot \nabla_\mx \ln(n(\mx))\right) \frac{d\mx}{ds}.\label{e:differential}
\end{equation}
This formulation makes it clear that the dynamics depend only on the gradient of $\ln(n)$ (and hence only on the gradient of the log-likelihood), so that both the refractive index and the likelihood can be renormalized by a constant factor without affecting the dynamics.  As well, in Eq.\ \ref{e:differential}, the rate of change of direction depends only on the component of the log refractive index gradient that is perpendicular to the direction of travel.  That is, the magnitude of the right-hand side is $\sin \theta |\nabla_\mx \ln(n(\mx))|$, with $\theta$ being the angle between the direction of travel and the refractive index gradient.

When integrating Eq.\ \ref{e:differential}, some care is necessary to maintain the unit length of $\frac{d\mx}{ds}$.  Indeed, because $\frac{d\mx}{ds}$ has unit length, we can interpret $\frac{d}{ds}\left(\frac{d\mx}{ds}\right)$ as the rate of change in $\theta$:
\begin{equation}
    \frac{d\theta}{ds} = -\sin\theta |\nabla_\mx \ln(n(\mx))|.\label{e:propagation}
\end{equation}

This equation specifies how the refractive index influences the angle of propagation, and hence how the luminosity changes according to Eq.\ \ref{e:radiance_boost}.  For readers who are less familiar with Eq.\ \ref{e:ray_traditional}, Eq.\ \ref{e:propagation} can also be derived simply from the differential form of Snell's law ($\frac{dn}{ds}\sin\theta = -n \cos\theta \frac{d\theta}{ds}$), with the identity $\frac{dn}{ds} = \cos\theta |\nabla_\mx n|$.  The main subtlety of Eq.\ \ref{e:propagation} is that the definition of $\theta$ changes during ray propagation.  Hence, the correct approach is to treat Eq.\ \ref{e:propagation} as an update to $\frac{d\mx}{ds}$ (e.g., as a ``kick'' step of a numerical integration method), and then to use the resulting $\frac{d\mx}{ds}$ to propagate the ray (e.g., as a ``drift'' step).  To maintain reversibility, we can integrate Eq.\ \ref{e:propagation} over the appropriate step length $\Delta s$:
\begin{equation}
    \tan\left(\frac{\theta_f}{2}\right) = \tan\left(\frac{\theta_i}{2}\right) \exp(-\Delta s |\nabla_\mx\ln(n(\mx))|), \label{e:reversible_propagation}
\end{equation}
where $\theta_f$ is the final angle and $\theta_i$ is the initial angle.  This equation is explicitly symmetric under a reversal of the propagation direction ($\theta \to \theta+\pi$) and the labels $\theta_f \leftrightarrow \theta_i$, as \modified{follows} from the identity $\tan((\theta+\pi)/2) = -\cot(\theta/2)$.

\begin{algorithm}[H]
\begin{algorithmic}
\Procedure{GRTSample}{$\mx_0, D, N, \Delta s, f, \mathcal{L}(\mx)$, optional: $W(\mx):=1$}
\State $\mv\gets$ random Gaussian in $D$ dimensions
\State $\Delta \ln L\gets 0$
\State $\Delta s \gets \sqrt{D}\Delta s$\Comment{Recommended---rescales step size with dimensionality}
\For{$k \gets 0, \dots, N$}
\State $\mv'\gets$ random Gaussian in $D$ dimensions
\State $\mv\gets \exp(-|f|)\mv + \sqrt{1-\exp(-2|f|)}\mv'$\Comment{Partial momentum refresh}
\If{$k<N$}
\State$\mx_{k+0.5} \gets \mx_{k}+0.5 \Delta s \cdot \mv/|\mv|$ \Comment{Drift step}
\State $\nabla\ln n \gets \frac{\nabla\ln (\mathcal{L}(\mx)/W(\mx))}{D-1}|_{\mx=\mx_{k+0.5}}$ \Comment{Log refract. index gradient}
\State$(\mv,\Delta \ln L)  \gets $\textsc{UpdateV}$(\mv, \nabla\ln n, D, \Delta \ln L, \Delta s)$ \Comment{Kick step}
\State$\mx_{k+1} \gets \mx_{k+0.5}+0.5\Delta s \cdot \mv/|\mv|$ \Comment{Drift step}
\EndIf
\EndFor
\State $r \gets $ uniform random number in $(0,1)$
\If{$\ln(r)>  \ln \left(\frac{\mathcal{L}(\mx_N)W(\mx_0)}{\mathcal{L}(\mx_0)W(\mx_N)}\right) -\Delta\ln L$} \Comment{Metropolis test}
\State $\mx_N \gets \mx_0$
\EndIf
\State \textbf{return}($\mx_N, W(\mx_N)$) \Comment{Return new sample and sample weight}
\EndProcedure
\vspace{3ex}
\Procedure{UpdateV}{$\mv, \nabla\ln n, D, \Delta \ln L, \Delta s$}
\If{$|\nabla \ln n|=0$ \textbf{or} $|\mv|=0$}
\State \textbf{return}$(\mv, \Delta \ln L)$ \Comment{Cases with no change in direction}
\EndIf
\State $\mathbf{\hat{v}} \gets \mv/|\mv|$
\State $\mathbf{\hat{n}} \gets \nabla\ln n / |\nabla\ln n|$ \Comment{Direction of refractive index gradient}
\If{$|\mathbf{\hat{v}}\cdot\mathbf{\hat{n}}|=1$}
\State $\Delta \ln L \gets \Delta \ln L+\Delta s\cdot|\nabla \ln n|\cdot(\mathbf{\hat{v}}\cdot\mathbf{\hat{n}})$
\State \textbf{return}$(\mv, \Delta \ln L)$ \Comment{Velocity parallel to index gradient}
\EndIf

\State $\theta_i \gets 2\arctan\left(\frac{|\mathbf{\hat{v}}-\mathbf{\hat{n}}|}{|\mathbf{\hat{v}}+\mathbf{\hat{n}}|}\right)$ \Comment{Initial angle between ray path and index grad.}
\State $\theta_f \gets 2\arctan\left(\frac{|\mathbf{\hat{v}}-\mathbf{\hat{n}}|}{|\mathbf{\hat{v}}+\mathbf{\hat{n}}|\exp(|\nabla\ln n |\Delta s)}\right)$ \Comment{Final angle}
\State $f_v \gets \sin(\theta_f)/\sin(\theta_i)$
\State $f_n \gets \cos(\theta_f)-f_v\cdot\cos(\theta_i)$
\State $\mv \gets f_v \mv + f_n |\mv|\mathbf{\hat{n}}$ \Comment{Direction update, maintains velocity normalization}
\State $\Delta \ln L \gets \Delta \ln L + (1-D)\ln(f_v)$ \Comment{Luminosity change}
\State \textbf{return}$(\mv, \Delta \ln L)$
\EndProcedure
\end{algorithmic}
    \caption{Generalized ray tracing sampler}\label{al:rts}
\end{algorithm}

The general ray tracing algorithm with arbitrary weighting function $W(\mx)$ is shown as Algorithm \ref{al:rts} in pseudocode.  Notes include:
\begin{enumerate}
\item The algorithm is explicitly independent of velocity normalization and is explicitly time-reversible.  Although we show a traditional drift-kick-drift leapfrog integrator here, any symmetric integrator may be used with the \textsc{UpdateV} function for kick steps.  Ideally, momentum refreshes should be done in between integration steps to avoid path heating per Section \ref{s:scattering}.
\item At each integration step, the old velocity vector is rescaled by a factor $f_v=\sin(\theta_f)/\sin(\theta_i)$ and added to the new velocity vector, corresponding to a phase-space compression factor of $f_v^{1-D}$, i.e., $\sin^{D-1}(\theta_i)/\sin^{D-1}(\theta_f)$, exactly matching the radiance boost in Eq.\ \ref{e:radiance_boost}.
\item The angle between the velocity and the refractive index gradient could be computed more transparently as $\theta = \arccos(\mathbf{\hat{v}}\cdot \mathbf{\hat{n}})$, but the version shown is more numerically stable if the usual function (\texttt{atan2}) is used with the numerator and denominator of the fraction.
\item The sampler could be equivalently recast in an unweighted form (i.e., returning $x_N$ with uniform weight) for arbitrary $W(\mx)$ by symmetric updates to the speed instead of moving a constant distance $\Delta s$ (e.g., Appendix \ref{a:velocity}), or by setting the Metropolis acceptance probability to the ratio of weights of the starting and final positions.
\item The Metropolis criterion could also be equivalently recast as a sample weight or an inverse velocity, corresponding to $\exp(\Delta \ln \mathcal{L}-\Delta \ln L)$, where the delta is taken across all prior steps; this would then allow a 100\% acceptance rate.  \label{acceptance}
\item Sample weighting could be equivalently recast as sample \textit{masking} or \textit{discretization}, wherein the number of copies of a particular point $\mx$ in the chain is drawn from a Poisson distribution with the Metropolis expectation value.  This also allows a ``perfect'' acceptance rate, except that some points in the chain are skipped.  For HMC, this can be a poor choice, since once a trajectory diverges from constant energy, it usually does not recover.  However, for ray tracing, even catastrophic errors are recoverable, as the light path will be redirected towards higher-likelihood regions as in Fig.\ \ref{f:overview}.\label{masking}
\item \modified{The velocity update is not symplectic, since it involves compression/expansion of the phase-space element according to the difference in the likelihood.  Alternate methods for retaining symplectic integration are discussed in Appendix \ref{a:symplectic}, but we do not adopt those here, since the symplectic methods are much harder to implement in practice.}
\end{enumerate}
\modified{We remind readers that the optimal efficiency (often defined as the optimal effective sample size per amount of compute time) may occur for acceptance rates lower than 100\%, and so optimal efficiencies are usually only obtained after a parameter search (e.g., as in Appendix \ref{a:hmc}).}

\subsection{Integrating prior methods into the framework of generalized ray tracing}

\label{s:prior}

The framework of generalized ray tracing is convenient for interpreting existing samplers, since it allows solving for the \modified{rate of} path curvature \modified{$\frac{d}{ds}\left(\frac{d\theta}{ds}\right)$} for an arbitrary sample weighting (equivalently, the speed of propagation along the path) and vice versa.  \modified{T}his lets us simply compute sample weighting for samplers that are primarily defined based on their paths (e.g., HMC) for straightforward comparison with other methods.

\subsubsection{Constant-Speed Samplers}
\label{s:constant_speed}

\modified{After initial submission of this paper, we learned that the simplest ray tracing method (i.e., unweighted with constant speed, and no continuous momentum refreshment) had identical dynamics as isokinetic sampling \citep{Evans83,Tuckerman01}, which is also the same as the second version of microcanonical Hamiltonian sampling \citep{Robnik23}.}\footnote{\modified{The original version of microcanonical HMC, published in \cite{Robnik22}, used a slightly different velocity update that is equivalent to setting  $W(\mx) = \mathcal{L}(\mx)^{1/D}$ in the ray tracing framework.}  Of note, \cite{Robnik22} noted the connection between geodesics and sampling density, calling it an ``interesting open question'' as to whether it could be used to inform sampler design. This paper is one answer.}

\modified{Each new derivation nonetheless introduced new advancements over the previous approaches.  For example, \citet{Robnik23} introduced continuous momentum refreshment for constant-speed sampling, which is helpful for achieving ergodicity with unadjusted Metropolis runs.  This paper includes several advances including a method for crossing arbitrary holes in parameter space (Section \ref{s:coverage}).  However, we find the broader advance in this paper to be the generalized framework of ray tracing, which allows deriving entirely new families of samplers with arbitrary dynamics and/or weighting (Section \ref{s:generalized}; see further generalizations in Section \ref{s:oversampling}).  We continue to refer to the simplest ray tracing sampler as ``ray tracing'' in this paper,\footnote{\modified{Reflecting that the same dynamical equations used by all methods in Section \ref{s:constant_speed} have a long history of prior implementation in both human and computer-based ray tracing approaches.}} nonetheless recognizing that readers are free to use any preferred term to describe the method.}

\subsubsection{Hamiltonian and Langevin Monte Carlo Methods}

\label{s:hmc_connection}

\begin{figure}
\vspace{-2ex}
\capstart
\centering
    \includegraphics[width=0.47\columnwidth,trim=0 0 7cm 0]{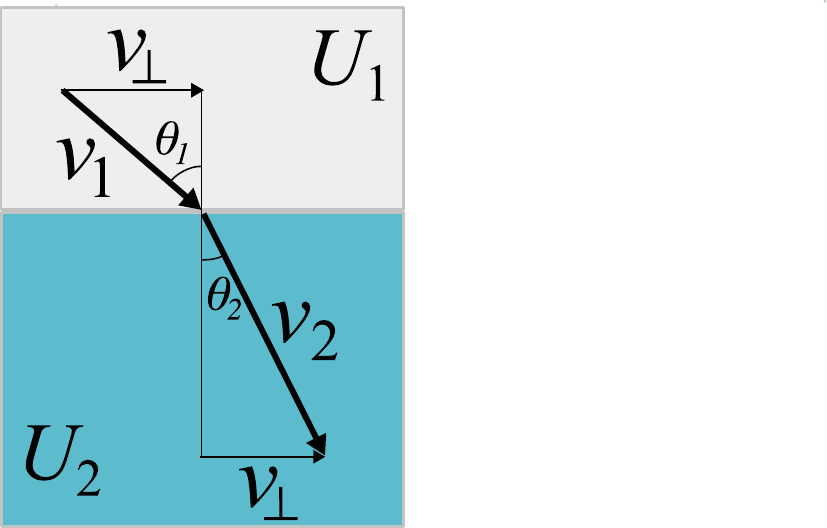}
    \caption{Path dynamics of Hamiltonian Monte Carlo.  A particle moving from potential $U_1$ to $U_2$ experiences a force normal to the potential boundary, causing it to accelerate.  The velocity component perpendicular to the force, $\mv_\perp$, is unaffected. Hence, we have the geometrical identity $|\mv_1| \sin\theta_1 = |\mv_\perp| = |\mv_2| \sin\theta_2$, and the path is identical to that of a light ray traveling in a medium with refractive index $n = |\mv|$.  Since the magnitude of $\mv$ is only a function of the potential energy for a given choice of total energy $E$, this results in a well-defined solution for $n$ for all particle paths at fixed total energy.}
    \label{f:hamil}
\end{figure}

We can reproduce traditional Hamiltonian dynamics \modified{naturally} using a \modified{generalized} ray tracing sampler.  The analogy to Fig.\ \ref{f:lightbeam} is a particle encountering a sudden jump in the potential energy $U = -\ln \mathcal{L}(\mx)$.  The change in specific kinetic energy, $T_2-T_1$, will be the opposite of the potential change, $U_2 - U_1$.  We can then solve for the change in direction due to the potential change by breaking the velocity into components.  By symmetry, the velocity $\mv_\perp$ perpendicular to the potential gradient $\nabla U$ will remain unchanged.  However, the velocity parallel to $\nabla U$ will be increased or decreased such that the direction of propagation becomes closer to the direction of the acceleration $-\nabla U$.  Hence, the geometry (Fig.\ \ref{f:hamil}) requires:
\begin{equation}
|\mv_1| \sin\theta_1 = |\mv_\perp| = |\mv_2| \sin\theta_2.
\end{equation}
This is immediately suggestive of the form of Snell's law, and the general solution for $n$ is:\footnote{It is tempting to expect $v\propto n^{-1}$, as for physical light rays.  However, the physics of particle vs.\ light transport are different---for example, the speed $v_\perp$ is not conserved for light rays propagating between two media.  The better analogy is with the Minkowski momentum of light, which indeed scales with $n$ as $p = h/\lambda = nh/\lambda_\mathrm{vac}$.}
\begin{equation}
 n(\mx) \propto |\mv(\mx)|  \label{e:hamil_n}
 \end{equation}
 by inspection, with the specific solution:
 \begin{equation}
      n(\mx) \propto |\mv(\mx)| \propto \sqrt{E+\ln \mathcal{L}(\mx)}, \label{e:hmc}
\end{equation}
for the usual specific kinetic energy ($T=\frac{1}{2}v^2$), with $E$ being the total specific energy.  This formulation as a ray tracing approach gives paths and weights (inverse velocities) that are identical to standard Hamiltonian Monte Carlo sampling.  For example, mechanical paths can only reach points that satisfy $E+\ln{L}(\mx)>0$.  This is also true of the ray paths: at positions $\mx$ where $\ln \mathcal{L}(\mx)=-E$, the gradient of the log refractive index is infinite, as $n=0$, corresponding to an impassible barrier.  Langevin Monte Carlo methods then follow, as they can be rewritten as single-step Hamiltonian Monte Carlo methods \citep{Girolami11}.  


The parameter-space sampling density of HMC is $L/v$, which for $n\propto|\mv|$ becomes $v^{D-2}$.  This results in a sampling bias at fixed energy, since some regions will be sampled preferentially compared to the likelihood function:
\begin{equation}
    \frac{L(\mx)/v(\mx)}{\mathcal{L}(\mx)} = \frac{(E+\ln\mathcal{L}(\mx))^{D/2-1}}{\mathcal{L}(\mx)}.
\end{equation}
This results in some interesting behavior in low-dimensional spaces.  Indeed, for $D=2$, HMC sampling at a single value of $E$ will yield a uniform(!) density, regardless of $\mathcal{L}(\mx)$, \modified{across the entire connected space} with $\ln \mathcal{L}(\mx)>-E$.  This is readily confirmed numerically by refreshing only the velocity direction but not the kinetic energy for an HMC sampler.

For high-dimensional spaces, the behavior is more regular.  We expect for a Gaussian-distributed velocity that $|v|\sim \sqrt{D}$, so typical kinetic energies (i.e., $E+\ln\mathcal{L}(\mx)$) will be $\sim D/2$.  Hence, we can reparameterize the bias equation in terms of a change in the likelihood $\Delta \ln \mathcal{L}$ away from the starting likelihood, giving:
\begin{equation}
    \frac{L(\mx)/v(\mx)}{\mathcal{L}(\mx)} \sim \left(\frac{D}{2}\right)^{D/2-1}\cdot\frac{(1+\frac{2}{D}\Delta\ln\mathcal{L}(\mx))^{D/2-1}}{\mathcal{L}(\mx)}. \label{e:hmc_asymp}
\end{equation}
In the limit of small $(\Delta \ln \mathcal{L})/D$, the rightmost fraction asymptotes to $\exp(\Delta\ln\mathcal{L})/\mathcal{L}(\mx)\propto 1$, resulting in sampling density proportional to the likelihood function, similar to the behavior of ray tracing.  \modified{This is confirmed with numerical tests in  Appendix \ref{a:hmc}.  The limit in Eq.\ \ref{e:hmc_asymp}} occurs after burn-in for large $D$, where the central limit theorem suggests that the typical set should have a log-likelihood width of $\Delta \ln \mathcal{L}\propto \sqrt{D}$.  However, during burn-in, $|\Delta \ln \mathcal{L}|$ can be large, with HMC undersampling more relative to the expectation for fair sampling (i.e., compared to ray tracing) unless the momentum is periodically refreshed.  This suggests that a very simple non-reversible method for approximating ray tracing with an existing HMC sampler is to resample the kinetic energy (but not the direction) at every integration step (following a Metropolis check), and to resample the velocity direction only after much longer trajectories.  Ray tracing, as implemented in Eq.\ \ref{e:reversible_propagation}, is effectively this energy resampling process taken to the continuous (and reversible) limit.

Alternate formulations of HMC that better explore multimodal regions (e.g., parallel tempering; see \citealt{Earl05} for a review) do so by reducing the velocity differences (and hence sample density differences) between high-- and low--probability regions, corresponding to reducing the refractive index gradient (see also Section \ref{s:tempering}).

\subsubsection{\modified{Affine-invariant ensemble} sampling}

The stretch step proposed in \cite{Goodman10} (and familiar as a common step in \citealt{emcee}) proposes a new position $\mx$ along a line connecting the current position $\mx_c$ and a target walker position $\mx_t$ from a complementary set of walkers.  This is equivalent to setting $n=\infty$ outside of a small cone with infinitesimally small solid angle $d\Omega$ originating from the target walker's position $\mx_t$ and including the current walker's position $\mx_c$.  Inside the cone, the sample weighting is given by the stretch step distribution size, namely $W(z) = \frac{1}{\sqrt{z}}$ for $z\in [\frac{1}{a}, a]$ and $0$ otherwise, with $a$ the usual step size parameter and $z(\mx) = |\mx-\mx_t|/|\mx_c-\mx_t|$.  The narrowing shape of the cone means that the net volume at a given $z$ would scale as $z^{D-1}dz$, correspondingly limiting the fraction of rays that would penetrate to low $z$, and matching the acceptance criterion of \cite{Goodman10}.  We could hence more simply recast a series of stretch steps with the same $\mx_t$ as 1D Gibbs sampling, with $\mathcal{L}(z) = z^{D-1}\mathcal{L}(\mx(z))$, and use the approach for Gibbs sampling in Section \ref{s:gibbs}.  

\subsubsection{Metropolis Sampling, Gibbs Sampling, and Monte Carlo Integration}

\label{s:gibbs}

Gibbs methods \citep{Geman84}, operate by sampling from the conditional distribution, often along single-parameter axes.  In the ray tracing framework, this corresponds to setting $W(\mx) =  \mathcal{L}(\mx)$ and $n=1$ everywhere.  As a result, rays travel in straight lines throughout parameter space, and the weights ensure sampling from the one-dimensional conditional distribution.  As expected, both energy and phase-space volume are always conserved with such ray paths (as $\nabla \ln n=0$), and so the probability conservation test in Eq.\ \ref{e:general_acceptance} is automatically satisfied.  

A Metropolis-Hastings \citep{Metropolis53,Hastings70} step corresponds to performing a single ray tracing step using the Euler method for integration, with size $\Delta s$ drawn from a desired distribution (typically Gaussian), with $n=1$ and $W(\mx) =  \mathcal{L}(\mx)$, and with the choice of unweighted samples based on the Metropolis criterion instead of a sample weight or a speed.

Last, Monte Carlo integrals can be thought of as Gibbs sampling with masked samples---that is, weights are set to 0 everywhere except for ray endpoints (i.e., where the direction of travel is refreshed), where they are set to $\mathcal{L}(\mx)$.

\subsubsection{Refractive Sampling}

While searching for prior work during paper preparation, we became aware of a not elsewhere published thesis chapter by Levi Boyles,\footnote{\url{https://www.proquest.com/docview/1564748120?pq-origsite=gscholar&fromopenview=true&sourcetype=Dissertations\%20&\%20Theses}} called ``Refractive Sampling,'' in which the authors used ray transport equations and derived similar acceptance criteria in 2014.  In their case, they adopted $\nabla \ln n = \frac{r}{\Delta s} \frac{\nabla \ln \mathcal{L}}{|\nabla \ln \mathcal{L}|}$, so that the direction of the refractive index gradient was tied to the direction of $\nabla \ln \mathcal{L}$, but not the magnitude, as they did not derive the conservation of basic radiance.  Instead, the authors evaluated using an effective refractive index ratio $r$ that was constant.  This corresponds to a strong overweighting of low-likelihood regions; unsurprisingly, the authors found that standard HMC performed substantially better for unimodal distributions, but that refractive sampling had the potential to better access multimodal distributions.

\subsubsection{Tempering}

\label{s:tempering}

Since generalized ray tracing separates the path dynamics from sample weights/velocities, it is straightforward to create fair samplers with path dynamics corresponding to a different effective temperature $\tau$---or even allowing the temperature to vary along the path.  For standard ray tracing dynamics, this is a simple substitution:
\begin{eqnarray}
    n & = & \mathcal{L}(\mx)^\frac{1}{\tau\cdot (D-1)}\\
    W & = & v^{-1} = \mathcal{L}(\mx)^\frac{\tau-1}{\tau}.
\end{eqnarray}
As expected, increasing the temperature corresponds to increasing the effective speed (or decreasing the effective weight) in unlikely regions, reflecting the fact that the paths will overpopulate regions with low likelihood.  We note that there is no issue with making $\tau$ a function of position or likelihood, to better explore long-tailed distributions or to better travel between modes of the distribution.  For example, setting $\tau\propto \ln \mathcal{L}$ is equivalent to Gibbs sampling, as follows from the identity $\ln(x^{1/\ln(x)})=1$.  In another example, if $\tau$ is a function of $\ln(\mathcal{L})$, then the spatial gradient of $\ln(n)$ used in Eq.\ \ref{e:reversible_propagation} becomes:
\begin{equation}
    \nabla_\mx \ln(n(\mx)) = \frac{\nabla_\mx \ln(\mathcal{L}(\mx))}{\tau\cdot (D-1)}\left[1 - \ln(\mathcal{L}(\mx))\frac{d\ln(\tau)}{d\ln(\mathcal{L}(\mx))}\right],
\end{equation}
which has the useful property that the direction of $\nabla_\mx \ln(n)$ is parallel (or anti-parallel) to the $\tau=1$ case.

The temperature $\tau$ could also be a function of time, provided that it is symmetric along the trajectory to maintain reversibility.  In this case, the effective weighting should be the product of weight ratios at constant temperature, as the weights are not conserved during temperature changes.  A disadvantage to this approach is that, by definition, the stationary distribution will change with time.  If the timescale for burn-in is longer than the timescale for temperature change, changing the temperature in this manner will lead to many rejected paths.

We note in passing that, despite the method above, HMC paths do not admit varying temperatures unless the starting and ending locations have the same likelihood.  This is because, for HMC, it is otherwise impossible to guarantee energy preservation at multiple temperatures simultaneously, leading to non-reversible paths \citep{Neal12}.

\subsection{\modified{Connecting Basic Radiance to the Oversampling Rate: Constructing Fair Samplers with New Dynamics}}

\label{s:oversampling}

\modified{In the foregoing sections, we treated basic radiance as a conserved quantity, and hinted at the connection to the sampling rate relative to the likelihood function.  Here, we make that connection explicit.  The determinant of the Jacobian of the transition as a function of path length $s$ is given by:}
\begin{equation}
    \mmodified{|J(s)| = \frac{v(s)}{L(s)}},\label{e:jacobian}
\end{equation}
\modified{where $L(s)$ is the radiance (Section \ref{s:radiance}) and $v(s)$ is the path speed (Section \ref{s:generalized}).  We can then write the general oversampling rate $\mathcal{R}$ (i.e., the ratio of the sampling rate to the likelihood) as:}
\begin{eqnarray}
    \mmodified{\mathcal{R}(s)}& \mmodified{=} &\mmodified{\frac{W(\mx(s))}{\mathcal{L}(\mx(s))|J(s)|}}   \nonumber\\
    & \mmodified{=} & \mmodified{R(s) \cdot \frac{n(\mx(s))^{D-1}}{\mathcal{L}(\mx(s))}\cdot \frac{W(\mx(s))}{ v(s)}}, \label{e:oversampling}
\end{eqnarray}
\modified{where $R(s)$ is the basic radiance (Section \ref{s:radiance}) and $W(\mx(s))$ is the weighting function (Section \ref{s:generalized}).  Up to this point, we have only considered systems that obey Snell's law, in which case $R(s)$ is constant, and the index of refraction $n(\mx)$ can be chosen to achieve fair sampling for any desired $W/v$ (Eq.\ \ref{e:weighted_index}).}

\modified{However, we can imagine a more general rule than Snell's law for updating the direction and/or velocity in response to the likelihood gradient. The derivation of Section \ref{s:radiance} results in a very simple rule for how the radiance changes in response to a change of angle with respect to the gradient, i.e., Eq.\ \ref{e:radiance_boost} for Snell's law.  For a more general direction update, taking the derivative of the \'etendue (Eq.\ \ref{e:etendue}) with respect to the path length $s$ provides an equally simple formula for generating a sampler for \textit{any} desired continuously variable oversampling rate $\mathcal{R}(s)$.}

\modified{It is helpful to show examples to demonstrate the power of this approach.  For example, one might ask whether there are any constant-speed samplers that have different dynamics than the isokinetic/MCHMC/simple ray tracing approach.  The answer is \textit{no} if one is limited to Snell's law.  However, the answer is \textit{yes} for a more general conservation law, as derived in Appendix \ref{a:further_generalizations}.  That is, there is a family of ``tumblers''---samplers that continuously rotate the path in the $\theta$ direction (i.e., the direction toward the likelihood gradient) at a rate that can be an arbitrary function of $\mathcal{L}(\mx)$.}

\modified{More practically, it is also simple to construct fair samplers that are allowed to leave the typical set between sampled locations (i.e., they are allowed to cross ``unfair'' locations on the way between two fair locations).  This is especially useful for multi-modal distributions as well as for high-dimensional distributions.  In the latter case, the typical set is often so narrow \citep{Betancourt17} that there is no freedom for universally fair samplers except to inch along a narrow sheet in parameter space.  Allowing the sampler to leave the typical set hence opens up a significant variety of new paths.  This type of sampler is derived as well in Appendix \ref{a:further_generalizations}.  Notably, this sampler has an advantage over previous multi-temperature samplers (as well as variable-temperature samplers) is that it does not need to tune a temperature distribution, nor does it need to ``burn in'' at any temperatures other than the desired $\tau=1$.  This makes a strong case that there are many more continuous-time gradient-based samplers than have been currently discovered that will prove useful for complex distributions.}

\subsection{Adaptive Sampling and Separability}

\label{s:adaptive}

Generically, all samplers benefit from adapting the transition function to the target distribution.  Existing approaches including adaptive covariance matrices as in  \cite{Haario01} and normalizing flow--enhanced approaches like \cite{Wong22} are hence expected to improve the performance of ray tracing.  

Adaptive covariance matrix approaches attempt to perform an affine transformation of the parameter space to remove parameter-space degeneracies.  For a $D$-dimensional independent and identically-distributed (i.i.d.) Gaussian distribution, typical ray paths will correspond to roughly circular orbits in the limit of large $D$.  Under an affine transformation, the distribution will become ellipsoidal.  Typical ray paths will still have an isotropic initial velocity distribution, resulting in helical paths along the ellipsoidal surface.  The greater the aspect ratio of the ellipsoid, the tighter the winding of the helix, and the more timesteps that will be necessary to find an independent sample.  Hence, in large-dimensional problems where measuring the full covariance matrix is intractable, the largest improvement will come from rescaling the space to lessen the dynamic range of the eigenvalues---i.e., targeting the largest and smallest principal components of the space.

With the above said, we do not attempt rescaling of the space for the neural network applications in this paper.  Because neural networks have a high degree of symmetry (e.g., adjacent nodes can be swapped without affecting the results of the network), renormalizing the space does not necessarily lead to improved sampling performance.  Instead, the filamentary nature of the likelihood space for neural networks means that there are highly local changes in the Hessian matrix that make a global rescaling less helpful.

We also note the related topic of separability.  Many samplers, including Hamiltonian Monte Carlo, Metropolis, and Gibbs sampling are \textit{separable}, in the sense that if the likelihood function can be written as the product over two independent subspaces ($\mx=(\mx_1,\mx_2)$ with $\mathcal{L}(\mx) = \mathcal{L}(\mx_1)\mathcal{L}(\mx_2)$), then the dynamics of exploration over the subspace $\mx_1$ are not affected by the dynamics of the exploration in the subspace $\mx_2$.  This is not generically true for ray tracing, since ray tracing dynamics have the constraint of constant \modified{speed}, which links the dynamics of any two subspaces together.  Hence, if the gradients are expected to be of significantly different magnitude for two different subspaces (e.g., different layers of a neural network), it may be advantageous to run ray tracing dynamics on each subspace independently---i.e., to have independent velocity vectors for each subspace and to update them in parallel according to Algorithm \ref{al:rts}.

\subsection{Practical Sampling of Neural Networks}

\label{s:practical}

High-dimensional problems often have large computational costs, necessitating some trade-offs to maximize sampling accuracy given the computing time available.  

 \subsubsection{Lack of a well-defined likelihood function}  
 \label{s:likelihood}
 This issue arises most often for regression tasks in which the network is fit to minimize a loss function (e.g., mean squared error; MSE), rather than to predict the distribution of the training data.  As the underlying label uncertainty is often assumed to be zero, the mean squared error cannot be interpreted as a $\chi^2$ or similar value in such cases.  However, one can still usefully ask about the posterior distribution of outputs for neural networks that perform within some tolerance $\Delta f_\mathrm{loss}$ of the best fit.  We can then construct a likelihood function that will lead to typical losses within $\Delta f_\mathrm{loss}$ of the best solution as:
 \begin{equation}
     \ln \mathcal{L}(\mx) \approx -f_\mathrm{loss}(\mx)\cdot(D_\mathrm{eff}/(2\Delta f_\mathrm{loss})), \label{e:approx_likelihood}
 \end{equation}
 where $D_\mathrm{eff}$ is the effective number of parameters, $f_\mathrm{loss}$ is the loss function, and $\Delta f_\mathrm{loss}$ is the desired tolerance.  This equation is exact for a Gaussian distribution, where the expectation value of the log likelihood should be $\sim D/2$ less than the best-fitting log likelihood value (equivalently, $\chi^2$ should be higher by $\sim D$).  Heuristically, the central limit theorem suggests that it should also be reasonable for other distributions in the limit of large numbers of dimensions, provided that at least a small fraction of dimensions are uncorrelated with each other.

 We note that $D_\mathrm{eff}$ will in practice only equal the total number of parameters if all the parameters are constrained by the data set.  Neural networks can have extensive symmetries and unused parameters, so the true effective parameter count can be significantly smaller (e.g., $3\%$ of the neural network parameter count in Section \ref{s:um}), and particularly so if the number of data points in the training set is much less than the parameter count (e.g., $0.06\%$ of the parameter count in Section \ref{s:resnet}).  Hence, some trial and error can be necessary to choose $D_\mathrm{eff}$ to achieve the desired loss tolerance.  \modified{The meaning of $D_\mathrm{eff}$ and a heuristic for choosing it are further discussed in Section \ref{s:deff}.}

\subsubsection{Lack of Convergence in Weight Space}

\label{s:function_space}
 
MCMC samplers are expected to converge asymptotically to the correct distribution for any function of the underlying parameter space, which for neural networks is also known as \textit{weight space}. However, for most machine learning applications, the distribution in weight space is not directly used, and instead the distribution in \textit{function space}---usually, the output of the neural network---is what is important.  Typically, posteriors in weight space for neural networks are narrow tubes or sheets, and so convergence in weight space is difficult \citep{Izmailov21}; but due to the many symmetries of neural networks, the effective convergence in function space is much more rapid.  \modified{For example, a well-known symmetry of ReLU neural networks is that output is unchanged when the input weights and biases of one layer are multiplied by a factor $y$, with the input weights and biases of the following layer multiplied by $y^{-1}$.  Hence, convergence in weight space without an additional prior is not possible, even as the function space is unaffected by the choice of normalization.}
 
\modified{For this reason,} it is common to apply a Gaussian prior in weight space of the form $\mathcal{L}\propto \exp(-\mx^2/\alpha^2)$ (e.g., \citealt{Izmailov21}).  An advantage of this is that the posterior distribution in weight space will be bounded and finite, which can improve convergence in weight space.  A disadvantage is that the choice of $\alpha$ is usually arbitrary, and if $\alpha$ is chosen to be too small, it can alter the meaning of the effective number of parameters $D_\mathrm{eff}$.  Hence, if using a weight space prior, it is typically necessary to try multiple values of $\alpha$ to make sure that the choice of $\alpha$ does not affect the distribution of the posterior in function space.  If not using a weight space prior (the approach we adopt here), the advantage is that a range of effective $\alpha$ is probed automatically, with the disadvantage that the weight space distribution may never converge.

\subsubsection{Stochastic Likelihood Sampling}

\label{s:stochastic}

Training neural networks is greatly accelerated by splitting the training data into multiple batches and using the loss gradients of individual batches to perform successive fitting steps instead of calculating the loss gradient of the entire training sample at each step.  Depending on batch sizes, this can speed up the rate of convergence by orders of magnitude, especially at early times when the network is far from a good solution.

Stochastic sampling of the likelihood gradient has been explored for traditional HMC in \cite{Chen14}, which found that a friction term is necessary to counterbalance the effective heating of using an imperfect gradient estimator.  \cite{Chen14} argued that such an approach was critical, since for arbitrarily long orbits, the amount of energy injected by stochastic sampling could be arbitrarily large, leading to arbitrarily bad mismatches between the sampling and the target distribution.  On the other hand, it is not always easy to estimate the correct noise caused by stochastic sampling, including the noise covariance, so it is difficult to accurately cancel the impact of stochastic sampling for HMC.

With ray tracing, however, the energy is explicitly preserved, even in the ray tracing formulation of Hamiltonian Monte Carlo (Section \ref{s:prior}).  The impact of stochasticity on the dynamics is instead limited to perturbations in the direction of travel.  To first order, this is in fact desired behavior(!), given that, in Section \ref{s:coverage}, we had to explicitly add continuous partial momentum refreshment to achieve ergodicity.

Of course, what we lose with stochastic sampling is perfect knowledge of the likelihood function.  The results are errors in calculating the Metropolis acceptance criterion in Eq.\ \ref{e:general_acceptance}, which result in 1) increased rejections, and 2) broadening of the typical set. 
To reduce the problem of increased rejections, we can interpret the stochastic likelihood as the true likelihood convolved with an aleatoric uncertainty.  What matters then is that the error in path integration remains subdominant to the uncertainty in calculating the likelihood.  We can approximate this with a stochastic acceptance criterion as:
\begin{equation}
    P(\mx\to\mathbf{y}) = \min\left(1, \left(\frac{n(\mx)^{1-D} L(\mx)}{n(\mathbf{y})^{1-D} L(\mathbf{y})}\right)^\frac{1}{\sqrt{1+\sigma_\mathrm{sto}^2}}\right), \label{e:batch_acceptance}
\end{equation}
where $\sigma^2_\mathrm{sto}$ is the variance in $\ln(\mathcal{L}_\mathrm{batch}/\mathcal{L}_\mathrm{true})$.  This criterion reflects the fact that we have imperfect information about whether the path is erroneously integrated or whether the likelihood was erroneously calculated.  Notably, because it is an acceptance criterion and does not otherwise affect the dynamics of the system, it will remove paths that deviate badly from the expected trajectories, but it will not lead to runaway growth of the loss function or to log sampling rates that deviate by $\gg \sigma_\mathrm{sto}$ from the true values.  In principle, step sizes should be lowered until acceptance rates per Eq.\ \ref{e:batch_acceptance} stop improving significantly.

With the above said, ray tracing performs so well with suppressing heating from stochastic gradients that it is easy to have $\sigma_\mathrm{sto} \gg \sigma_{\ln(\mathcal{L}_\mathrm{true})}$ while still achieving low error on the resulting true likelihood distribution (as in Section \ref{s:gaussian_sg}).  In such cases, Eq.\ \ref{e:batch_acceptance} loses meaning as a test for acceptance of individual MCMC steps, since it will simply randomly discard valid samples from the chain.  A simpler alternative in such cases is to set a likelihood threshold (e.g., $\pm 2\sigma_\mathrm{sto}$) for individual steps to reject badly deviating paths, and then to verify that decreasing the step size does not change any moments of the distribution.

\subsection{Sampling Recipe}
\label{s:recipe}

We have found the following steps helpful for sampling the neural networks described in Sections \ref{s:um}--\ref{s:1B}:
\begin{enumerate}
    \item Perform burn-in of a single walker at a low temperature (or use an optimizer like Adam) to find a location near the typical set.  For HMC burn in, frequent momentum refreshes are needed to remove excess kinetic energy.
    \item If necessary, choose the likelihood scaling in Eq.\ \ref{e:approx_likelihood} according to the desired loss tolerance $\Delta f_\mathrm{loss}$.
    \item Adjust the step size, starting with a reasonable guess (e.g., $\Delta s\sim 0.03 D^{1/2}$), and running until the loss distribution stabilizes.  Try reducing the step size until the averaged loss (or inverse log-likelihood) per epoch stops decreasing.  It may be necessary to adjust the likelihood scaling to maintain the desired $\Delta f_\mathrm{loss}$.
    \item \modified{Optionally a}djust the refresh rate.  Lower refresh rates lead to \modified{less path variance} (less diffusion)\modified{.}  However, lowering the refresh rate too far can lead to too much time spent exploring the same trajectory, so eventually autocorrelation times can increase\modified{, and path heating from stochastic gradients will occur (more of an issue for HMC than ray tracing)}.  Ray tracing generally has a wide window for similar results with different refresh rates, \modified{and path stochasticity is often determined by gradient stochasticity for all but the most extreme values of the refresh rate}.
    \item Perform full sampling run: initialize the desired number of walkers with random initial conditions, perform burn-in on each walker, and then perform approximate MCMC sampling with fixed hyperparameters.
    \item Validate with a convergence test in the function space (e.g., the values of the network predictions on the validation set), as well as verifying the loss on the test set.
\end{enumerate}

\section{Applications}

\label{s:applications}

In this paper, we focus on the applications to parameter spaces with large numbers of dimensions ($>1000$).  This is because global and hybrid global/local samplers like nested sampling and normalizing flows are often better choices for lower-dimensional problems.  For such high-dimensional problems, the only practical algorithms are gradient-based, including Hamiltonian Monte Carlo (HMC) and ray tracing.  We first examine sampling dynamics for a 10000-dimensional Gaussian distribution in Section \ref{s:gaussian}, and then apply the ray tracing method to neural networks with 1400 to 1.5 billion parameters in Sections \ref{s:um} to \ref{s:1B}.  Additional comparisons with HMC for standard distributions appear in Appendix \ref{a:hmc}. Herein, all autocorrelation times are computed \modified{as described in Section \ref{s:gaussian}}.

\subsection{Ray Tracing on Gaussian Distributions}

\label{s:gaussian}

\begin{figure}
\vspace{-10ex}
\capstart
    \hspace{-5ex}
     \begin{overpic}[width=1.15\columnwidth]{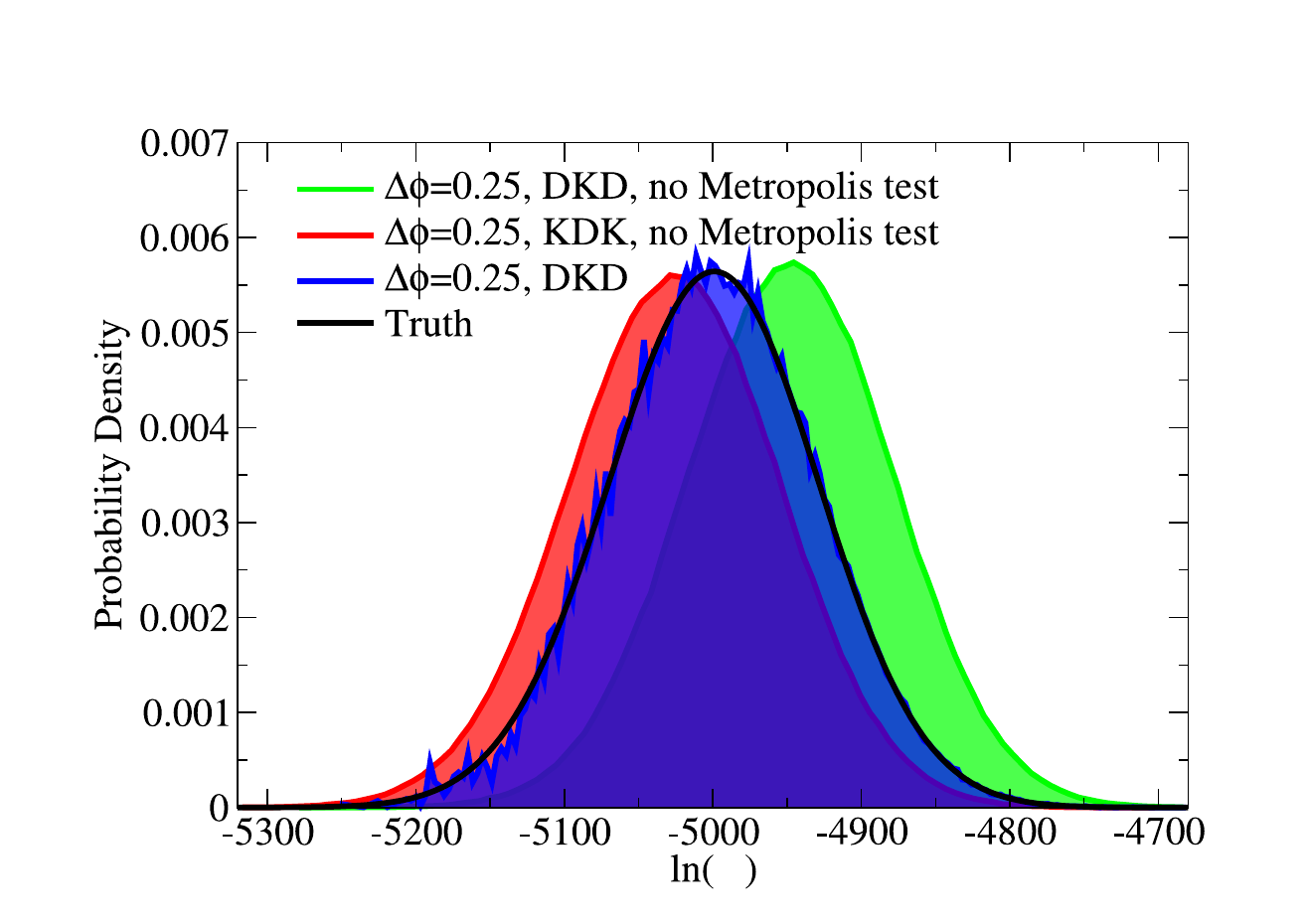}
        \put(54.4,3.2){$\mathcal{L}$} 
    \end{overpic}   
    \vspace{-4ex}
    \caption{This figure shows ray tracing samples for a 10000--dimensional Gaussian likelihood ($\ln \mathcal{L}=-0.5|\mx|^2$), using a moderate step size equivalent to a change in propagation direction of $\langle \Delta\phi\rangle =0.25$ at each step.  The non-Metropolis cases both converge to stable distributions that are biased, so a Metropolis test can be helpful to achieve larger step sizes when exact likelihoods are used.}
    \label{f:metro_test}

\vspace{-4ex}
\capstart
    \hspace{-5ex}
     \includegraphics[width=1.15\columnwidth]{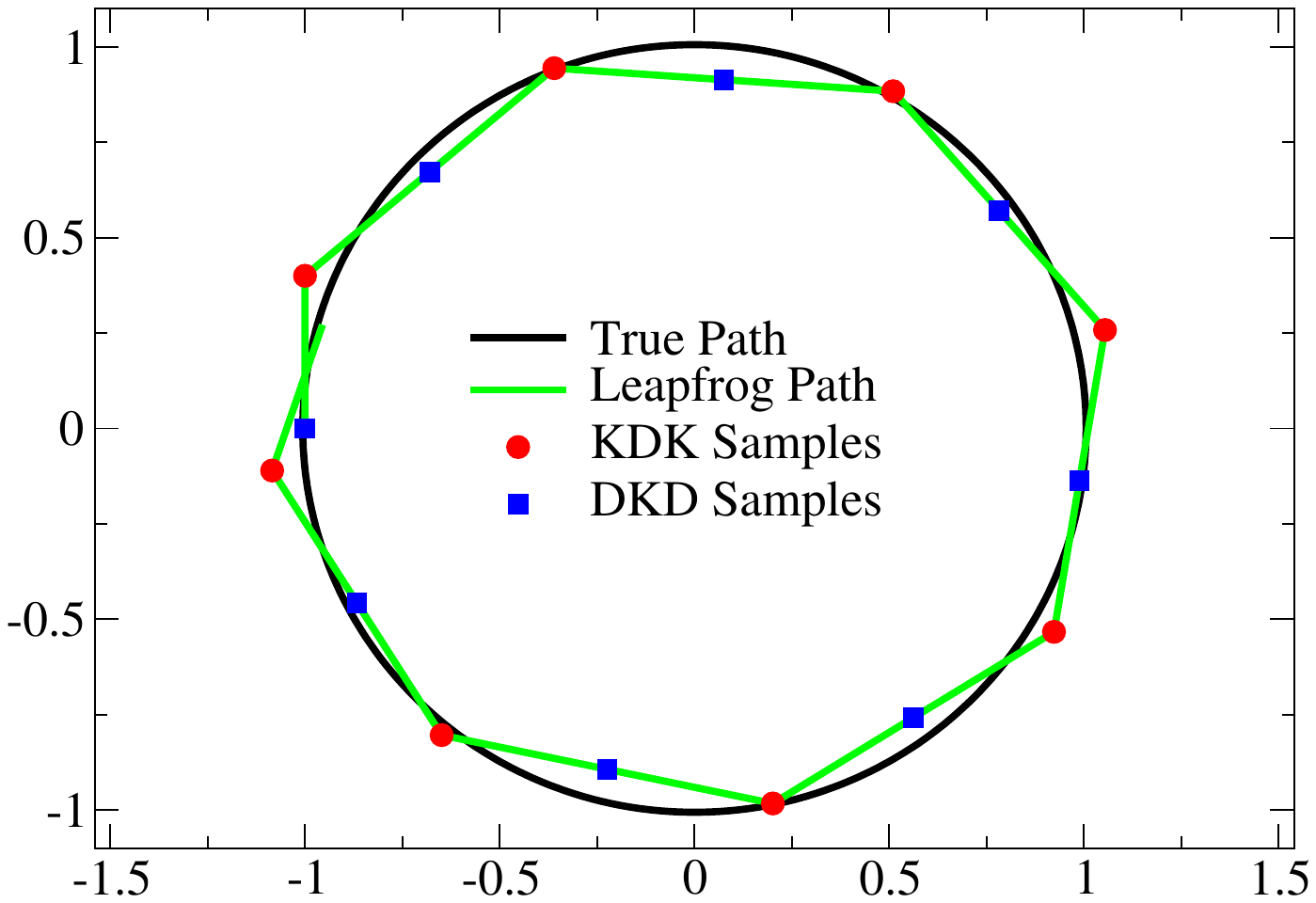}
    \vspace{-4ex}
    \caption{Leapfrog path ($\Delta \phi = 0.8$) vs.\ true path for the potential $U = x^2$.  Kick-drift-kick integrators sample the \textit{red points}, which are typically outside the true path, whereas drift-kick-drift integrators sample the \textit{blue points}, which are typically inside the true path.  At in-between locations, the path radii are correct, but the velocities are not tangent to the true path.  Randomly choosing between KDK and DKD steps is one way to have correct expectation values for both the positions and velocities, with the tradeoff of larger path variance.}
    \label{f:integrator}

   \capstart
    \vspace{-4ex}
   \hspace{-5ex} 
\includegraphics[width=1.15\columnwidth]{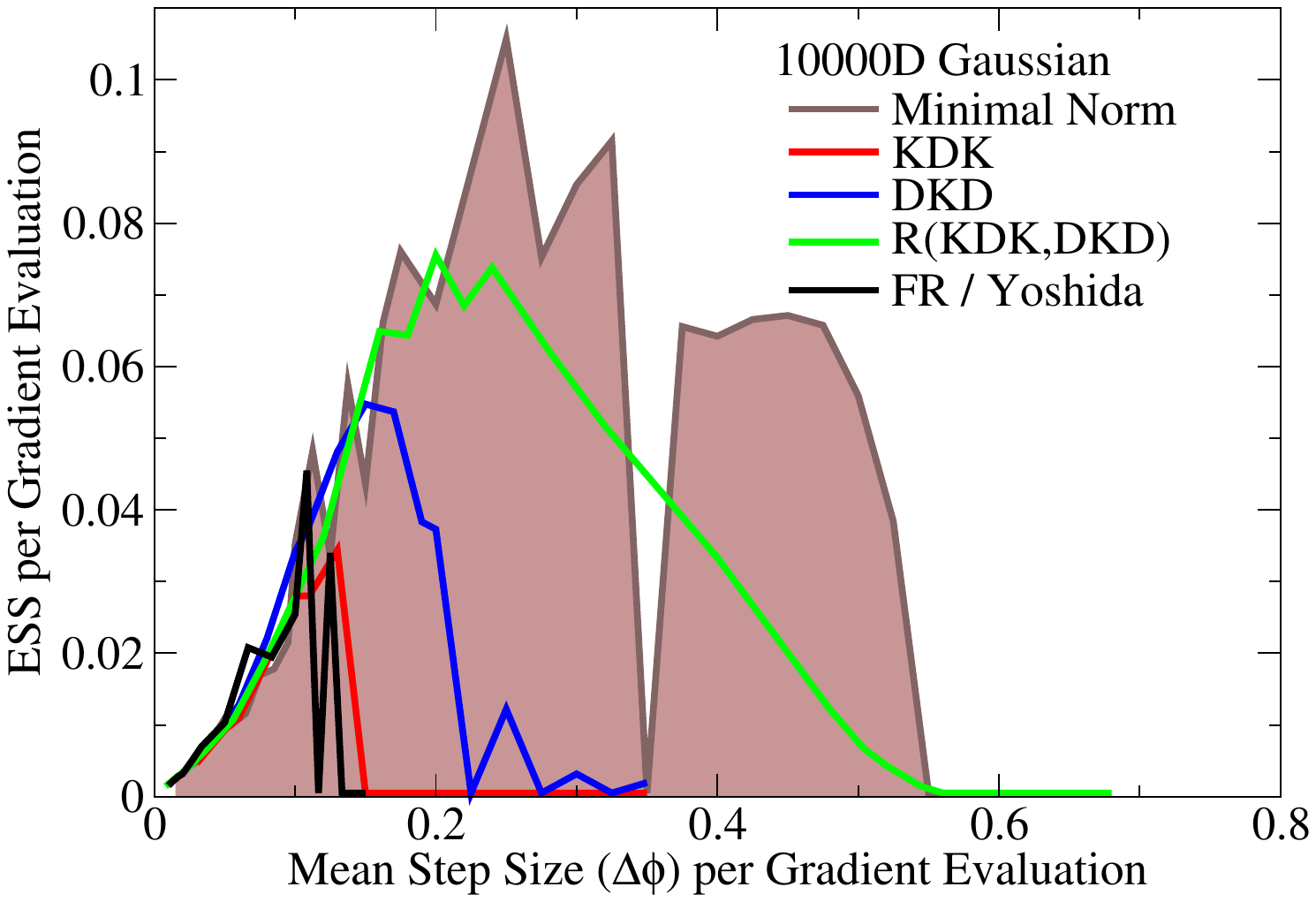}
    \vspace{-4ex}
    \caption{Effective sample sizes (ESS) per gradient evaluation for kick-drift-kick (KDK), drift-kick-drift (DKD), \cite{Omelyan03} Minimal Norm second-order (Minimal Norm), and \cite{Forest90}/\cite{Yoshida90} integrators, all for the same 10000--dimensional Gaussian likelihood.  All show a characteristic sharp dropoff with increasing step size (shown in units of the average angular change $\Delta \phi$).  We find that randomly choosing between KDK and DKD steps smooths the dropoff substantially, and the \cite{Omelyan03} integrator improves performance even further.}
    \label{f:ess_gaussian}
\end{figure}

\subsubsection{Perfect Gradient with Imperfect Integrators}

In this section, we consider ray tracing dynamics on a 10000--dimensional independent and identically distributed (i.i.d.) Gaussian distribution ($\ln\mathcal{L} = -0.5|\mx|^2)$, with a perfect gradient calculator.  We consider both the Metropolis-corrected and uncorrected cases, given the existence of both uncorrected and corrected microcanonical HMC methods \citep[i.e.,][]{Robnik22,Robnik25}.  

When perfect gradients are available, we find that it may be more efficient to use a Metropolis-corrected integrator \modified{when high accuracy is desired}, since the step sizes can be much larger.  
We show, for example, the likelihood distributions for kick-drift-kick (KDK) and drift-kick-drift (DKD) leapfrog integrators in Fig.\ \ref{f:metro_test}.  Without a Metropolis test, the KDK integrator converges to a lower-likelihood distribution, and the DKD integrator converges to a higher-likelihood distribution as compared with the true distribution, which is recovered accurately when a Metropolis test is used.  Although KDK and DKD integrators will trace identical paths, they will sample different locations along that path, as shown in Fig.\ \ref{f:integrator}, with KDK sampling at larger radii than DKD.  Intuitively, KDK will sample at the vertices of the polygonal path (which tend to be at lower likelihoods), while DKD will sample at the midpoints of the edges of the polygonal path (which will tend to be at higher likelihoods), as shown in Fig.\ \ref{f:integrator}.  

For each integration method and step size we test, we start 500 independent chains from a Gaussian distribution with $4\times$ the standard deviation as the true likelihood function.  We then discard the first 1500 steps as burn-in, the earliest 1000 of which have no Metropolis test applied.  The latter choice is because the ray tracing technique functions effectively as a gradient descent algorithm when it is far away from the typical set.  We refresh the path direction and apply a Metropolis test every $N = \sqrt{D}\cdot \pi/(2\Delta s)$ steps, corresponding to traveling an angle of $\pi/2$ within the typical set, which is optimal for Gaussian distributions.  To evaluate performance, we use the metric of effective sample size (estimated using the ratio of the chain length to the autocorrelation time) per gradient computed.  To be conservative, we take the autocorrelation time to be the maximum of 1) the average autocorrelation time of the individual parameter space dimensions, 2) the average autocorrelation function of the absolute values of the individual parameter space dimensions \modified{(i.e., folding; see also \citealt{Vehtari21})}, and 3) the autocorrelation time of $\ln(\mathcal{L})$.  \modified{We use the \citealt{Sokal1996MonteCM} autocorrelation time estimator, and for multiple chains, we calculate the autocorrelation of individual chain samples with respect to the multi-chain average and standard deviation.} This ensures proper mixing both in the phases and in the energies of the resulting distribution.  

We show the resulting performance for several step sizes and symplectic integrators, including KDK, DKD, the minimal norm second-order integrator of \cite{Omelyan03}, and the fourth-order integrator of \cite{Forest90} and \cite{Yoshida90} in Fig.\ \ref{f:ess_gaussian}.  To provide a better comparison across different integration orders, we report the mean step size per gradient evaluation ($\Delta \phi \equiv \pi/(2NG))$, where $N$ is the path step count as above and $G$ is the number of gradients per full integration step), instead of the total step size across all integration stages.  

At low step sizes, acceptance rates are high due to low cumulative path errors, and so all integrators show a universal rise due to a lower number of gradients needed for a given path length.  The KDK, DKD, and Forest-Ruth / Yoshida integrators all show a steep drop-off at low step sizes, because these methods often fail to reach the typical set by the time burn-in is over.  For DKD and KDK steps, the cause is overshoot or undershoot (Fig.\ \ref{f:integrator}), and for the Forest-Ruth / Yoshida integrator, the cause seems to be the large positive step sizes, as the method contains a negative step in the middle.  The simplest method to improve the performance of DKD/KDK steps is to randomly choose between the two along the path, which offers better effective sample sizes over a wider range of steps.  However, the \cite{Omelyan03} method offers the best performance over the widest range of step sizes, as it minimizes both velocity and position errors relative to other methods.

We expect the situation to be different at much higher dimensionality, where the errors of fourth-order methods improve more rapidly than the errors of second-order methods as the step size shrinks.  Following \cite{Neal12}, we can estimate that both the step size and the effective sample size per gradient will scale as $D^{-1/(2k)}$, with $k$ the order of the integration method, for i.i.d.\ Gaussian distributions.  Hence, the factor of $\sim 3$ advantage in effective sample size of the \cite{Omelyan03} second-order over the Forest-Ruth / Yoshida fourth-order integrators is expected to disappear around a few tens of millions of dimensions.  Empirically, we find this to be the case at $\sim$30 million dimensions.  We note that \cite{Omelyan03} includes an 11-stage fourth-order algorithm that may have an even higher performance than the Forest-Ruth / Yoshida integrator and hence a lower crossover dimensionality.

As discussed in Appendix \ref{a:hmc}, we find that ray tracing and HMC have roughly equivalent performance when tuned appropriately, across a range of standard tests including Gaussian, Rosenbrock, and Cauchy distributions.  The main performance benefit of ray tracing discussed in this paper comes from its resilience to heating in stochastic gradients, discussed in the following section.

\subsubsection{Stochastic Gradients}

\label{s:gaussian_sg}

\begin{figure}
    \vspace{-4ex}
    \capstart
    \hspace{-8ex}
    \begin{overpic}[width=1.125\columnwidth]{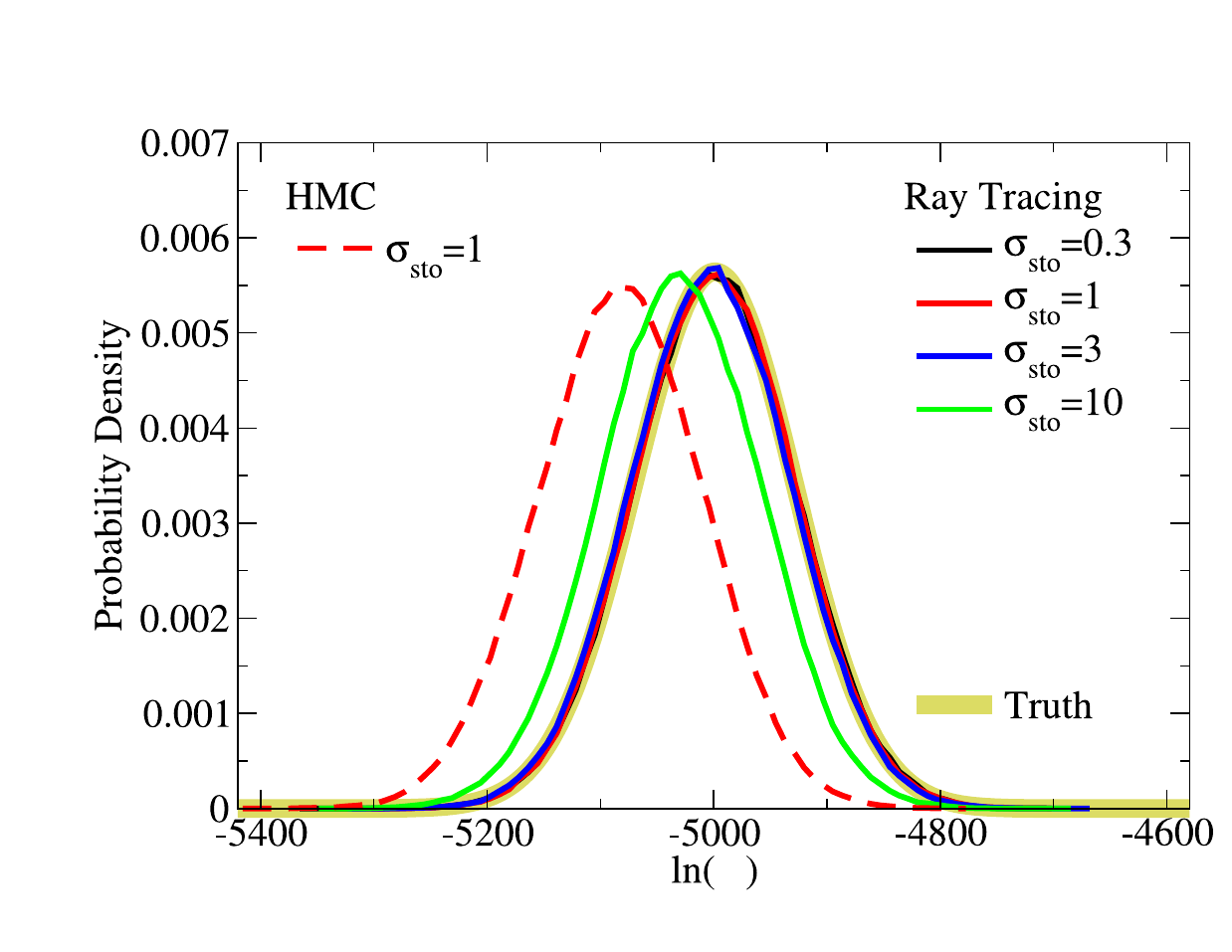}
        \put(57.85,5.7){$\mathcal{L}$} 
    \end{overpic}   
    \vspace{-6ex}
    \caption{True likelihood distribution for the ray tracing algorithm using stochastic gradients (Eq.\ \ref{e:sto_grad}) with a 10000-dimensional Gaussian distribution and a step size of $\Delta \phi = 0.03$.  Minimal deviation from the true distribution (thick sand-colored line) is observed until $\sigma_\mathrm{sto}=10$, corresponding to perturbing the center of the Gaussian by an amount 10 times larger than its standard deviation(!)  By comparison, Hamiltonian Monte Carlo shows worse performance already by $\sigma_\mathrm{sto}=1$, i.e., with $100$ times smaller perturbation variance.}
    \label{f:sto_gradient_dist}

    \vspace{-4ex}
    \capstart

   \hspace{-8ex} \begin{overpic}[width=1.125\columnwidth]{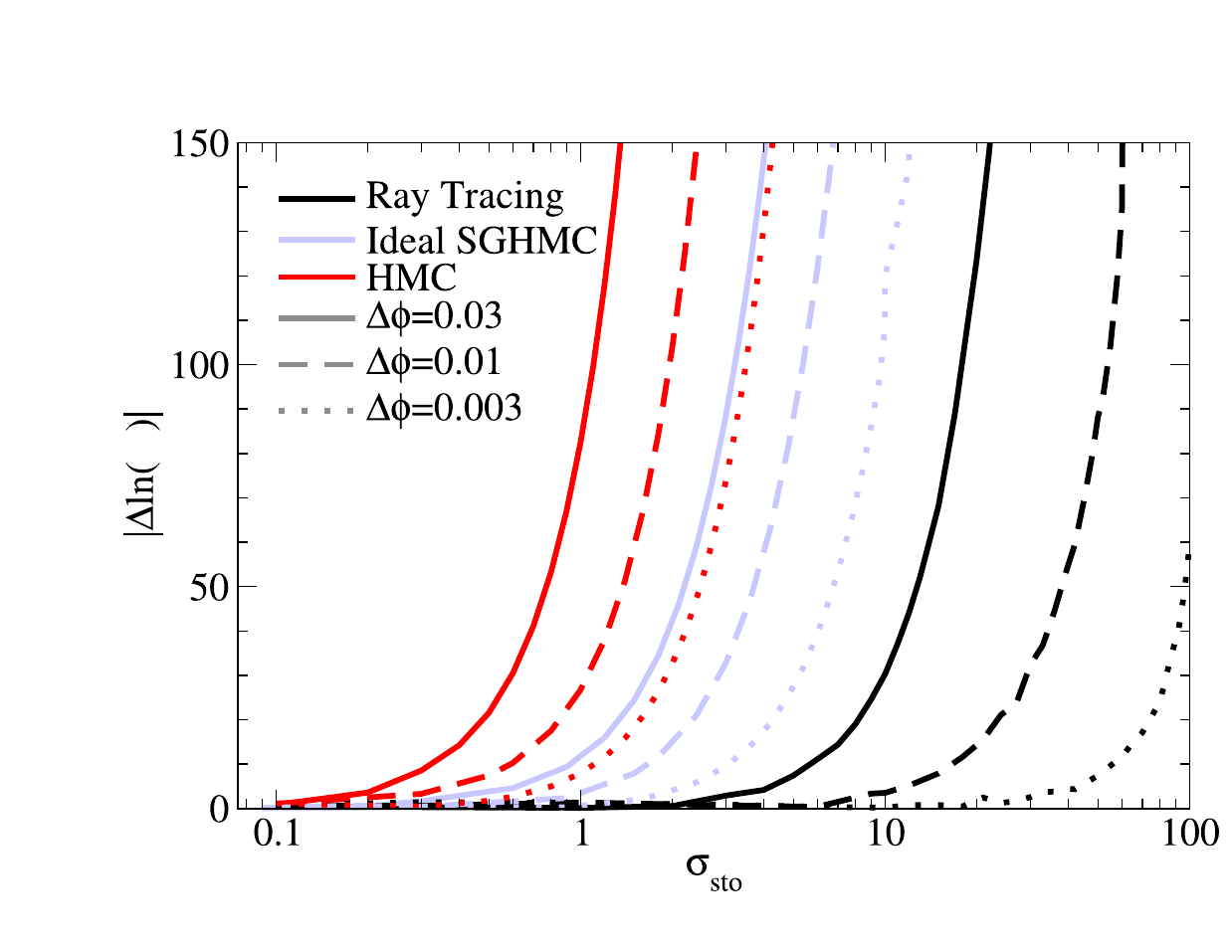}
    \put(10.5,39.6){\rotatebox{90}{$\mathcal{L}$}} 
    \end{overpic}   
    \vspace{-6ex}
    \caption{Comparison between the resilience of ray tracing\modified{,} traditional Hamiltonian Monte Carlo (HMC)\modified{, and idealized Stochastic Gradient HMC (SGHMC)} to noise in stochastic gradients, again for a 10000-dimensional Gaussian distribution.  The vertical axis shows the deviation in the median log likelihood ($\ln \mathcal{L}$) as a function of the gradient noise (parameterized by $\sigma_\mathrm{sto}$ in Eq.\ \ref{e:sto_grad}).  At a moderate step size of $\Delta\phi=0.03$, ray tracing can tolerate $16\times$ larger $\sigma_\mathrm{sto}$ than HMC ($\sim 250 \times$ more variance) for an equivalent accuracy in the likelihood posterior distribution.  \modified{Stochastic Gradient HMC tolerates only $3\times$ more $\sigma_\mathrm{sto}$ than HMC, under the unrealistic ideal condition that the gradient noise is perfectly known.}  As the step size decreases, the advantage of ray tracing increases: the stochasticity resilience\modified{s} of \modified{both} HMC \modified{and SGHMC} scale as only $\mathcal{O}(\Delta \phi^{-1/2})$, whereas the resilience of ray tracing scales as $\mathcal{O}(\Delta\phi^{-1})$.}
    \label{f:sto_gradient_median}
\end{figure}

When training neural networks, it is often too costly to have a perfect likelihood function, and so the tradition is to sample from a stochastic gradient of the log-likelihood---i.e., a distribution that has the expectation value of the true log-likelihood gradient with some scatter.  For a Gaussian distribution, we can approximate this in a few different ways, the simplest being:
\begin{equation}
    \mathcal{L}_\mathrm{sto}(\mathbf{x}) \equiv \exp\left(-\frac{1}{2}[\mathbf{x}-\mathbf{R}(\sigma_\mathrm{sto})]^2\right),\label{e:sto_grad}
\end{equation}
where $\mathbf{R}(\sigma_\mathrm{sto})$ is a random Gaussian distribution with variance $\sigma_\mathrm{sto}^2$.  As required, $\langle \nabla \ln \mathcal{L}_\mathrm{sto}(\mathbf{x}) \rangle = -\mathbf{x}$.  However, the expected magnitude $\langle |\nabla \ln \mathcal{L}_\mathrm{sto}(\mathbf{x})| \rangle$ scales proportionally to $\sqrt{1+\sigma_\mathrm{sto}^2}$ for points in the typical set, so the distribution of $\mathcal{L}_\mathrm{sto}$ is very different from the distribution of the true likelihood function even for small values of $\sigma_\mathrm{sto}$.

As with the previous section, we choose $D=10000$.  We use a smaller step size ($\Delta \phi =0.03$), and do not use a Metropolis test, because the expected uncertainty in evaluating the likelihood is typically far higher than the error from evaluating the path, as is shown later in this section.  (In other words, had we used a Metropolis test, we would be randomly discarding samples, as opposed to removing any biases).  As well, we use the R(KDK,DKD) integrator, finding that this seems optimal for both ray tracing and Hamiltonian Monte Carlo.  

Results for the true likelihood distribution as a function of $\sigma_\mathrm{sto}$ are shown in Fig.\ \ref{f:sto_gradient_dist}.  The simplicity of the figure belies how shocking it is.  Only at $\sigma_\mathrm{sto}=10$ does the likelihood distribution become visibly different from the true distribution for ray tracing.  At this large value of $\sigma_\mathrm{sto}$, only 1\% of the variance in the gradient is correlated with the actual location of the typical set, and typical values of the stochastic likelihood are a factor $\exp(-500000)$ smaller than the values of the true likelihood.  In comparison, HMC shows worse performance with perturbations that have $100\times$ less variance ($\sigma_\mathrm{sto}=1$).


The performance benefit is equally stark in Fig.\ \ref{f:sto_gradient_median}, which shows the error in the median of the log-likelihood distribution, compared to the true value, as a function of $\sigma_\mathrm{sto}$.  Both HMC and ray tracing show errors that grow linearly in the variance of the stochastic gradients:
\begin{equation}
|\Delta \ln\mathcal{L}| = (\sigma_\mathrm{sto}/\sigma_c)^2,  
\end{equation}
where $\sigma_c$ is a characteristic resilience size for stochastic gradient perturbations.  For HMC in this application, $\sigma_c^2 = 0.012$, whereas for ray tracing, $\sigma_c^2 = 3.2$, more than 250$\times$ larger.  
This conclusion becomes even stronger as the step size is reduced.  For HMC, the characteristic perturbation resilience size scales as $\sigma_c\propto 1/\sqrt{\Delta\phi}$, whereas for ray tracing, it scales as $\sigma_c \propto 1/\Delta\phi$ (Fig.\ \ref{f:sto_gradient_median}).  Given that the expected variance $\sigma_\mathrm{sto}^2$ will scale as $\mathcal{O}(B^{-1})$, with $B$ the mini-batch size, the total compute time will be independent of batch size $B$ for HMC (i.e., $\mathcal{O}(1)$): decreasing the mini-batch size by a factor of 100 requires decreasing the step size by a factor of 100, for no overall computational gain.  However, for ray tracing, the total compute time will scale as $\mathcal{O}(\sqrt{B})$: decreasing the mini-batch size by a factor of 100 would only require a factor 10 smaller step size, yielding the same effective sampling size with 10x less computational cost.  Hence, ray tracing benefits from smaller mini-batch sizes (analogous to improved performance for stochastic gradient descent methods), whereas HMC does not.  

We can understand the relative noise resilience intuitively.  For both ray tracing and HMC, taking smaller steps (i.e., more steps for a given trajectory length) will reduce the effective gradient noise per unit distance traveled.  However, as mentioned in Section \ref{s:stochastic}, using stochastic gradients will heat the velocity component for HMC at every step \citep{Chen14}, boosting the trajectory to higher energies (i.e., lower likelihoods).  Hence, for HMC, larger step counts will partially cancel the improvement from lower average gradient noise per unit distance traveled.   In contrast, the kinetic energy of the ray tracing path is fixed, so a longer trajectory simply means better averaging over the stochastic directions of the gradients, with no memory for previous heating. 

\modified{We also compare directly to stochastic gradient HMC (SGHMC; \citealt{Chen14}), the previous state-of-the-art method, in Fig.\ \ref{f:sto_gradient_median}.  The essential approach of SGHMC is to subtract the average expected kinetic energy gain from the stochastic gradient, which requires an estimate of the gradient stochasticity.  This is typically impossible to obtain exactly.  However, for this stochastic Gaussian test problem, we know the gradient stochasticity \textit{a priori}, and so we can compare with an ``ideal'' SGHMC that subtracts the exact noise expectation.  We find that SGHMC is more resilient to noise than pure HMC, but only by a factor of $\sim 3$ in $\sigma_\mathrm{sto}$, and SGHMC retains the same noise scaling as HMC.  The performance gap compared to ray tracing is simple to understand: SGHMC subtracts the mean kinetic energy bias due to gradient stochasticity, whereas (as above) ray tracing removes kinetic energy fluctuations exactly at every single step.  This leads to dramatically better noise resilience for ray tracing than for SGHMC, because SGHMC still remains susceptible to additional noise at every step.  Given the difficulty with making a fair comparison to SGHMC in more complex problems (i.e., due to challenges in estimating the gradient stochasticity) and SGHMC's similar performance scaling as HMC, we only compare with traditional HMC in further sections.}

We note that robustness to stochastic gradients can be achieved for some other sampling methods in Section \ref{s:prior} by using weights with generalized ray tracing and keeping $|\mv|$ constant.  This leads to accurate paths, with inaccurate (stochastic) weights.  Nonetheless, the correct weights can be recomputed for a subsample of the full chain in a postprocessing step.  Given that the autocorrelation time is often large for parameter exploration of neural networks, the weights may only need to be calculated at large intervals, leading to an overall performance gain.  In addition, weights may be initially computed at low accuracy, so that more expensive high-accuracy calculations can be saved for the locations that are expected to contribute most to the overall weight of the path.  As a result, we expect that traditional HMC sampling could be made to perform almost identically to ray tracing \modified{if implemented as a constant-speed sampler}.

\begin{figure*}
\vspace{-10ex}
\capstart
\hspace{-3ex}\includegraphics[width=1.1\columnwidth]{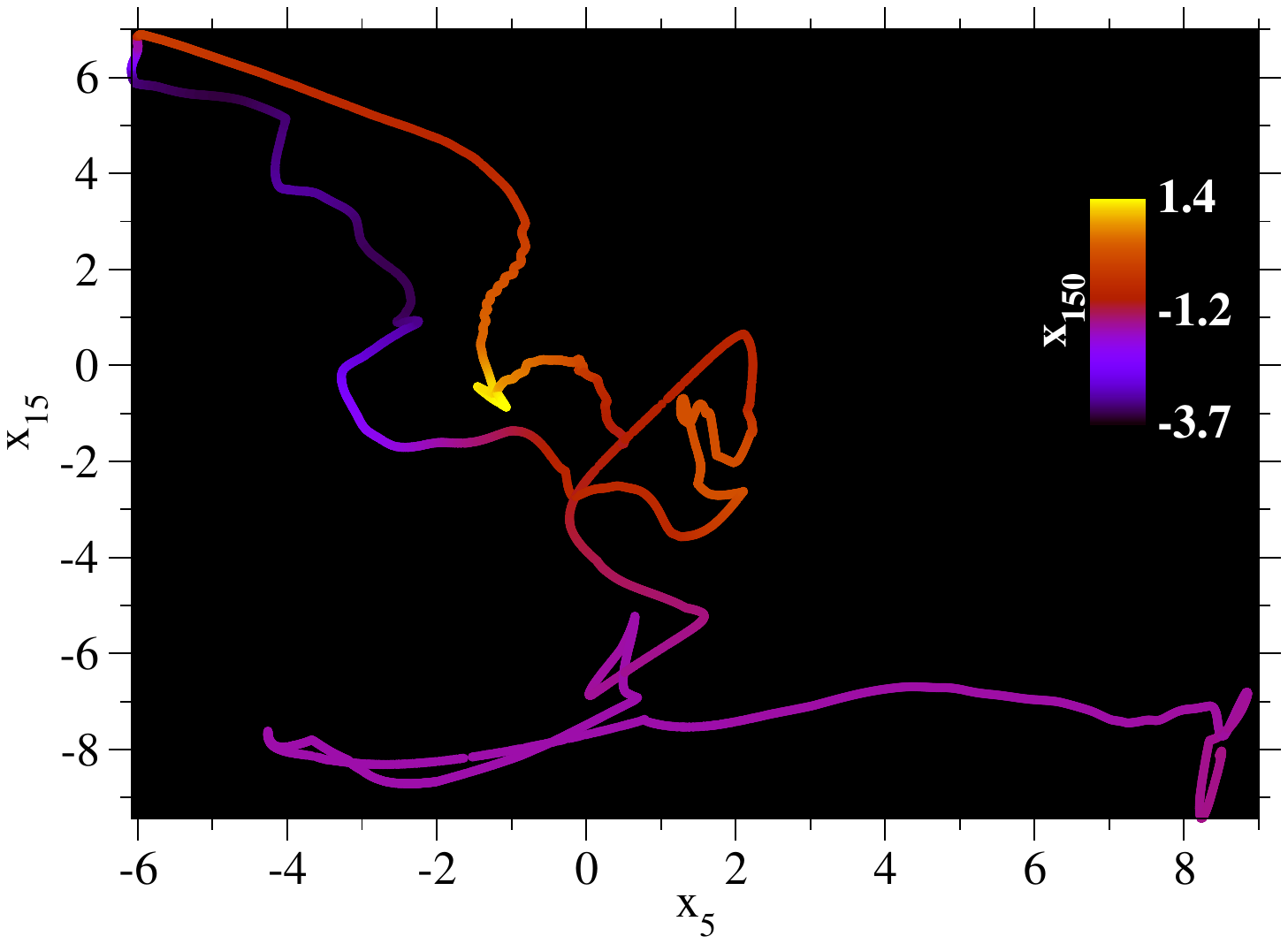} \hspace{-3ex}\includegraphics[width=1.1\columnwidth]{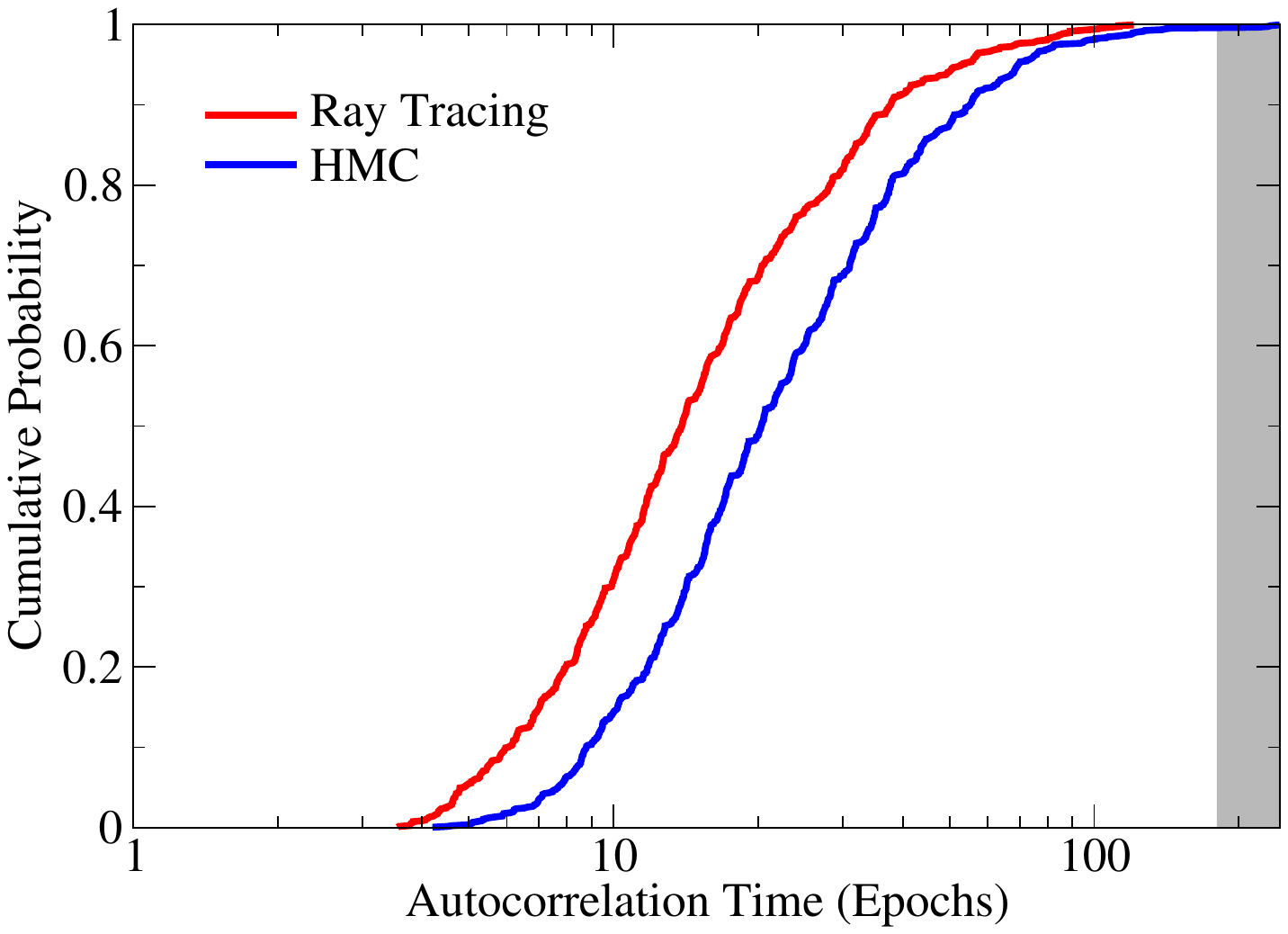}\\[-4ex]
    \caption{\textbf{Left}: A 3D projection of a sample trajectory in weight space from ray tracing in a 1433-dimensional neural network (a multi-layer perceptron) in Section \ref{s:um}; dimensions 5, 15, and 150 are shown.  The path resembles a tangled piece of spaghetti, but the path never crosses itself, due to the many dimensions available for exploration.  Frequent twists and turns are evident, indicating that the likelihood space is not well-approximated by any simple shape, as assumed by, e.g., variational inference.  For this sample trajectory, a small step size of $\Delta t=2\times 10^{-5}$ was used (10$\times$ smaller than our main analysis) with no momentum refresh to show the geometry of the target distribution as accurately as possible. \textbf{Right}: Stochastic HMC had longer autocorrelation times than ray tracing for the same multi-layer perceptron application.  This figure shows the cumulative distribution of autocorrelation times for validation set outputs (expressed in units of training epochs) for both HMC and ray tracing.  The grey shaded region shows where the chain length is insufficient to guarantee convergence.  Ray tracing achieved 100\% convergence, and HMC achieved 99.6\% convergence across validation outputs. In both cases, the first 100 epochs were discarded as burn-in.}
    \label{f:spaghetti}
    \label{f:acor}
    \capstart
    \hspace{-3ex}\includegraphics[width=1.1\columnwidth]{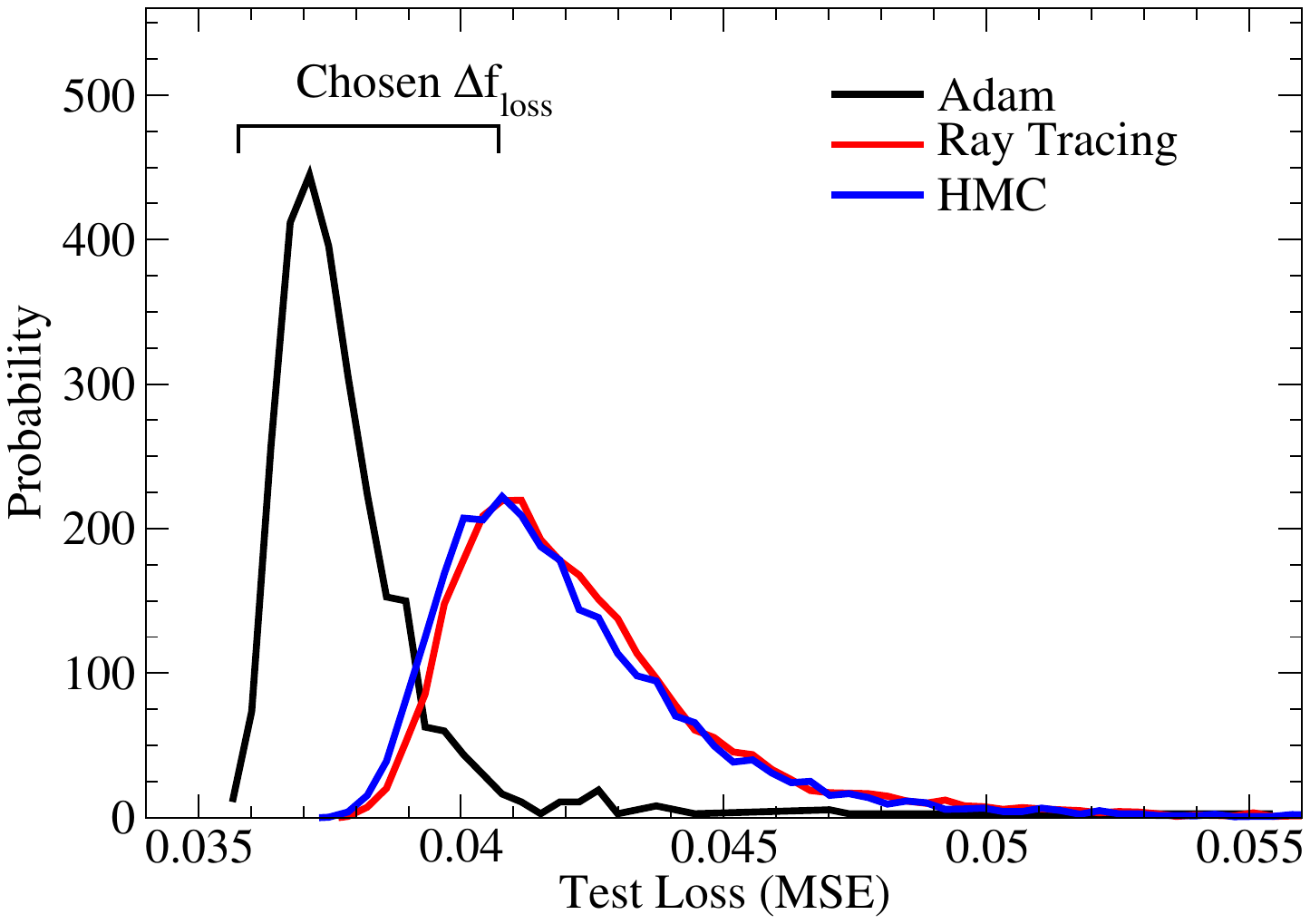}     \hspace{-3ex}\includegraphics[width=1.1\columnwidth]{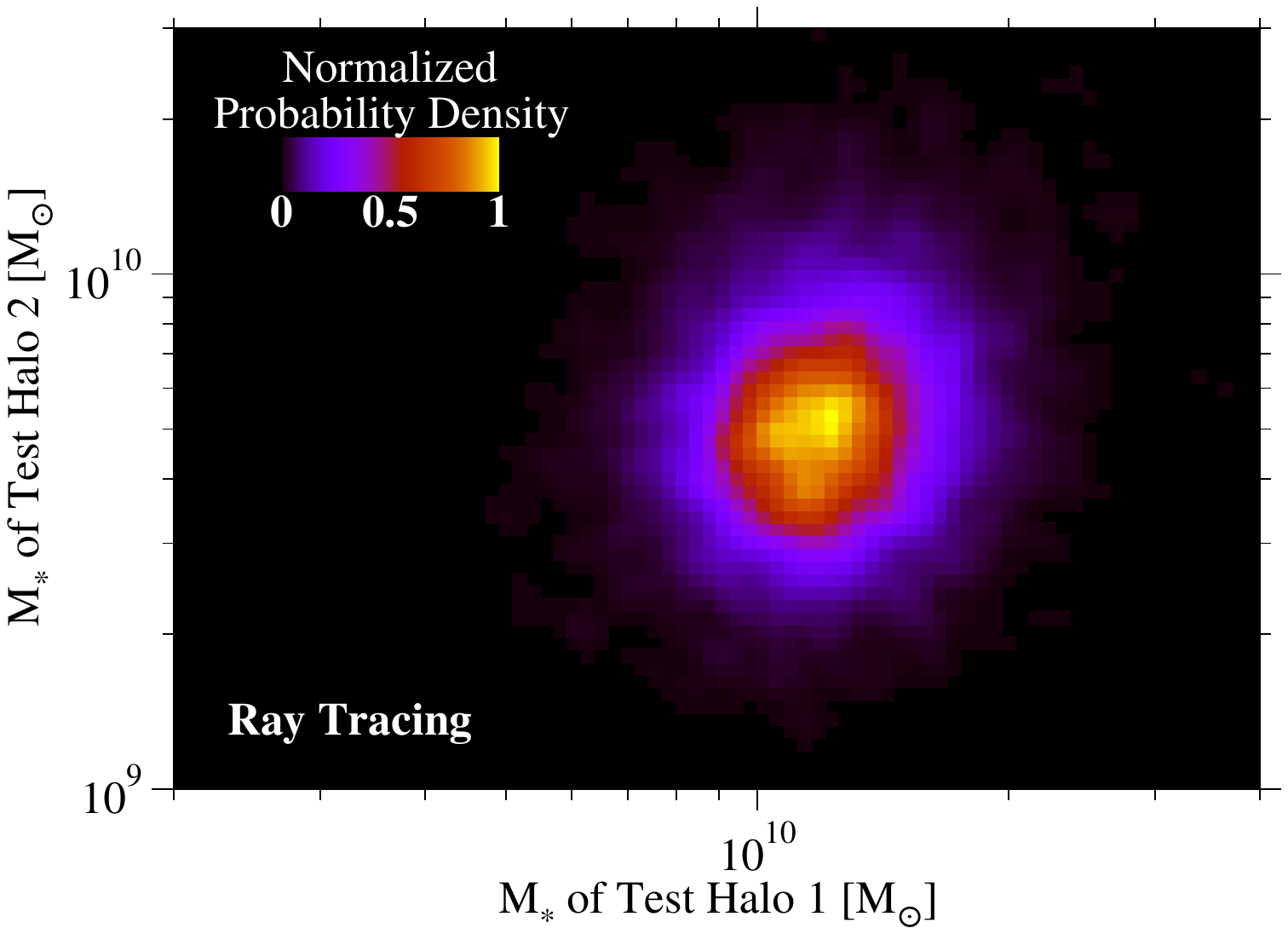}\\[-8ex]
    
    \hspace{-3ex}\includegraphics[width=1.1\columnwidth]{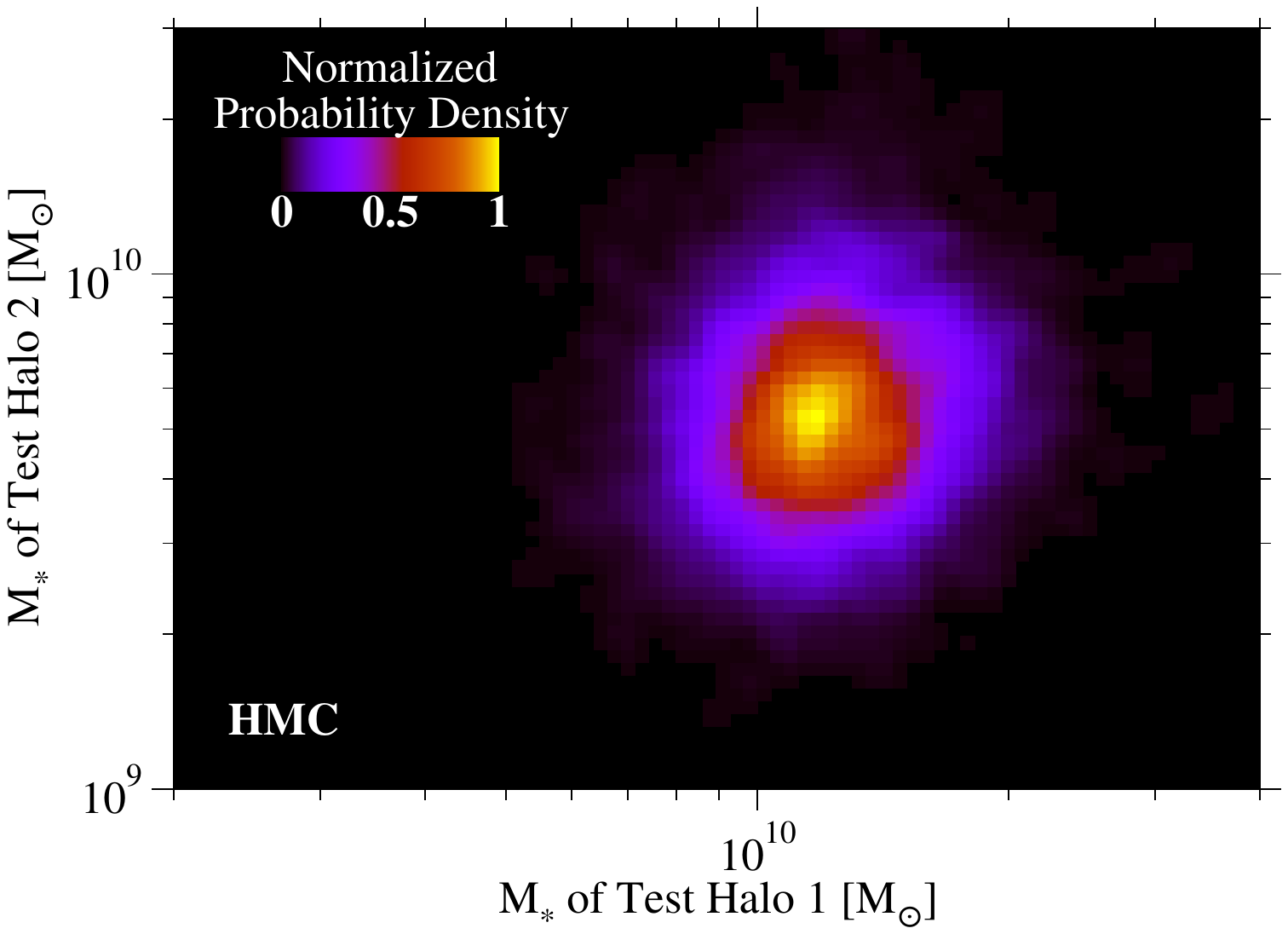}     \hspace{-3ex}\includegraphics[width=1.1\columnwidth]{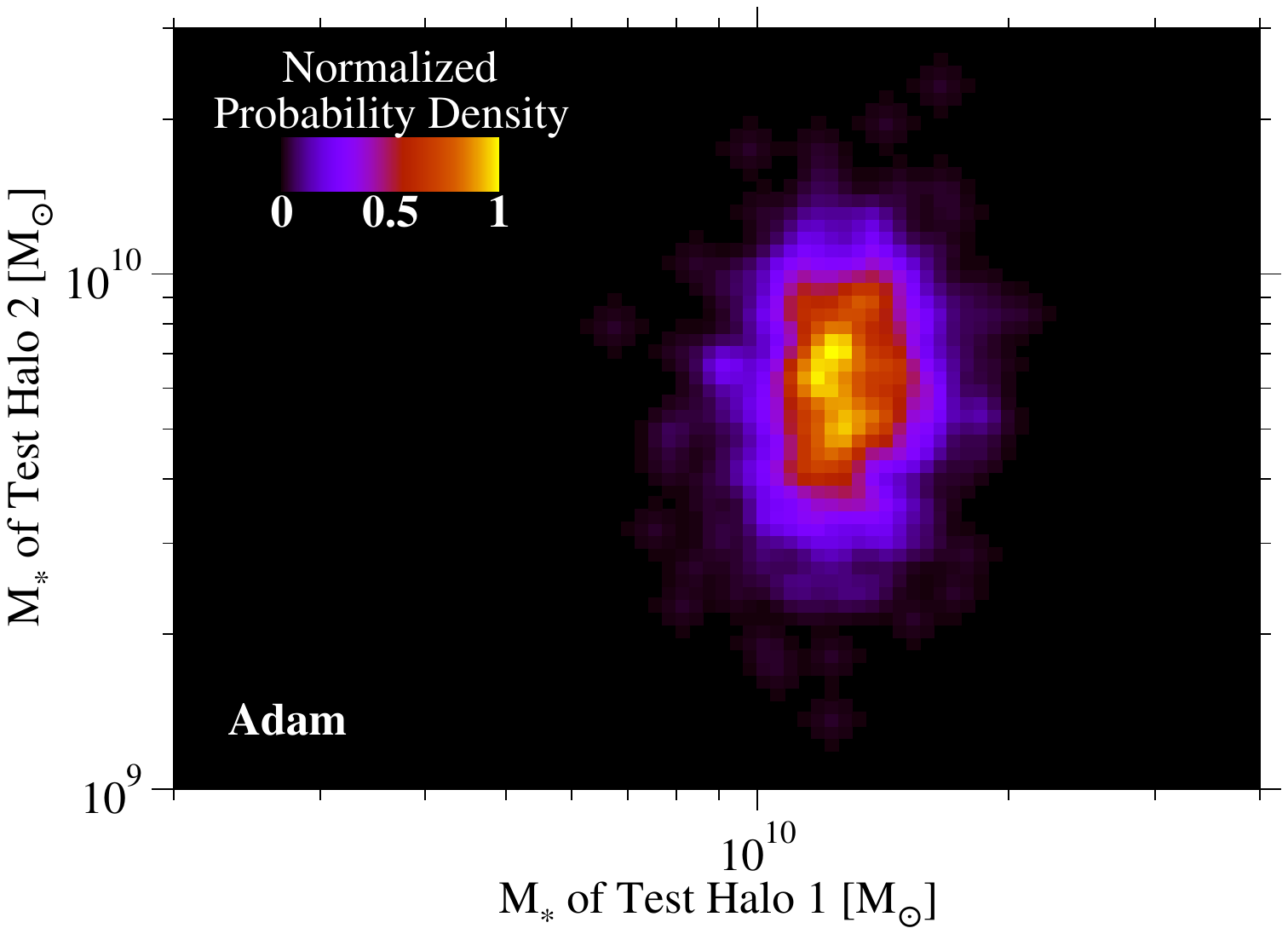}\\[-4ex]
    \caption{\textbf{Top left}: distributions of the loss function (mean-squared error; MSE) on the validation set for the multi-layer perceptron application in Section \ref{s:um}.  Adam has the lowest loss, as it searches for loss minima.  Ray tracing and HMC both sample from the posterior distribution of models within a given tolerance of the minimum \modified{Adam} loss (here chosen to be $\Delta f_\mathrm{loss}=0.00\modified{51}$, or $\sim 10\%$ of typical MSE), and so have higher loss values. \textbf{Top right}, \textbf{bottom panels}:  distribution of model predictions for two halos in the validation set, for ray tracing, HMC, and Adam.  HMC and ray tracing were run with identical parameters (i.e., mean step sizes and momentum refresh rates).}
    \label{f:distributions}
\end{figure*}

\subsection{Comparing Stochastic Ray Tracing and Stochastic HMC for a Multi-Layer Perceptron with 1433 Parameters}

\label{s:um}

Many scientific applications of neural networks involve using them as fitting functions, for example to accelerate a slow code or to interpolate values within a large dataset \modified{(see \citealt{Mcculloch43,Rosenblatt56} and \citealt{Amari67} for early papers, as well as citations to the review of \citealt{Bishop94} for more recent works)}.  
In this section, we build a small example neural network to accelerate the output of the \textsc{UniverseMachine} code \citep{BWHC19}, an empirical model that generates observed galaxy properties \modified{from} gravity-only simulations.  The network we use is simple enough for its posteriors to be explored both with HMC and with ray tracing, allowing us to compare the two approaches.


 To fit the \textsc{UniverseMachine} outputs, we use a simple multi-layer perceptron (MLP), a type of neural network where the layers are fully connected (i.e., each neuron receives inputs from all the outputs of the previous layer).  For this example application, the number of training data samples ($\sim 800$k) is much larger than the number of network parameters ($D=1433$).  We also chose a mini-batch size (20000) that was large with respect to the network parameter count, so that the expected stochasticity was relatively small, enabling HMC sampling to be used. Full details of the network inputs, outputs, architecture, hyperparameters, and hardware used are in Appendix \ref{a:um1560}.

We found that it was straightforward to perform approximate MCMC sampling with both ray tracing and HMC on either a laptop computer or a consumer GPU following the recipe in Section \ref{s:recipe}.  For both samplers, we chose the loss tolerance to be $\Delta f_\mathrm{loss}=0.00\modified{51}$, corresponding to an increase in RMSE of $\sim 10$\% and a loss scaling of 5000 (i.e., $-\ln\mathcal{L}(\mx)=5000 f_\mathrm{loss}(\mx)$).  Per Eq.\ \ref{e:approx_likelihood}, this implies an effective number of dimensions of $D_\mathrm{eff}\sim \modified{51}$ for the network.  To place step sizes for HMC and ray tracing on equivalent footing, we calculate an effective timestep size, $\Delta t \equiv \Delta s/\sqrt{D}$, for ray tracing, corresponding to the distance traveled divided by the mean HMC exploration speed. 

We find, as expected from past work \citep{Izmailov21}, that sampler mixing is generally excellent in function space and is poor in weight space for both methods.  Although both samplers converged to equivalent function space distributions, the parameter norm $|\mx|$ increased without bound for both HMC and ray tracing.  We also found that the likelihood function in weight space was not well-approximated by a Gaussian distribution.  Besides the fact that the parameter norm continued to increase indefinitely, a Gaussian weight-space distribution would have led to helical orbits (Section \ref{s:adaptive}); instead, trajectories showed much more complex likelihood geometries (Fig.\ \ref{f:acor}, left panel).
For our main analysis, we used the same hyperparameters for both samplers (step size $\Delta t=2\times 10^{-4}$ and momentum refresh rate $f=5\Delta t$), which resulted in somewhat longer autocorrelation times for HMC, as shown in Fig.\ \ref{f:acor}.  

We found that ray tracing and HMC gave similar loss distributions (Fig.\ \ref{f:distributions}, top-left panel).  Marginal distributions appear similar between HMC and ray tracing for model predictions on individual halos in the validation set (Fig.\ \ref{f:distributions}).  We found that both HMC and ray tracing showed steep dependence of acceptance rates on step sizes, as expected from Fig.\ \ref{f:ess_gaussian}---while acceptance rates were $\sim 98-99$\% with the step sizes above, increasing them by a factor of 2 invariably resulted in one or more walkers getting stuck for a significant fraction of its path length, leading to a clear spike in the posterior distributions for the validation set.

\begin{figure*}[p!]
   \vspace{-20ex}
\capstart
    \hspace{-7ex}
     \includegraphics[width=1.15\columnwidth]{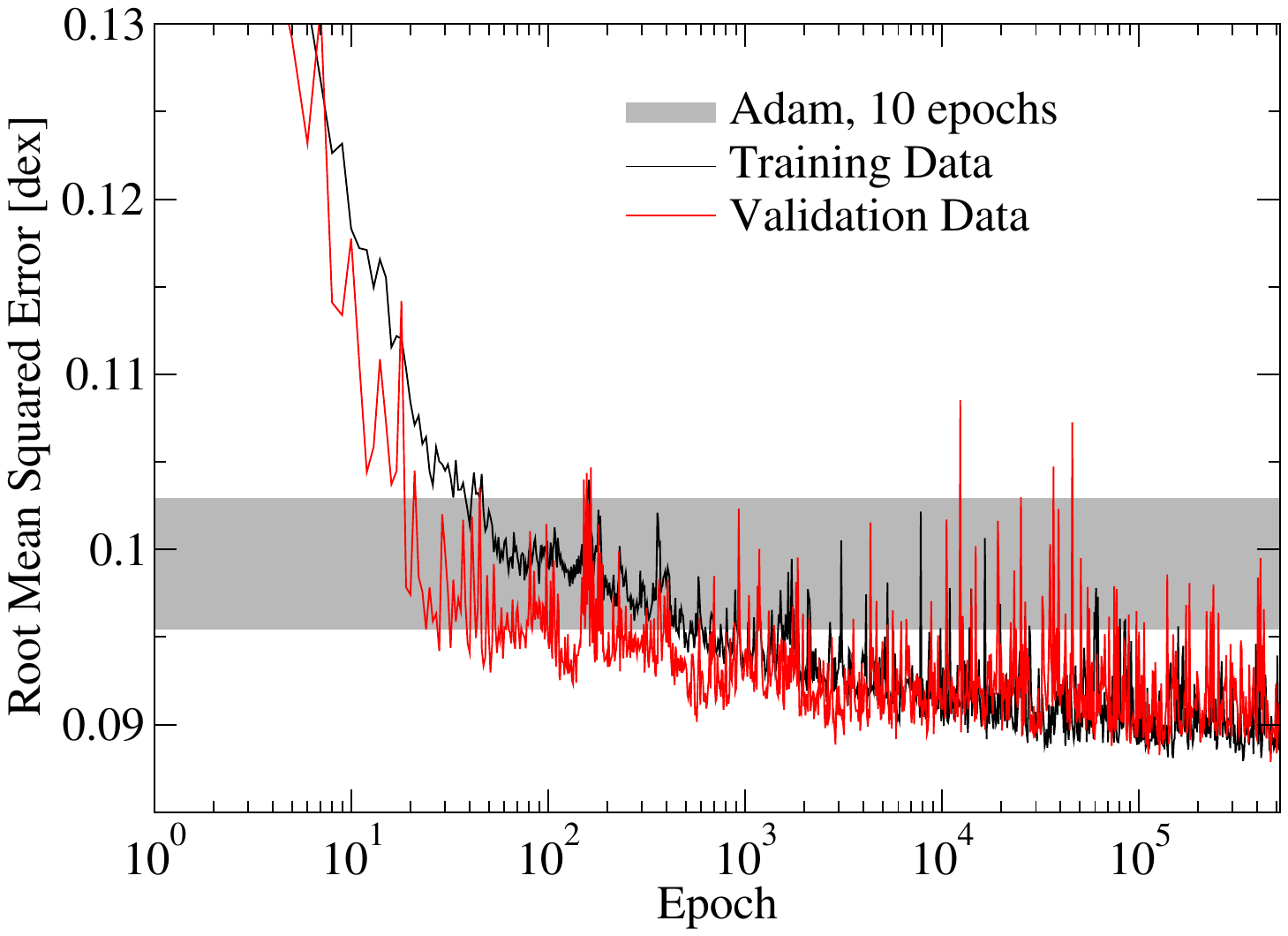}\hspace{-5ex}
     \includegraphics[width=1.15\columnwidth]{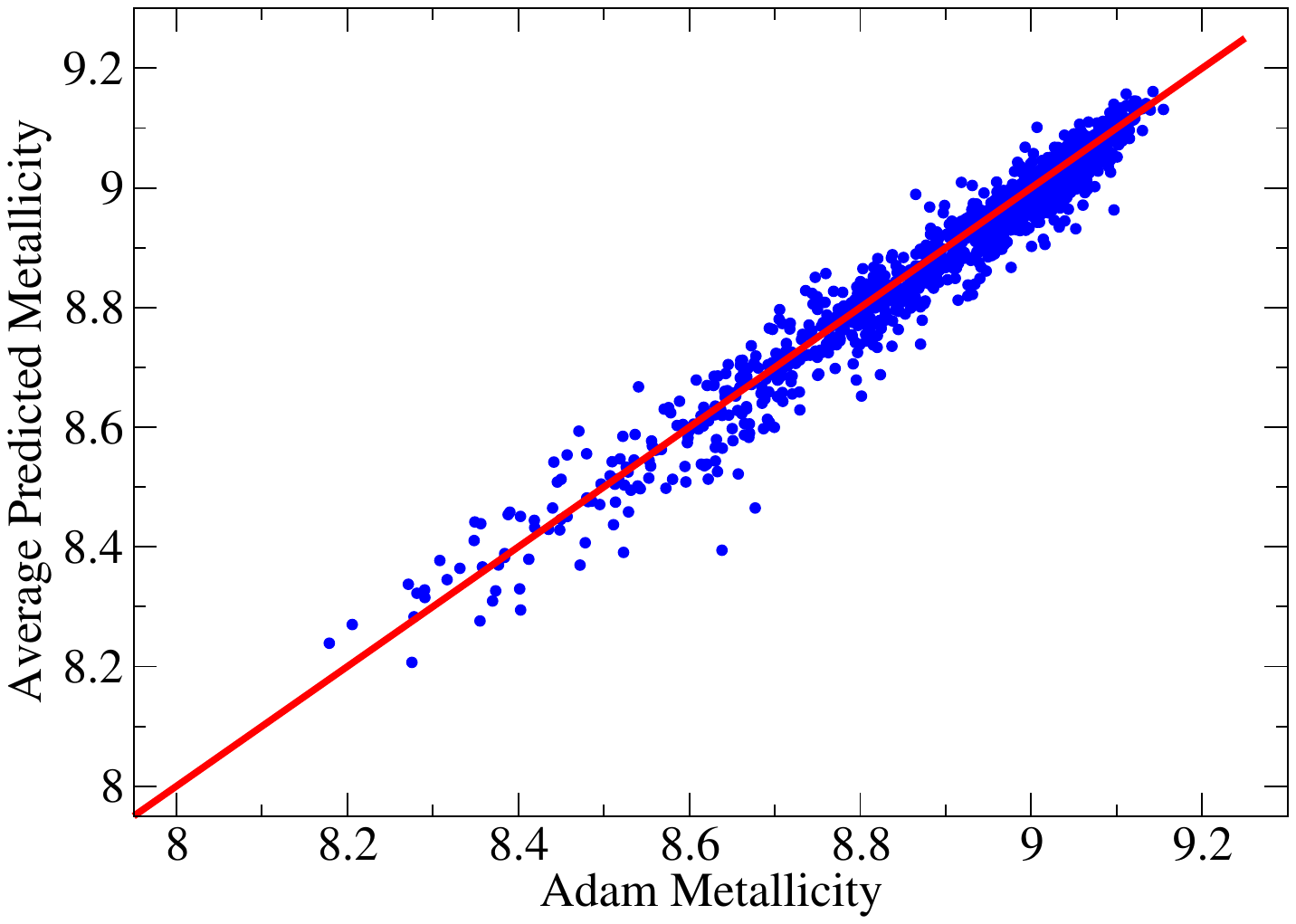}\\[-10ex]
     \vspace{0ex}
    \hspace{-7.1ex}
\includegraphics[width=1.15\columnwidth]{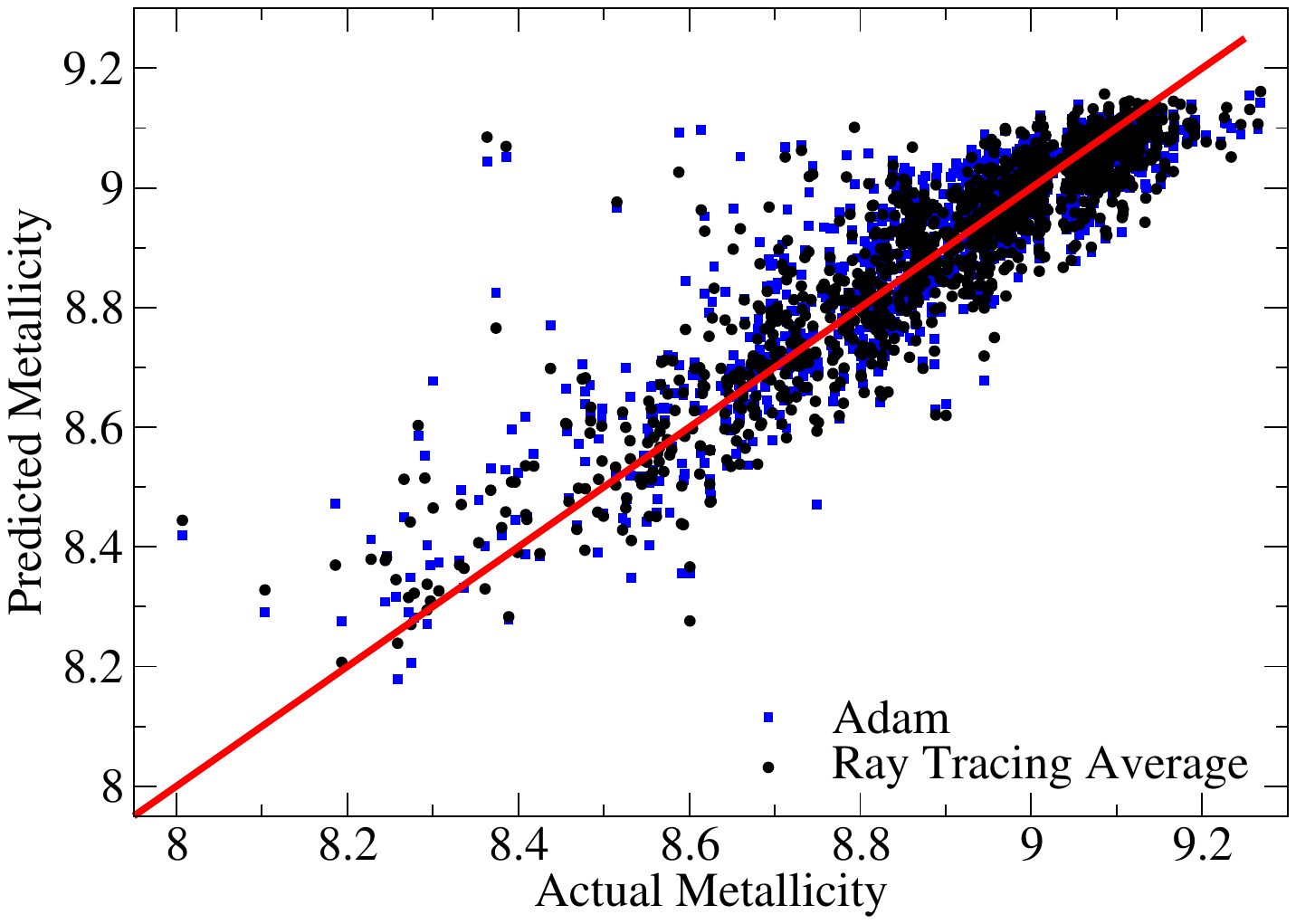}\hspace{-5ex}
     \includegraphics[width=1.15\columnwidth]{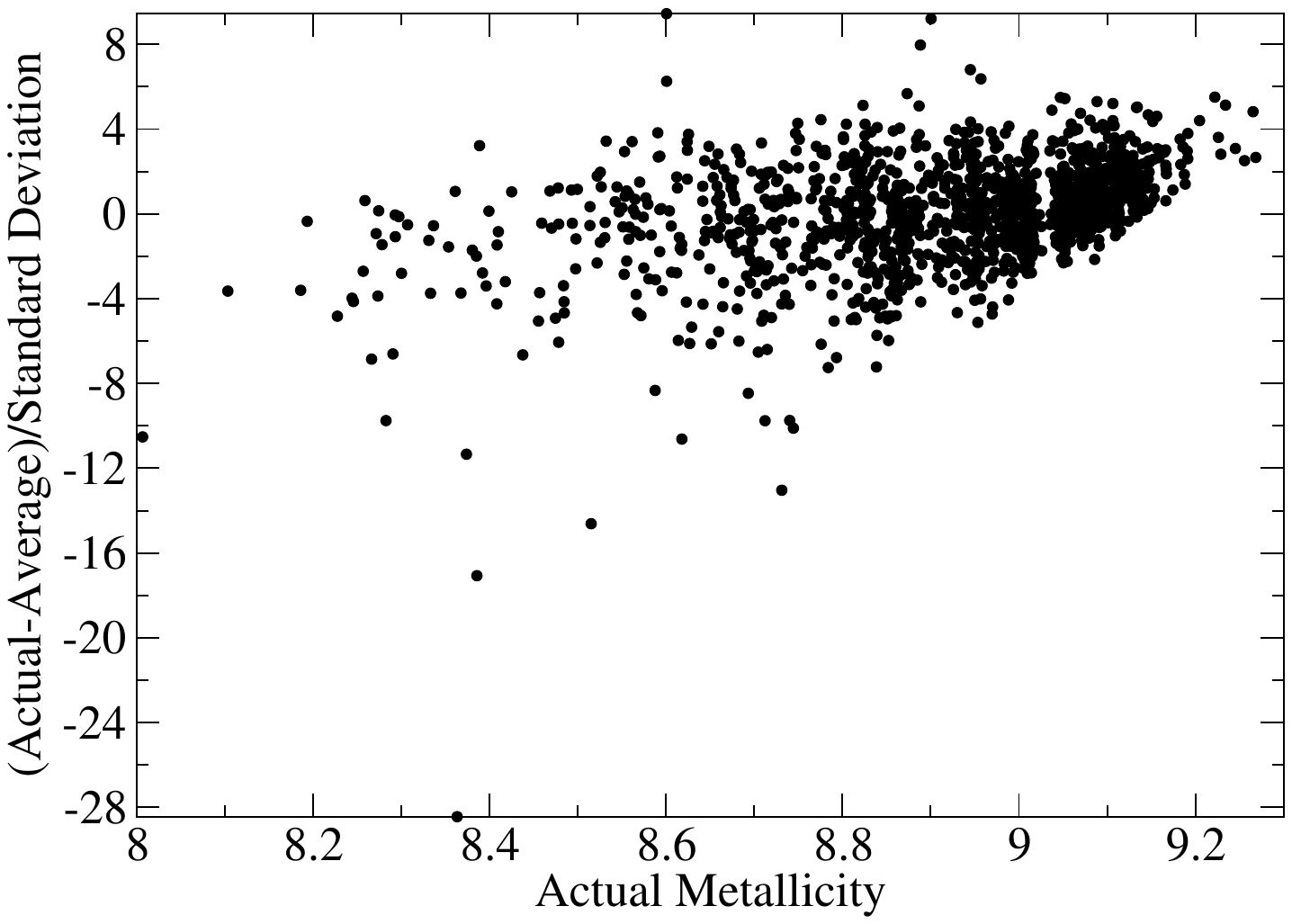}\\[-6ex]
    \caption{\textbf{Top left}: Root mean square error (RMSE) for training vs.\ validation data, using ray tracing on a ResNet-34 network with 22 million free parameters and a training data input size of $\sim$12000 (Section \ref{s:resnet}).  Even after sampling for $>$500000 epochs, training and validation loss are similar, suggesting resilience to overfitting.  The grey shaded region shows the 68\% range of Adam predictions after 10 epochs of training (to match the \citealt{Wu19} methodology). \textbf{Top right}: Adam returns similar overall metallicity estimates as MCMC sampling on the validation data.  \textbf{Bottom left}: Average predicted metallicity for ray tracing and predicted metallicity for Adam vs.\ true metallicity for the galaxies in the validation data set.  Several outliers are notable, as also found in \citet{Wu19}; as expected from the top-right panel, Adam and ray tracing result in similar outliers.  \textbf{Bottom right}: Typical offsets between true metallicities and average predicted metallicities are well outside the sample standard deviation.  This is expected (and good) behavior: it indicates that the uncertainty in the best prediction is much less than the variance of the data around the best fitting manifold, so that increasing the training sample size is less likely to lead to improved predictions (see text).  Typical posterior distribution widths are $\sim$0.02 dex for network predictions, compared to an RMSE of $0.09$ dex for the distribution of actual metallicity errors.  All gas-phase oxygen metallicities are reported in standard log units ($12+\log_{10}(O/H)$).}
    \label{f:resnet}

     \vspace{-4ex}
\capstart   \hspace{-7ex}\includegraphics[width=1.15\columnwidth]{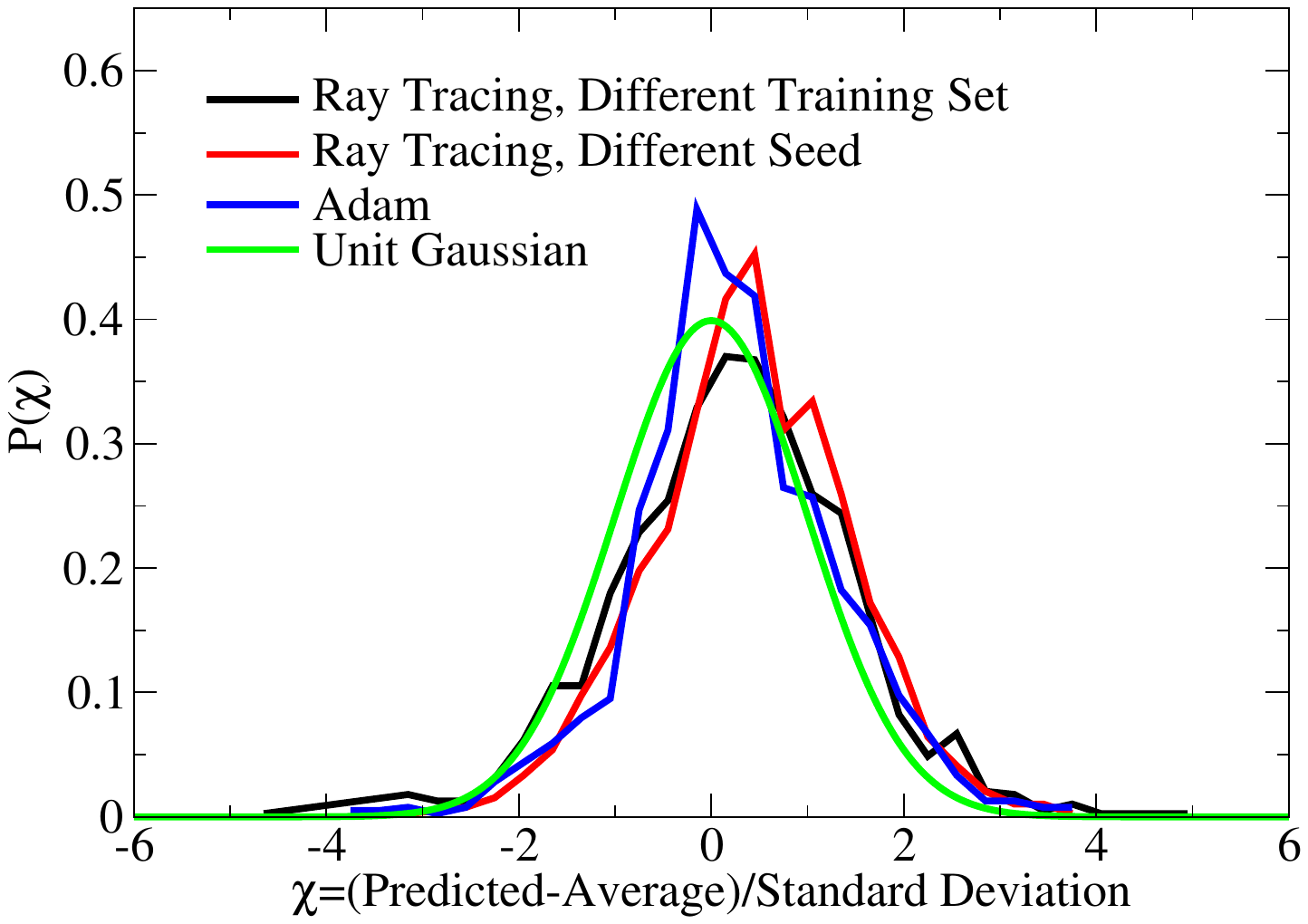}\hspace{-5ex}\includegraphics[width=1.15\columnwidth]{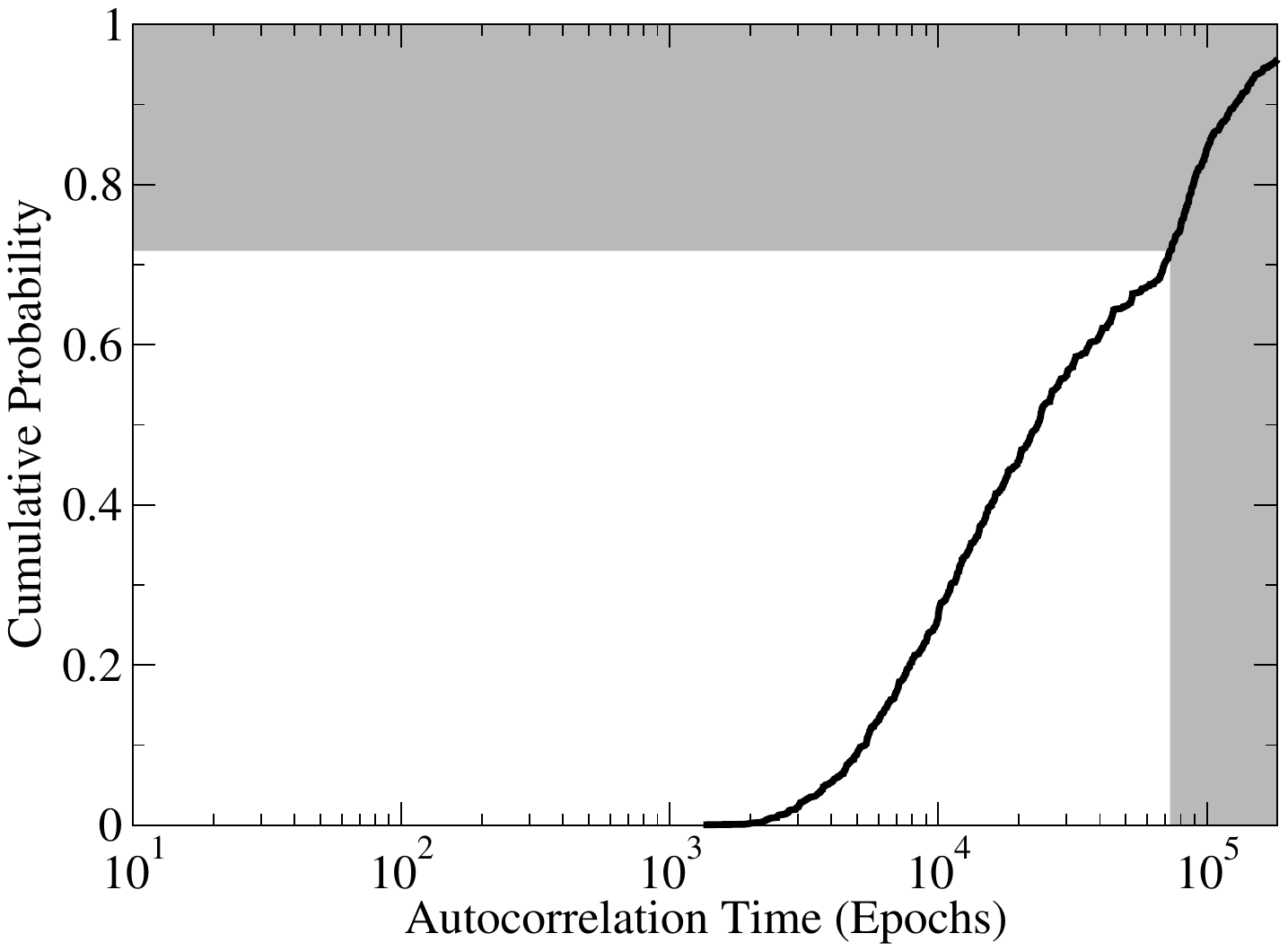}\\[-5ex]
     \caption{\textbf{Left}: The distribution of standardized differences (i.e., difference in predictions divided by the posterior standard deviation) of galaxies in the validation set for 1) a model generated using 5000 epochs of ray tracing on an identically-sized but non-overlapping training set drawn from the same data distribution, 2) a model generated using 5000 epochs of ray tracing using the same training set but a different random starting position, and 3) the original ResNet-34 weights after 10 epochs of Adam.  All are essentially similar to a unit Gaussian distribution, consistent with the interpretation that the Bayesian network posterior is the uncertainty in the best prediction to make given the input data.  \textbf{Right}: Autocorrelation time distribution for output metallicities for galaxies in the validation set.  The gray shaded region indicates where autocorrelation times are long relative to sampling time, and so represent unconverged results.  Roughly \modified{70}\% of the validation galaxies have converged metallicity predictions, with about \modified{30}\% of the validation galaxies showing unconverged results.}
     \vspace{-20ex}
     \label{f:resnet_hist}
\end{figure*}

For comparison, the Adam optimizer \citep{Kingma14} had lower average losses and tighter posterior prediction distributions (Fig.\ \ref{f:distributions})---as expected, since we chose the loss tolerance to be nonzero.  At the same time, we note that Adam typically explored in a tight region around the initial position ($\langle |\mx_\mathrm{final}-\mx_\mathrm{initial}|/|\langle \mx_\mathrm{initial}|\rangle \sim 0.3$), whereas the samplers were able to explore locations that were very different from the initial positions ($\langle |\mx_\mathrm{final}-\mx_\mathrm{initial}|/|\langle \mx_\mathrm{initial}|\rangle \sim 3.8$ for HMC and $\sim 3.6$ for ray tracing).  This may be one reason why the posterior uncertainties of HMC and ray tracing are broader than that expected by inflating the Adam posteriors according to $\Delta f_\mathrm{loss}$.  For example, in Fig.\ \ref{f:distributions}, the posteriors for HMC and ray tracing show a 50\% broader range than the Adam posterior for the first halo in the validation set, whereas the 10\% increase in MSE would imply a broader range of 5\% on average.  Section \ref{s:deep_ensembles} contains a broader discussion on the relative merits of deep ensembles vs.\ Bayesian neural networks, and when the choice of each is appropriate.


\subsection{\modified{Approximate} Stochastic Ray Tracing for a Convolutional Neural Network with $\sim$22 Million Parameters}
\label{s:resnet}

A second common scientific application is re-using an existing neural network and transferring it to a new problem after fine-tuning.  Here, we repeat a simplified version of the method in \cite{Wu19}, who used a convolutional neural network (ResNet-34; \citealt{He16}) to predict spectroscopic gas-phase metallicities (i.e., the fractional occurrence of heavy elements relative to hydrogen) for galaxies from input 3-color \textit{gri}-band photometric images from the Sloan Digital Sky Survey Data Release 14 \citep{DR14}. 

\cite{Wu19} found that they could estimate metallicities from galaxy images with a relative uncertainty of $\sim$22\% ($\sim$0.085 dex) compared to the true values.  A natural scientific question is whether the success of their method (and the biases inherent in it) are inherited from the underlying ResNet-34 pre-training, or whether it came from the overall architecture of the network.

Because ResNet-34 networks have $\sim$22 million parameters, and \cite{Wu19} used only $\sim$$10^5$ training images, prolonged training would result in overfitting.  However, with an appropriate MCMC sampler, overfitting is not a problem.    Networks that contain more information (e.g., networks that overfit to a given dataset) have lower entropy and thus occupy a much smaller volume in parameter space (see also the discussion in Section \ref{s:discussion}).  Since learning the underlying manifold (e.g., features in an image that correlate with metallicity) requires dramatically less information than memorizing input--output pairs, the expected likelihood of an overfitted network is vanishingly small.  The caveat to this intuition is that, in many problems, the appropriate likelihood function may not be known \textit{a priori} (e.g., if $D_\mathrm{eff}$ is not known), and so has to be found by trial and error---e.g., by comparing the training loss to the validation loss, as is standard practice.

Full details of the network architecture, hyperparameters, and hardware are presented in Appendix \ref{a:resnet}.  We found that it was straightforward to perform approximate MCMC sampling \modified{in function space (see Section \ref{s:function_space})} with ray tracing on a single consumer GPU following the recipe in Section \ref{s:recipe}, but due to the small mini-batch size (32 images) used, the gradient variance was too high to attempt to sample with HMC.  As expected, training with Adam indefinitely led to overfitting; given that there were many more network parameters than training set images, we expect that the ``best'' achievable training loss would be a perfect memorization of the training set metallicities ($f_\mathrm{loss}=0$).  Following the methodology in \cite{Wu19}, we hence truncated Adam fits after 10 epochs.

Using ray tracing, we could achieve a typical loss similar to or better than the best 10-epoch Adam fits using a loss scaling of $4\times 10^4$, corresponding to $\Delta f_\mathrm{loss}=0.18$ (measured empirically during sampling) and implying an effective number of parameters of $D_\mathrm{eff}\sim$14,000, similar to the number of images in our training set ($\sim$12000).  Results for training vs.\ validation loss are shown in the top-left panel of Fig.\ \ref{f:resnet}.  Aside from the fact that we are performing MCMC on a 22-million parameter neural network, such a plot would be unremarkable.  As expected from the intuition that overfitted networks should occupy vanishingly small regions of parameter space, we were able to sample for $>$500000 epochs without overfitting, with similar validation and training loss throughout.

\begin{figure*}
\capstart
Shortest Autocorrelation Times:\\
\includegraphics[width=\textwidth]{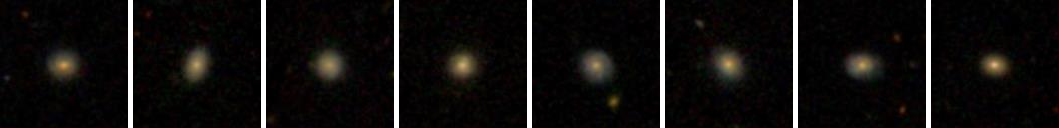}\\[1ex]
Longest Autocorrelation Times:\\
\includegraphics[width=\textwidth]{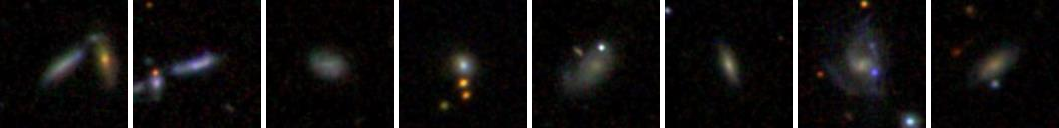}
\caption{Examples of galaxies with the shortest (top row) within-chain autocorrelation times ($<$100 epochs) and the longest (bottom row) within-chain autocorrelation times ($>$$10^5$ epochs).  Galaxies with long autocorrelation times are typically outliers in some sense: they often have unusual colors, mergers, or contamination with stars, leading to unconstrained metallicities.  Colors shown are from SDSS \textit{gri}-band images.}
     \label{f:resnet_acor}
\end{figure*}

Typical scatter between the Adam results and the average of ray tracing sampling was small (5\%, or $\sim$0.02 dex; Fig.\ \ref{f:resnet}, top-right panel) relative to overall network error (22\%, or $\sim$0.09 dex for both Adam and ray tracing; Fig.\ \ref{f:resnet}, bottom-left panel).  This suggests that starting Adam optimization from the ResNet-34 pretrained seed (as in \citealt{Wu19}) did not significantly lower the errors, but it also did not bias the results.

In general, the standard deviations in predicted metallicities across the MCMC posterior are small, of order 5\% (0.02 dex).  As with the comparison to Adam, this is much smaller than the RMSE on the training and the validation sets ($\sim 0.09$ dex; see Fig.\ \ref{f:resnet}, bottom-right panel), which is desired behavior.  This is straightforward to understand.  To minimize the RMSE, the network attempts to estimate $\langle Z(I)\rangle$, i.e., the average metallicity $Z$ for a given image $I$.  Even though there is scatter in $Z$ at fixed $I$ (e.g., from spectroscopic measurement error), the uncertainty in the average metallicity $\langle Z(I)\rangle$ will go to zero in the limit of infinite training data.  More generally, we would like the uncertainty in predictions to be dominated by metallicity scatter at fixed $I$ (irreducible/aleatoric uncertainty in $Z$ for a fixed image $I$) as opposed to the uncertainty in the average metallicity $\langle Z(I)\rangle$ (epistemic uncertainty, reducible with more training data).  Hence, small posterior uncertainties compared to overall RMSE is desired behavior that indicates that we have sufficient training data for most galaxies.

To confirm that our interpretation is correct (i.e., that $\langle Z(I)\rangle$ is well-constrained by the training dataset), we compared predictions from our main run to ray tracing with 1) a different starting seed and 2) a different identically-sized but non-overlapping training set, both after 5000 epochs of training.  We also compared the predictions to a deep ensemble of 1000 Adam runs trained for 10 epochs each starting from the original ResNet-34 weights.  The distribution in the predictions from the main run compared to all three of these alternate predictions is essentially a unit Gaussian (Fig. \ref{f:resnet_hist}, left panel), with (small) deviations due to correlated errors in the shape of the best-fitting manifold for $\langle Z(I)\rangle$.

We show the distribution of autocorrelation times for the metallicities of galaxies in the validation set in Fig.\ \ref{f:resnet_hist}, right panel.  \modified{For a three-chain run, w}e find that $\sim$\modified{70}\% of the galaxies in our validation set have autocorrelation times short enough to be well-sampled, and $\sim$\modified{30}\% do not.  This percentage \modified{is somewhat lower than} past efforts to sample from ResNet-type networks.  For example, \cite{Izmailov21} found 90\% convergence using HMC for Bayesian neural network posteriors on validation set images from CIFAR-10.\footnote{\url{https://www.cs.toronto.edu/~kriz/cifar.html}}  In comparison, \cite{Izmailov21} used a smaller network (ResNet-20, with 270000 parameters instead of $>$22 million in our ResNet-34 network), smoother activation functions (SiLU, instead of ReLU in our network), and more processing power (512 GPUs instead of our single GPU).

It may be tempting to ask why convergence is not achieved for all galaxies in the validation set.  However, the correct question is really the opposite: with 22 million network parameters and only $\sim$12000 training set images, it is surprising that convergence is obtained for \textit{any} of the galaxies in the validation set, given how unconstrained most network parameters are.  Some resolution to both questions can be found in Fig.\ \ref{f:resnet_acor}, which shows galaxies in the validation set with short autocorrelation times (top row) and long autocorrelation times (bottom row).  Visual examination suggests that the galaxies with short autocorrelation times are relatively homogeneous, common galaxies, so that many examples of systems on the same manifold exist in the training set, and it is plausible that the network samples give reasonable uncertainties for their metallicities.  In contrast, the galaxies with long autocorrelation times have unusual colors, mergers, and extensive contamination with stars, leading to fewer corresponding reference points in the training set from which a metallicity could be inferred.  These aspects could be remedied by targeted augmentation of the training set and image precleaning, respectively; however, these aspects are beyond the scope of this paper to address.

\begin{figure}
 \vspace{-4ex}
    \capstart \hspace{-7ex}\includegraphics[width=1.15\columnwidth]{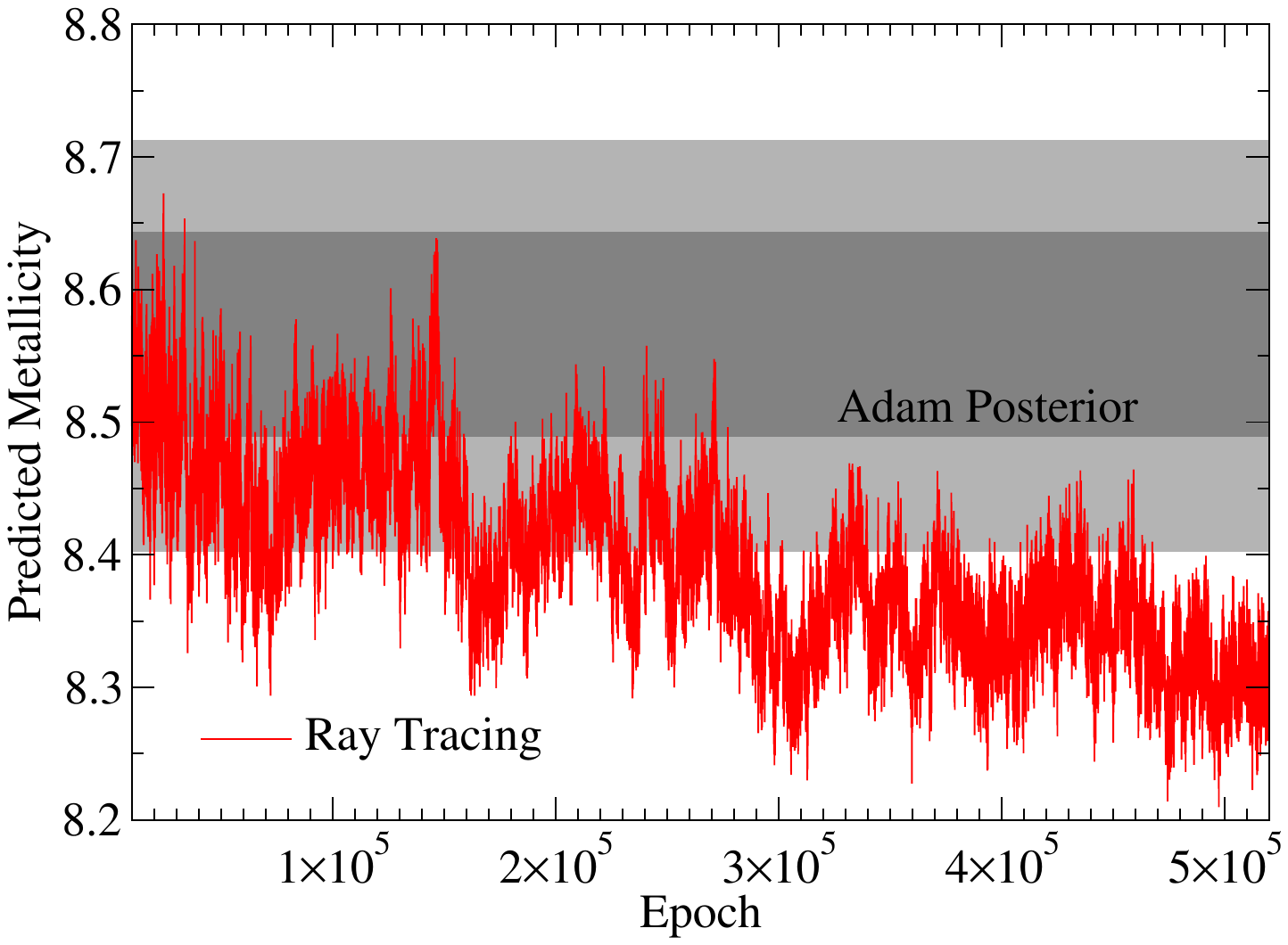}\\[-50ex]
     \phantom{\hspace{56ex}}\includegraphics[width=0.17\columnwidth]{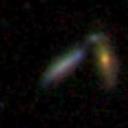}\\[35ex]
     \caption{The sequence of predicted metallicities for a galaxy in the validation set (shown in the center of the inset) with an exceptionally long autocorrelation time.  In practice, the metallicity of the galaxy is unconstrained given the training set and the model architecture, and so the samples do not converge to any fixed distribution over the fraction of the posterior set that has been sampled.  Of note, the distribution returned by Adam (here shown as the 68\% [\textit{dark grey}] and 95\% [\textit{light grey}] confidence intervals for 1000 independent runs) does not capture the true posterior because Adam---like any optimizer---must be run for a limited number of epochs to avoid overfitting.}
     \label{f:long_trace}
\end{figure}

The fact remains that, when a given input does not have many comparable examples in the training set, and when the model is underfitted, the model output is likely to be unconstrained.  As an example, we show the galaxy with the longest autocorrelation time in Fig.\ \ref{f:long_trace}, for which the network output has not converged even after 500000 epochs.  For this galaxy, the distribution of predicted metallicities from our deep ensemble of Adam-trained networks does not encompass the range from ray tracing, as Adam runs must be terminated early to avoid overfitting.  Indeed, the epistemic uncertainty in predicted metallicities for this galaxy ($>$0.2 dex) substantially exceeds both the typical value for other galaxies (0.02 dex) and the typical RMSE in metallicity predictions (0.09 dex).

\subsection{\modified{Approximate} Stochastic Ray Tracing on a Large Language Model with 1.5 Billion Parameters}
\label{s:1B}

\begin{figure*}
\capstart
\hspace{-5ex}\includegraphics[width=1.15\columnwidth]{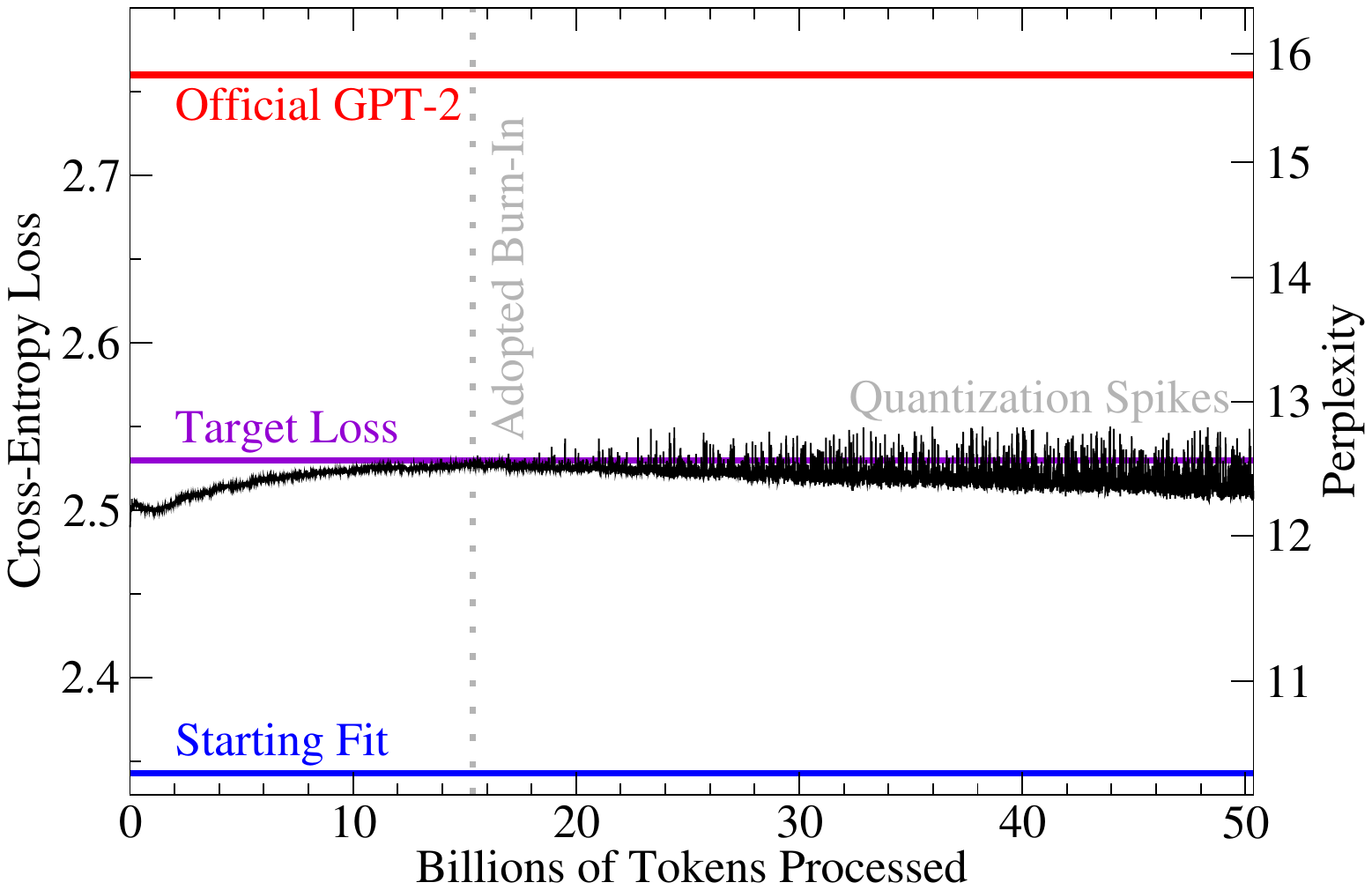}\hspace{-5ex}\includegraphics[width=1.15\columnwidth]{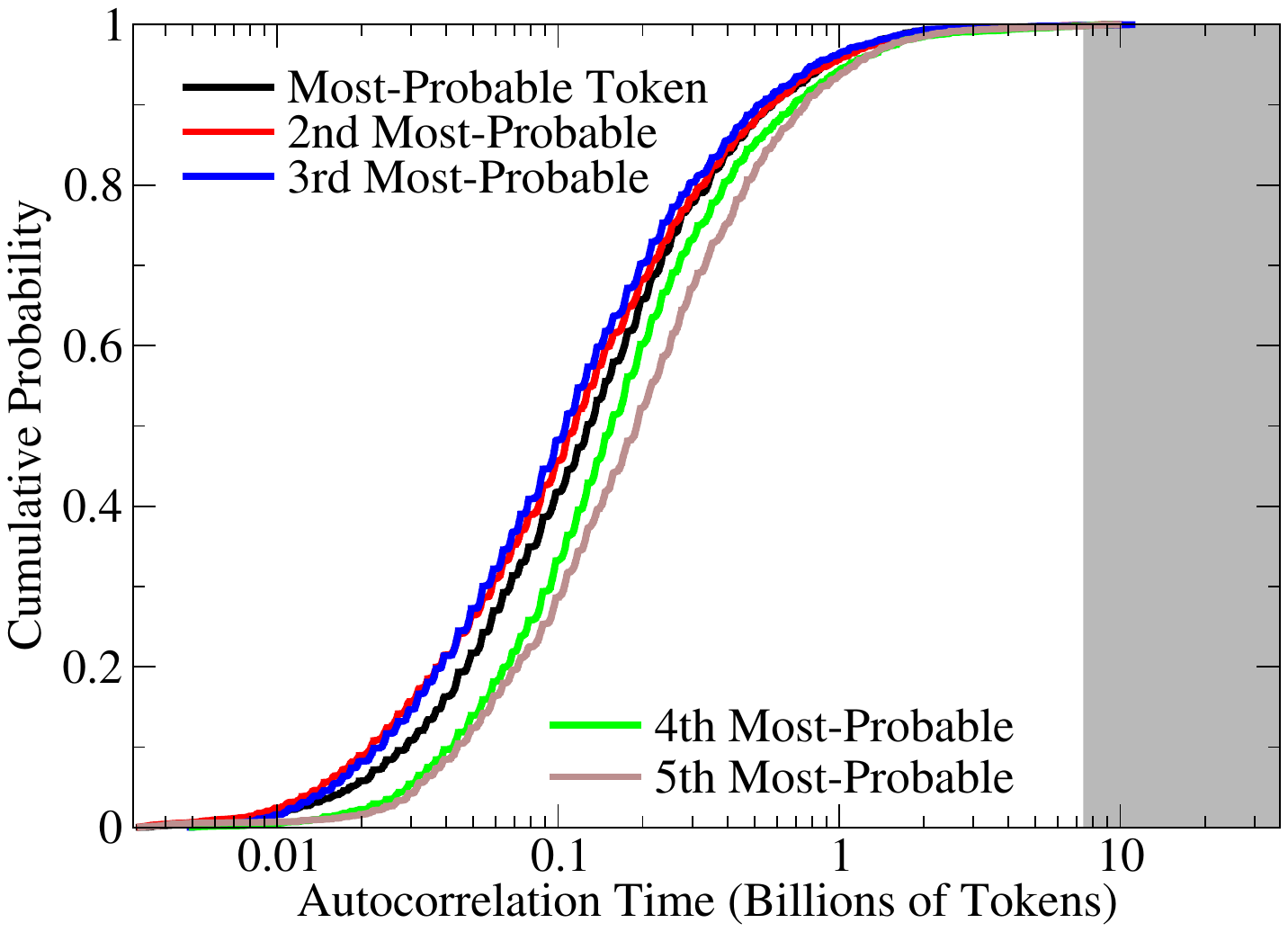}\\[-3ex]
     \caption{\textbf{Left}: Validation loss (cross-entropy) as a function of total tokens processed.  For reference, perplexity (i.e., the exponential of cross-entropy loss) is also shown.  We chose a target loss ($\sim$2.53; purple line) in between the official GPT-2 weights applied to our training set (red line) and the initial trained seed (blue line), corresponding to $\mathcal{L}(\mx) = 4\times 10^{9}f_\mathrm{loss}(\mx)$, and used a step size of 5 (corresponding to $5\times 1024=5120$ tokens).  After long training times, we see quantization spikes---a sharp excursion to high validation loss, followed by an immediate return to the target loss---caused by using BF16 (16-bit) floating point values for parameters.  \textbf{Right}: Cumulative distributions for autocorrelation times of model predictions for  the validation set \modified{from a single MCMC chain}.  For each of the first 5,120 tokens in the validation set, we recorded the probabilities of the top five predicted next tokens, and computed the autocorrelation times of these probabilities as shown above.  Median autocorrelation times for GPT-2 token predictions with ray tracing are around $10^8$ tokens.  \modified{The two-chain result shows identical convergence (Appendix \ref{a:convergence};  Fig.\ \ref{f:vehtari_llm}) albeit at lower autocorrelation time resolution.}}
     \label{f:gpt2}
\vspace{3ex}     
\capstart
\normalsize
Validation text example:\\[-4ex]
\begin{framed}
\input{graphs/val}
\end{framed}
\vspace{-1.5ex}
Color Key: \textcolor[hsb]{0.700,1.000,1.000}{Blue: high entropy}; \textcolor[hsb]{0.900,1.000,1.000}{Purple: long autocorrelation time}; \textbf{\textcolor[hsb]{0,1.000,1.000}{Bold Red: long autocorrelation time and high entropy}}.
\caption{Bayesian exploration of parameters for GPT-2 allows coloring validation text by both entropy and autocorrelation time.   \textcolor[hsb]{0.700,1.000,1.000}{Blue: \textbf{high entropy} ($>4$) but short autocorrelation} (colored text is unpredictable given training set data); \textcolor[hsb]{0.900,1.000,1.000}{Purple: \textbf{long autocorrelation} ($>500$M tokens) and low entropy} (colored text is predictable, but the network struggles to learn the correct probability distribution); \textcolor[hsb]{0,1.000,1.000}{\textbf{Bold Red}: \textbf{long autocorrelation} and \textbf{high entropy}} (colored text is unpredictable, and the network struggles to learn the correct probability distribution).  Existing optimization methods can tell which words are high-entropy, but only Bayesian methods (such as ray tracing) have access to autocorrelation times, which indicates where information flows inefficiently into and out of the network.  There is relatively little overlap between words with long autocorrelation times and words with high entropy.  The validation text shown is from the \texttt{fineweb-edu} dataset, using text originally from \citet{Krashen2000}.}
\label{f:gpt2_example}
\end{figure*}

\begin{figure}
\vspace{-4ex}
    \capstart
\hspace{-10ex}\includegraphics[width=1.15\columnwidth]{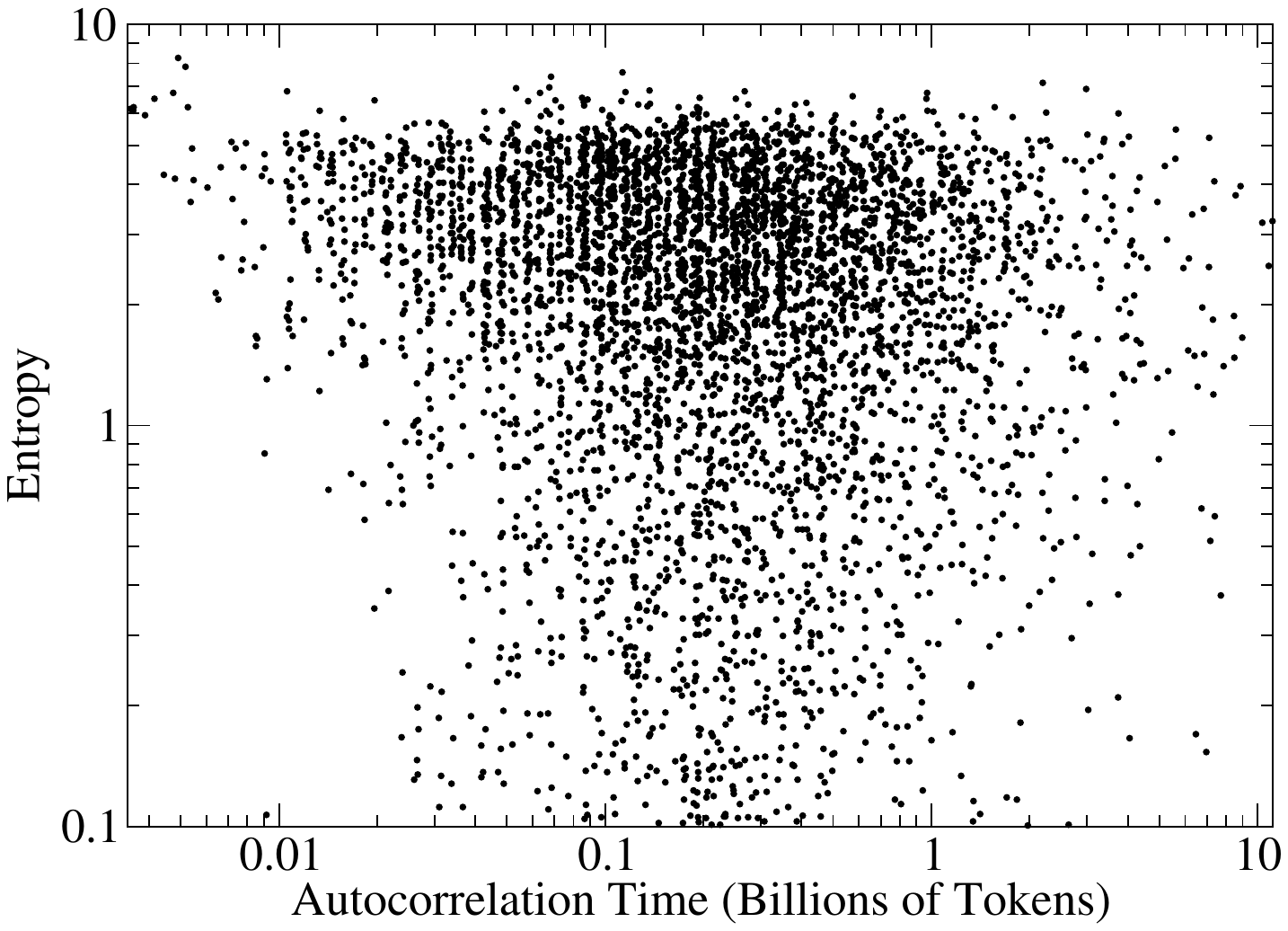}\\[-4ex]
    \caption{Autocorrelation time and entropy are uncorrelated across our validation dataset. This figure shows entropy vs.\ autocorrelation time for the most-probable next token, in units of billions of tokens; there are roughly 2$\times 10^5$ steps per billion tokens.   More efficient network architectures than GPT-2 would result in more efficient information flow into the network for a given amount of tokens processed, which would manifest as lower autocorrelation times.}
    \label{f:gpt2_autocorrelation}
 \capstart
    \vspace{5ex}
    Prompt: ``\textit{A patient recently returned from an overseas trip and is suffering from fever, headache, nausea, joint pain, fatigue, coughing, and abdominal pain.  The patient should be diagnosed with }''\\[-10ex]
    
    \hspace{-10ex}\includegraphics[width=1.15\columnwidth]{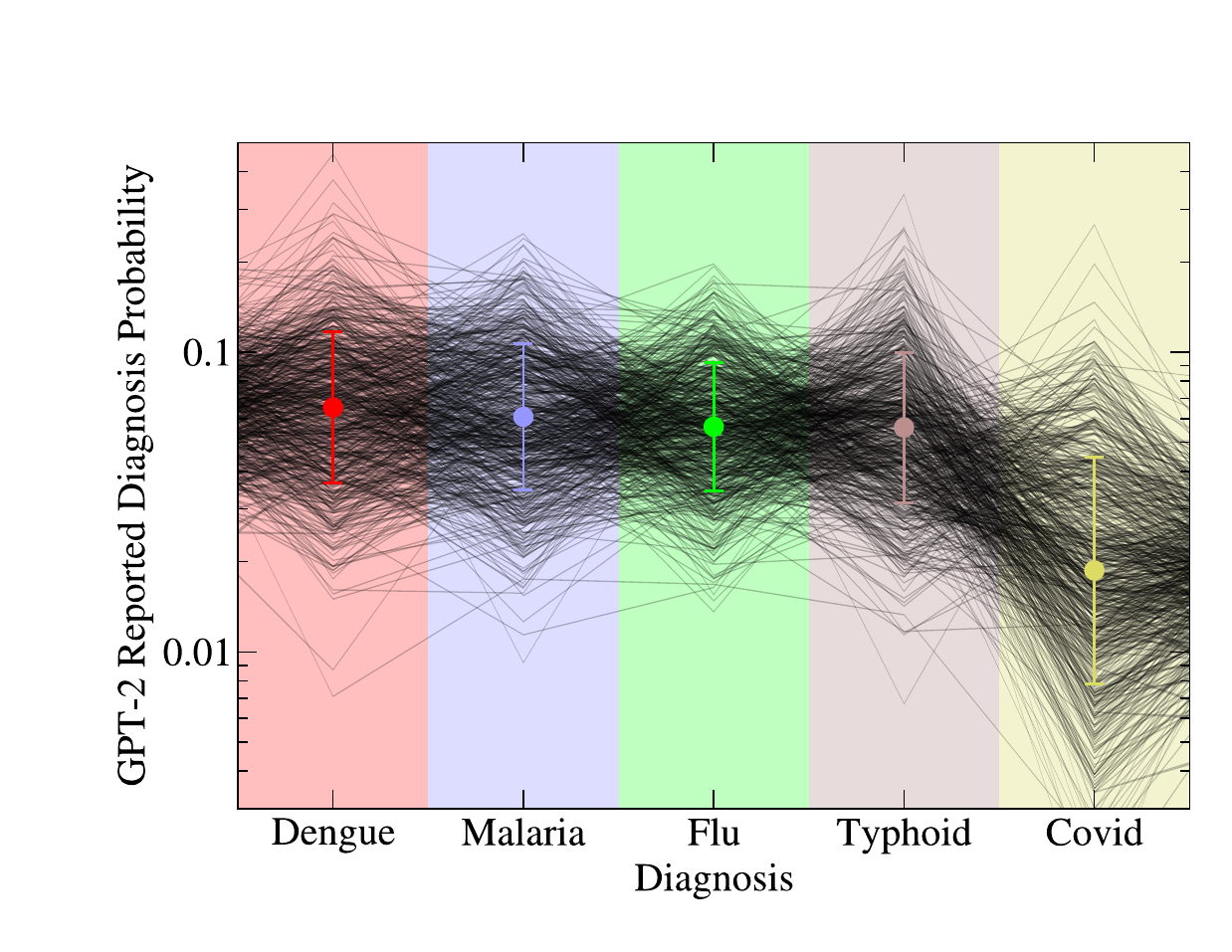}\\[-4ex]
    \caption{Bayesian sampling allows quantifying the uncertainty associated with a given input data vector, in this case, a query about a patient’s diagnosis given a list of symptoms (shown at the top of the figure).  This figure shows the distribution of answers across different diagnoses (i.e., the GPT-2 Reported Diagnosis Probability, or $P($diagnosis$|$model,symptoms$)$).  Data points and error bars show the averages and standard deviations across all sampled models, and grey lines show individual model realizations.    The symptoms in the prompt were drawn from the Mayo Clinic page for typical symptoms of malaria.  Different neural networks gave a wide variety of answers, with malaria and four other illnesses with overlapping symptoms among the most common responses.  Although 23\% of networks gave malaria as the most preferred diagnosis (i.e., more likely to be a response than any other illness), most of the sampled networks gave another preferred diagnosis.  Hence, trusting the likelihoods from a single realization (even one trained on the large number of medical journal papers in the \texttt{fineweb-edu} dataset), may mislead practitioners away from the correct underlying distribution of diagnoses.}
    \label{f:gpt2_posterior}
\end{figure}

The frontier of large neural networks has been in transformer-based networks \citep{Vaswani17} including large language models (LLMs), which now reach above 1 trillion parameters \citep[e.g.,][]{Du21,OpenAI23}.  Large language models are seeing increasing usage in astronomy and other scientific fields \citep[e.g.,][]{Fouesneau24,Ting25,Zheng25}.  As access to the GPU clusters required to train such models is not widespread, we demonstrate instead that the ray tracing algorithm scales well to the 1.5 billion-parameter GPT-2 LLM \citep{radford2019language}, which is one of the largest models that can be trained on consumer hardware.  Text prediction networks, like GPT-2 and later variants, operate by predicting the next token (i.e., text segment) in a passage given the passage's previous $N_\mathrm{context}$ tokens ($1024$, for GPT-2).  The exact meaning of ``token'' depends on the LLM, but for GPT-2, tokens are largely complete words and morphemes (i.e., fragments of words that carry meaning, such as ``con'' or ``ing'').  

In this section, we perform approximate Bayesian sampling on GPT-2, \modified{ again in function space (see Section \ref{s:function_space})}.  As usual, full details of the hardware and network architecture are in Appendix \ref{a:gpt2}.  Although the GPT-2 weights and network geometry are public, the training data are not, so we use a 100-billion token subset of the \texttt{fineweb-edu} training set \citep[as implemented in \texttt{llm.c}\footnote{\url{https://github.com/karpathy/llm.c}}]{Penedo24}.  We begin from a public pretrained network with the GPT-2 architecture\footnote{\url{https://huggingface.co/karpathy/gpt2_1558M_final3_hf}} on the \texttt{fineweb-edu} dataset.  For the loss, we use the standard cross-entropy metric, defined as:
\begin{equation}
    f_\mathrm{loss}(\mx) = \langle -\ln(p_\mx(y|C))\rangle,
\end{equation} 
where the average is taken over all output tokens $y$ in the training set, with $p_\mx(y|C)$ the probability that a given model with parameter values $\mx$ will return the correct token $y$ given the preceding context tokens $C$.  We adopted a likelihood of $\mathcal{L}(\mx) = 4\times 10^9 \cdot f_\mathrm{loss}(\mx)$, corresponding to $\Delta f_\mathrm{loss}=0.1875$ (assuming $D_\mathrm{eff}=1.5\times 10^{9}$).  Intuitively, this lets us explore models that have an accuracy within about $\exp(\Delta f_\mathrm{loss})\sim 20$\% of that of the seed model (which we label $\mx_0$).    We adopt the burn-in time to be the point at which our target loss ($f_\mathrm{loss}(\mx_0)+\Delta f_\mathrm{loss}\sim 2.53$) is reached, which occurs after roughly 3 million steps ($\sim$15 billion tokens).

As shown in the left panel of Fig.\ \ref{f:gpt2}, we can sample from the posterior distribution of GPT-2, maintaining roughly constant validation loss, confirming that even the small mini-batch size used ($5\times N_\mathrm{context}$) does not lead to runaway trajectory heating.  We do find that the validation cross-entropy loss shows brief spikes to values as high as 20 for a single step, followed by an immediate return to the typical loss ($\sim$2.53).  Further tests showed that these spikes occur even when the step size is reduced (e.g., by a factor 5, which would be expected to result in $\sim$ 25$\times$ less heating).  We hence attribute these spikes to the quantization of the network, which uses a 16-bit floating point format (BF16) instead of the 32-bit format (F\modified{P}32) used for the other networks in this paper.  \modified{Using F\modified{P}32 or F\modified{P}64 would presumably have resulted in fewer quantization spikes (i.e., holes in parameter space), at the cost of significantly longer runtimes and higher memory use, both of which would have precluded using consumer hardware for this analysis.}  We adopt the masking approach in Section \ref{s:pseudocode}, effectively skipping the 4\% of samples where the validation loss exceeds 2.55 (corresponding to our target loss plus twice the expected uncertainty in the validation loss).

Along with the quantization spikes, we also note that the loss shows a very gradual decline below the target value, such that the final model in the chain has a 2\% better chance of predicting tokens in the validation set than the first model where the target loss is reached.  Qualitatively, this suggests that the initial seed $\mx_0$ was not at the global minimum, and so the network as a whole is slowly migrating towards a different likelihood basin around a better minimum.  We note, nonetheless, that we are interested here in an \textit{approximate} sampling of the posterior distribution---and that networks that perform within 2\% of each other are substantially interchangeable in practice.

The right panel of Fig.\ \ref{f:gpt2} shows the autocorrelation times for a subsample of the network predictions on the validation set.  Specifically, for each token in the first mini-batch of the validation set (5120 tokens total), we sample the predicted probabilities for the five most-probable next tokens.  This gives a $5\times$5120-dimensional projection of the network's full output.  We report the autocorrelation times for each dimension in units of billions of tokens, i.e., how many tokens are required to be processed through the ray tracing algorithm before the dimension reaches an independent sample. Fig.\ \ref{f:gpt2} shows that the median autocorrelation time for predictions of GPT-2 is around 10$^8$ tokens.  For comparison, this is about the same number of words as in 1000 books, or about the expected number of books read by a single adult in their lifetimes.\footnote{\url{https://www.pewresearch.org/internet/2016/09/01/book-reading-2016-appendix-a/}}  However, there is a long tail to very high token counts ($\gg$$10^9$), which suggests that 1) some concepts are rarely mentioned in the training set, and/or 2) the network is too inflexible to adapt easily to new interpretations and requires repeated exposures to the same information (see also Section \ref{s:information}).     The algorithm reaches (with the caveat above that we are performing approximate sampling) greater than 99.\modified{7}\% convergence across the dimensions probed\modified{, using two chains with different initial velocity seeds}.  This is substantially higher than for the ResNet-34 network in the previous section because our training set size is much larger than the network parameter count.

One key finding of this analysis is that entropy and autocorrelation times are uncorrelated.  This is shown in the example validation set text in Fig.\ \ref{f:gpt2_example} as well as more quantitatively in Fig.\ \ref{f:gpt2_autocorrelation}.    Here, we define entropy in the usual sense as $H(C,\mx) = -\langle \ln(P(T|C,\mx))\rangle$---i.e., the negative average log-probability for the next token $T$ at a given model location $\mx$ and for given context tokens $C$.  Large entropy values mean that there could be many possibilities for the next token $T$, whereas low entropy values mean that there are relatively few possibilities.  For a human text-predictor, high entropy means either 1) the human has not been exposed to a topic, and so has no basis to know what may come next, 2) there are many possible directions to take the phrase, and/or 3) there are many possible word choices for the same content.  For a machine predictor, the first explanation does not apply (due to exceptionally large training sets), and so entropy is essentially synonymous with ambiguity, which depends on the specific content and language (English, in this case).  Autocorrelation times, in contrast, are a measure of information flow rates into and out of the network, which depend on network structure, parameter count, and the distribution of content topics in the training set.  As a result, it is plausible that the two concepts measure orthogonal quantities.  

Returning to more applied topics, large language models are routinely used to answer queries across scientific fields.  An open question is how uncertain the responses from large language models are, which is relevant to how much trust should be placed in their responses.  This is especially critical for fields such as medical diagnostics \citep[see][for a review]{Zhou25}.

Bayesian sampling provides a way to answer this question, so we perform a simple experiment.  We sample the approximate posterior distribution of GPT-2 networks, and then ask each sample network to complete a sentence asking for a diagnosis for patient symptoms.  For the patient symptom list, we use symptom examples from the Mayo Clinic page for typical symptoms of malaria.\footnote{\url{https://www.mayoclinic.org/diseases-conditions/malaria/symptoms-causes/syc-20351184}}  As each sample network outputs a range of potential diagnoses, we draw 10$^4$ samples from each network, and classified the diagnoses by disease type.  This enables us to extract an uncertainty on the reported network likelihoods.  As an example, a single network might place the chance at 10\% of the time that the patient has malaria---but different networks give a \textit{range} of responses for what this chance should be, from 3\% to 20\%.  By analogy, this experiment would be similar to asking multiple doctors for their opinions on what a patient's underlying disease most likely is.

As shown in Fig.\ \ref{f:gpt2_posterior}, we found that different networks (like different doctors) preferred different diseases, and gave a wide range of estimated confidences (standard deviations of 0.2--0.4 dex).  This is a toy example, in that the networks were trained on medical text in the \texttt{fineweb} data set, as opposed to actual patient symptom---diagnosis pairs.  However, it shows that, at the scale of GPT-2, individual network prediction likelihoods can be widely variable.

We repeat, as above, that this example shows only approximate sampling of GPT-2 on a single consumer-level GPU, which is much less powerful than the hardware used to train GPT-2 (or, in fact, any more modern LLM).  This suggests the hypothesis that, in general, the hardware used to train a given LLM will also be sufficient to perform approximate Bayesian sampling of its posterior distribution, which will be tested by future research.

\section{Discussion}

\label{s:discussion}


\subsection{Pedagogy on the Meaning of Bayesian Sampling}

We feel that a key impact of ray tracing will be enabling widespread approximate Bayesian sampling for neural networks, having demonstrated that sampling with ray tracing is possible for problems with more than a billion parameters.  Since not all readers will have previously done Bayesian sampling of neural networks, some pedagogy may be helpful.  

At a basic level, the meaning of Bayesian networks depends on the network outputs.  In many past scientific papers, neural networks have been used to predict a single ``best'' output prediction given the input data.  In these cases, Bayesian sampling will result in the distribution of possible ``best'' predictions reflecting the underlying dataset.  In the limit of infinite training data, the predicted distribution will converge to an infinitesimally thin manifold, since there is only a single description of the training data that minimizes the loss.  This is similar to the case of linear regression---in the limit of infinite data, there is no uncertainty in the location of the line that best fits the data.  Hence, the sampled predictions for single-output networks should not be interpreted as the distribution of potential outcomes, but instead as the uncertainty in the \textit{average value} of the potential outcomes for similar inputs (see also Section \ref{s:resnet}).

When, the distribution of potential outcomes is desired instead, this requires building and training a network that generates a probability distribution for its output.  Such networks are also known as neural posterior estimators (NPEs), with common examples including normalizing flows, diffusion models, and large language models.  \modified{As noted in the introduction, the output of an NPE is the posterior $P(\mathbf{y}|\mathbf{D},\mathbf{P}, \mx)$, where $\mx$ are the model parameters.}  In these cases, Bayesian sampling \modified{is necessary to convert these conditional distributions into the desired posterior, via the law of total probability: $P(\mathbf{y}|\mathbf{D},\mathbf{P}) = \sum_\mx P(\mathbf{y}|\mathbf{D},\mathbf{P}, \mx)P(\mx|\mathbf{D},\mathbf{P})$.  The error from assuming that the direct NPE output approximates the true posterior $P(\mathbf{y}|\mathbf{D},\mathbf{P})$ can be significant}, as shown \modified{for large language models} in, e.g., Fig.\ \ref{f:gpt2_posterior}.  This is \modified{also} helpful in cases where wrong-but-confident outputs from a single model would cause problems, as Bayesian sampling would then reveal the true diversity of posterior estimates consistent with the training data \modified{(see also Section \ref{s:resnet})}.

\subsection{Sampling Temperature: Deep Ensembles vs.\ Samplers}

\label{s:deep_ensembles}

The appropriate temperature for sampling and uncertainty quantification has led to some disagreement in the literature over the best approach.  Deep ensembles \citep[e.g.][]{Lakshminarayanan16} are effectively zero-temperature sampling, while \cite{Wenzel20} argue that ``cold posteriors'' (temperatures $\tau<1$) are preferred and \cite{Izmailov21} argue for normal Bayesian posteriors ($\tau=1$).  We advocate that $\tau=1$ is the most defensible answer, but note that the correct likelihood function changes depending on the desired outcome.

It's helpful to consider a concrete example.  For a typical supervised learning task to predict a single output value with mean-squared error loss, the log-likelihood might be:
\begin{equation}
    \ln(\mathcal{L}(\mx)) = -\sum_i\frac{\left(\mathbf{f}_\mx(\mathbf{y}_{i})-\mathbf{v}_i\right)^2}{2\sigma_i^2}+G(\mx),
\end{equation}
where $\mathbf{y}_{i}$ are the input training data, $\left(\mathbf{f}_\mx(\mathbf{y}_i)\right)$ are the network outputs, $\mathbf{v}_i$ are the training labels, $\sigma_i$ are the errors in the training labels, and $G(\mx)$ is the (unknown) generalization loss---i.e., the loss that would be present if the training set included all possible samples with the correct weights.

For many supervised learning tasks, there is no label uncertainty ($\sigma_i=0$), and the log-likelihood is infinitely negative.  Hence, sampling in a proper Bayesian sense ($\tau=1$) is equivalent to sampling a rescaled, finite log-likelihood (e.g., $\ln(\mathcal{L}_s(\mx)) = -\sum_i \left(\mathbf{f}_\mx(\mathbf{y}_i)-\mathbf{v}_i\right)^2$) at zero temperature.  Deep ensembles effectively use this approach, under the heuristic that the global minimum (if $G(\mathbf{x})$ were known) is likely to be contained within the range of local minima reached during training with different seeds; conformal prediction \citep[e.g.,][]{Angelopoulos21} can then regularize this heuristic estimate.

However, in many scientific fields, there \textit{is} label uncertainty (e.g., due to measurement uncertainty) that has to be taken into account.  In addition, the marginal coverage guarantees of conformal prediction are often not sufficient, as scientists often need \textit{conditional coverage}: that is, we wish to know the uncertainty for a given measurement (for example, the uncertainty in cosmological parameters for the single Universe that we can observe), as opposed to the uncertainty across all potential measurements (e.g., universes in a simulated test set).  Similar needs occur in other fields, e.g., medicine and self-driving cars, where inaccurate conditional coverage can be life-threatening---that is, we wish to avoid wrong-but-confident outputs.   Bayesian neural networks in the traditional sense, with $\tau=1$, are currently the only method that provides conditional coverage while simultaneously admitting the label uncertainty necessary in many scientific fields. 


\subsection{Information Flow and Improving Network Architectures}

\label{s:information}

In the limit of an arbitrarily large number of dimensions, MCMC samplers effectively explore a constant-loss surface, or equivalently, a constant-information surface.  In the latter framing, during sampling, information is constantly flowing in and out of the neural network, with each new ingestion of information being accompanied by another piece of information flowing out.  This gives a heuristic interpretation of why MCMC sampling results in uncertainty quantification, as it is in effect equivalent to bootstrap resampling of the training set (which can also be done with deep ensembles; see \citealt{LoaizaGanem25}).

Information flow also provides a way to compare different network architectures.  Heuristically, ``better'' architectures can absorb information more easily (and more easily forget/replace it when better information arrives).  We could hence define a mean log information flow rate as:
\begin{equation}
    \langle \ln \dot{I}\rangle = -\langle E \rangle - \langle \ln \tau\rangle,
\end{equation}
where $\langle E\rangle $ is the average log-loss (e.g., the usual cross-entropy loss for a large language model), and $\langle \ln \tau\rangle$ is the average log autocorrelation time of, e.g., the network outputs for the validation set in units of bits.\footnote{In this framing, an output that is static should be treated as having an indefinitely long autocorrelation time.}  This is very simple to calculate---e.g., for the GPT-2 model in Section \ref{s:1B}, it is about $-32$.

Since this measure is quantitative, it can be used on-the-fly to reconfigure network architecture during training.  At a basic level, a range of network architectures could be generated using a genetic algorithm, trained in parallel, and the architectures with the best information flow rates could be used as the basis of a new generation, and the process iterated, akin to natural selection of the architectures that are most resource-efficient.  This method may be helpful in finding architectures that perform better in data-limited fields, including large language models and image/video generation models.

\subsection{\modified{Interpreting the Effective Number of Parameters $D_\mathrm{eff}$}}

\label{s:deff}

\modified{We interpret the effective number of parameters ($D_\mathrm{eff}$, Eq.\ \ref{e:approx_likelihood}) to be the dimensionality of the manifold of network parameters that fit the training set to the desired loss tolerance.  If $D_\mathrm{eff}$ is much less than the number of network parameters, it implies that a smaller network could be used to fit the same data, because the likelihood gradient is not changed by motion along many directions in network parameter space.   For example, in Appendix \ref{a:sensitivity}, we show that the implied $D_\mathrm{eff}$ is stable for networks varying by over $10\times$ in parameter count applied to the task in Section \ref{s:um}.  Vice versa, if $D_\mathrm{eff}$ is of the same order as the number of network parameters (as in Section \ref{s:1B}), it implies that the network parameter space is largely constrained by the data, which may imply that a larger network would achieve a lower loss (e.g., as in \citealt{OpenAI23}).}

\modified{Similar intuition applies to comparing $D_\mathrm{eff}$ to the training set size.  If $D_\mathrm{eff}$ is similar to the training set size, it implies that each training point is constraining a different dimension of the best-fitting manifold, which raises the concern that the training set size is too small to give robust constraints on all network parameters (as in Section \ref{s:resnet}).  Vice versa, if $D_\mathrm{eff}$ is much less than the training set size, it implies redundancy across data points, and---provided that the training set is a representative sample---suggests that the training set is sufficient to constrain all dimensions of the underlying manifold, even if imperfectly.}

\modified{Given the above, we hypothesize the following simple heuristic equation for the value of $D_\mathrm{eff}:$}
\begin{equation}
    \mmodified{D_\mathrm{eff} \sim \min\begin{bmatrix} \textrm{Number of network parameters} \\ \textrm{Number of elements in training set} \\
    \textrm{Problem dimensionality}
    \end{bmatrix}}.
\end{equation}

\subsection{Generalization of Bayesian Networks}

We caution that Bayesian sampling does not always imply improved generalization (see also \citealt{Izmailov21}).  This is because there are at least two key types of generalization that are relevant: 1) \textit{extrapolation}, that is, generalization to inputs on the same manifold as the data but outside the training set, and 2) \textit{problem transfer}: that is, generalization to inputs on different manifolds.  Bayesian sampling plausibly gives reasonable extrapolation, though it will not magically be better than deep ensembles---the reason being that, by definition, there is no information in the training set on how the extrapolation should be done.  The only advantage that Bayesian sampling may have here is that it is less prone to overfitting than optimization algorithms.

We also expect that Bayesian sampling will not magically lead to better problem transfer.  This is very simple to understand: a network that does well on multiple input manifolds contains more information than a network that does well on only a single input manifold.  Hence, even though both networks will perform equally well on the training sample, the volume of parameter space that contains multi-function networks will be exponentially suppressed compared to the volume that contains single-function networks.

It is not always easy to guess whether a given generalization will be extrapolation or problem transfer, since our intuition may lead us astray.  For example, while we might intuit that adding noise to an image should fall into the extrapolation category, our brains have been trained since birth to recognize our surroundings in varying levels of light.  In contrast, if networks have only been trained on low-noise images in which single pixels are informative, they may not be equipped to deal with the de-localized information present in noisy images.  We can also give a more human example of dialects.  If you were raised on American English, and a colleague told you that she thought things were sorted and Bob's your uncle, but the head boffin was a few sandwiches short of a picnic basket and then everything started going pear-shaped---that would be on a different manifold from what you were expecting English to be.  Different dialect manifolds can therefore contain mutually incomprehensible content for speakers of the ``same'' language, and so speakers used to one manifold can be helpless to generalize to another.

The straightforward recommendation for generalization is to use a domain adaptation approach (e.g., Universal Domain Adaption, \citealt{You19}), which can both test whether a new data set is distinguishable from the training data set, and can remove information from the training data set that distinguishes it from the target data set (and vice versa), so that only common information is used.  Example past applications to astronomical problems have included galaxy morphology classification \citep{Ciprijanovic23}, strong lensing \citep{Agarwal24}, and estimating cosmological parameters \citep{Roncoli23}.

\subsection{Misalignment and Hallucinations}

Misalignment refers to a difference between training goals and what a model actually learns to do.  One consequence is hallucinations, in which a model that is trained to provide helpful responses instead invents false information.  Bayesian neural networks have increased chances for misalignment and hallucinations (often considered negative traits) and for creativity (often considered a positive trait), because they represent the variety of approaches necessary to perform well on the training set.  At the same time, it is unlikely that different models will be misaligned in exactly the same way, unless overfitting has occurred (e.g., in deep ensembles).  

We note simply that we have lived with misalignment among humans for a long time.  The modern human solution for misalignment has been to avoid giving full decision-making power to any single person, implicitly expecting that different misalignments between different people will average out.  In that sense, when misalignment is a concern, it may be better to average decision-making power across different samples from a Bayesian network.  Similarly, we hypothesize that it is less likely that different samples will have the same hallucinations, whereas it is more likely that non-hallucinated information is shared across multiple models (see also Fig.\ \ref{f:gpt2_posterior}).  Future Bayesian exploration of large language models will test these hypotheses.

\subsection{\modified{Performance of Constant-Speed Samplers vs.\ HMC for Non-Stochastic Gradients}}

\label{s:performance_claims}

\modified{Multiple papers \citep[e.g.,][]{Bayer23,Crespi25,Sommer25} have claimed enormous efficiency advantages for constant-speed samplers as compared to traditional HMC samplers with perfect (non-stochastic) likelihood gradients for high-dimensional problems.  We believe that these efficiency differences are real, but we expect that they are largely due to implementation choices instead of dynamical differences between HMC and constant-speed samplers.}

\modified{First, HMC samplers are expected to have almost perfectly constant speed for high-dimensional problems.  HMC velocities are typically drawn from a Gaussian distribution, which has $P(\mv)\propto \exp(-\mv^2/2)$.  By the central limit theorem, the expected velocity variance $\sigma^2_\mv$ drops as $D^{-1}$, and so HMC in practice will behave like a constant-speed sampler.  This is derived more rigorously in Section \ref{s:hmc_connection}, which shows that under typical conditions, the sampling density of HMC asymptotes to that of a constant-speed sampler.  As shown in Appendix \ref{a:further_generalizations}, equal sampling density implies identical dynamics for samplers that can be written in terms of an index of refraction $n(\mx)$, including both HMC and constant-speed samplers.  That is to say, for problems with non-stochastic likelihoods, HMC and constant-speed samplers take \textit{identical} paths in the limit of high dimensionality $D$ for typical distributions.  This is borne out by the tests in Section \ref{s:um} and Appendix \ref{a:hmc}, where we find relatively little difference in sampling efficiency between HMC and constant-speed sampling after step size and path length optimization.}

\modified{With the above said, there are more variables than the choice of sampler dynamics, including the choice of tuning method, continuous momentum refreshment, and the question of whether to use a Metropolis test.  HMC tests have typically been performed with a Metropolis test and using the No U-Turn dynamic tuning method \citep[NUTS;][]{Hoffman11}.  As shown in \cite{Robnik24}, removing the Metropolis test offers a large performance improvement for \textit{both} HMC and constant-speed samplers, since it changes the definition of how the sampling accuracy is evaluated.  In particular, if it is simply necessary to have correct marginal distributions within a given tolerance, then the sampling efficiency of both HMC and constant-speed samplers can be made constant with dimensionality instead of dropping as $D^{-1/4}$, which gives significant improvements for high-dimensional problems.}

\modified{The choice of tuning also has an impact, which depends on the problem.  J.\ Robnik (priv.\ comm.) estimates an impact of roughly a factor of two from using NUTS vs.\ an optimized constant sampling method. Lastly, the different step sizes for HMC likely impact its performance for certain pathological distributions (e.g., Neal's Funnel; \citealt{Neal03}) where the likelihood variance does not decrease with increasing dimensionality (per J.\ Robnik and U.\ Seljak, priv.\ comm.).}

\modified{Given this context, we can summarize how Bayesian sampling has been advanced by recent literature.  \cite{Izmailov21} represents a successful, brute-force approach to using HMC to explore neural network likelihoods for networks with $<3\times 10^5$ parameters, using perfect likelihood gradients.  \cite{Sommer25} uses constant-speed sampling for neural networks with $10^3-10^5$ parameters, again with perfect likelihood gradients.  We attribute their obtained order of magnitude speedup over HMC to their choice of tuning and lack of Metropolis adjustment, rather than sampler dynamics, which we expect to be very similar to HMC.  In this paper, we explore networks of $>10^9$ parameters, made possible by the use of stochastic likelihood gradients, which provide speedups of multiple orders of magnitude over perfect likelihood gradients, provided that constant-speed sampling is used instead of HMC.}

\subsection{Future Directions}

While the general problem of efficient exploration of neural network weight space remains unsolved, the techniques developed here suggest some promising directions.  The basic problem is that neural networks have a filametary structure, such that the most direct path between two points on the typical set is over a large potential barrier---i.e., locations that should never be traveled by any \modified{continuously fair} sampler.  Hence, the basic solution  to weight space exploration must involve exiting the typical set and re-entering it.  \modified{This involves the sampler becoming unfair (as it leaves the typical set) and then fair again (as it returns) along its trajectory in a reversible way.  Exactly such a sampler is presented in Appendix \ref{a:further_generalizations} (the ``likelihood arc'' sampler), which will be tested on neural networks in future work (Behroozi et al., in prep.).}  In terms of the information stored in the network, this kind of sampler functions somewhat like a diffusion model---i.e., information leaves the network during the first half of its trajectory, and then different information is learned in the second half, bringing the sampler to a new place in the typical set.

\section{Conclusions}

\label{s:conclusions}

In this paper, we developed a \modified{family of} gradient-based MCMC method\modified{s} that perform ray tracing with a spatially-varying refractive index.  Our main conclusions are:
\begin{itemize}
\item The \modified{simplest} method, \modified{equivalent to isokinetic sampling,} performs well for sampling function spaces of neural networks, including those with more than a billion parameters, even on consumer hardware (Section \ref{s:applications}).  These results are possible because the method is extraordinarily robust to stochastic gradient errors, showing orders of magnitude lower bias than Hamiltonian Monte Carlo (HMC) at the same step size (Section \ref{s:gaussian_sg}).
\item The method\modified{s} provide fair sampling, even for imperfect integrators, and can pass through arbitrary likelihood and parameter space barriers (Section \ref{s:methods}).  The latter ability resolves a key challenge with past methods.
\item The \modified{family} generalizes straightforwardly, encompassing many other existing methods (HMC, \modified{first-generation} microcanonical HMC, Gibbs, stretch-step, Metropolis, etc.) with the appropriate choice of weighting (Section \ref{s:prior}).  This allows straightforward comparison of the relative time each method spends exploring high-probability vs.\ low-probability regions.
\end{itemize}

\section*{Acknowledgements}
PB was funded by a Packard Fellowship, Grant \#2019-69646, as well as an NSF EAGER grant, \#2404989.  PB thanks Oddisey Knox for asking about a homework problem involving a spatially varying refractive index, and PB thanks his father for encouraging him to pursue an idea that PB initially thought would be no more efficient than existing methods.  PB also thanks Andrej Karpathy for open-sourcing the \texttt{llm.c} and \texttt{nanoGPT} implementations.  Finally, PB thanks Adrian Bayer, Haley Bowden, \modified{Johannes Buchner}, Suchetha Cooray, Carol Cuesta-Lazaro, ChangHoon Hahn, Andrew Hearin, Ben Horowitz, Marc Huertas-Company,  Christian Jespersen, Elisabeth Krause, Guilhem Lavaux, Surhud More, Radford Neal, \modified{Jacob Robnik}, \modified{Uro\v{s} Seljak}, Masahiro Takada, Leander Thiele, Yixin Wang, John Wu, and Haowen Zhang for helpful conversations and comments that improved this paper.  This research was supported in part by grant NSF PHY-2309135 to the Kavli Institute for Theoretical Physics (KITP).

\textbf{Large Language Model Use Disclosure:} We used ChatGPT from September 2024 until final paper submission to assist with code drafting and debugging; all suggestions were checked by hand for correctness.  Similarly, we used ChatGPT to suggest methods to draw random samples from Rosenbrock and Cauchy distributions in Appendix \ref{a:hmc}, as well as to derive the log-likelihood distribution for Cauchy distributions; as above, all derivations were checked by hand for correctness.  Writefull and ChatGPT suggested improvements to the text of the paper, which were again checked by hand.  Finally, as described in Section \ref{s:1B}, we sampled from the posterior distributions of GPT-2 like models to predict successive tokens, to estimate autocorrelation times and epistemic uncertainties for GPT-2 like models.




\bibliographystyle{mnras}
\bibliography{refs} 



\appendix
\section{Lambertian Emissivity Without Full Surface Normal Information}

\label{a:lambertian}

Supposing a ray bundle has intersected with a differential area $dA$ of a parameter space boundary (e.g., as in Fig.\ \ref{f:lightbeam}), we wish to choose a new direction $\hat{\mathbf{s}}$ according to a Lambertian emission distribution.  As in Section \ref{s:radiance}, this corresponds to uniform radiance $L$, or equivalently, an output power $P=L\cos(\theta)dAd\Omega$ that is proportional to the cosine of the angle $\theta$ between $\hat{\mathbf{s}}$ and the surface normal vector $\hat{\mathbf{n}}$.

If we were to choose $\hat{\mathbf{s}}$ according to a random direction $\mathbf{\hat{v}_1}$, this would correspond to uniform power in all directions by symmetry, $P=C dAd\Omega$, which would result in the surface not having uniform radiance (i.e., its apparent surface brightness would vary with angle as $\sec(\theta)$).  Conveniently, we need only find the location of the boundary in a two-dimensional subspace to produce a random ray with uniform radiance, according to the algorithm in this section.

First, we choose a second random direction $\mathbf{\hat{v}_2}$, which together with $\mathbf{\hat{v}_1}$ defines a plane.  We can, without loss of generality, rotate $\mathbf{\hat{v}_2}$ in this plane such that it is perpendicular to $\mathbf{\hat{v}_1}$.  We can then decompose $\mathbf{\hat{v}_1}$ and $\mathbf{\hat{v}_2}$ into components parallel to the surface normal and parallel to the boundary:
\begin{eqnarray}
    \mathbf{\hat{v}_1} &\equiv& c_1 \mathbf{\hat{n}} + c_2 \mathbf{\hat{b}_1}\\
        \mathbf{\hat{v}_2} &\equiv& c_3 \mathbf{\hat{n}} + c_4 \mathbf{\hat{b}_2},
\end{eqnarray}
where $\mathbf{\hat{b}_1}$ and $\mathbf{\hat{b}_2}$ are parallel to the boundary.  We can then project the surface normal vector into the $\mathbf{\hat{v}_1}-\mathbf{\hat{v}_2}$ plane and normalize:
\begin{eqnarray}
\mathbf{\hat{n}_{v}}\equiv \frac{\mathbf{\hat{n}}\cdot \mathbf{\hat{v}_1}+\mathbf{\hat{n}}\cdot \mathbf{\hat{v}_2}}{||\mathbf{\hat{n}}\cdot \mathbf{\hat{v}_1}+\mathbf{\hat{n}}\cdot \mathbf{\hat{v}_2}||}=\frac{c_1 \mathbf{\hat{v}_1}+c_3\mathbf{\hat{v}_2}}{\sqrt{c_1^2 + c_3^2}}
\end{eqnarray}
Hence, we find that the dot product with the unprojected surface normal is:
\begin{eqnarray}
    \mathbf{\hat{n}_{v}}\cdot \mathbf{\hat{n}} = \sqrt{c_1^2 + c_3^2} \ge c_1,
\end{eqnarray}
where the right-hand side satisfies the inequality $\sqrt{c_1^2 + c_3^2}\le 1$ by construction.

The direction $\mathbf{\hat{n}_{v}}$ is hence as or more aligned with the surface normal $\mathbf{\hat{n}}$ than the original direction $\mathbf{\hat{v}_{1}}$.  As well, we can compute $\mathbf{\hat{n}_{v}}$ without full knowledge of $\mathbf{\hat{n}}$, because we need only determine the direction perpendicular to the boundary in the restricted $\mathbf{\hat{v}_1}-\mathbf{\hat{v}_2}$ plane, which is a 2D problem instead of a $D$-dimensional problem.  Now, let $\phi$ be the angle between two randomly chosen vectors in $(D+1)$-dimensional space.  We then choose $\mathbf{\hat{s}}$ by rotating $\mathbf{\hat{n}_{v}}$ in the $\mathbf{\hat{v}_1}-\mathbf{\hat{v}_2}$ plane by an angle $\frac{\pi}{2}-\phi$.  

We can now verify that the distribution of $\mathbf{\hat{s}}$ satisfies Lambertian emissivity.  By construction, the dot product of $\mathbf{\hat{s}}$ with the surface normal vector will be:
\begin{eqnarray}
    \mathbf{\hat{s}}\cdot \mathbf{\hat{n}} = \sin(\phi) \sqrt{c_1^2+c_3^2}.
\end{eqnarray}
Although this looks complex, we can interpret it intuitively.  We can generate $\mathbf{\hat{v}_1}$ and $\mathbf{\hat{v}_2}$ via a random rotation of the Euclidean basis vectors in $D-$dimensional space.  Hence, $c_1$ and $c_3$ are the first two components of a random unit vector; equivalently, $\sqrt{c_1^2+c_3^2}$ is the projected length of a random unit vector in $D$ dimensions after projection onto the $\mathbf{\hat{x}}-\mathbf{\hat{y}}$ plane.  We can interpret $\sin(\phi)$ as the distribution of projected lengths of $(D+1)$-dimensional unit vectors in $D-$dimensional space.  Hence, combining the operations, the distribution of $\mathbf{\hat{s}}\cdot \mathbf{\hat{n}}$ is simply the distribution of projected lengths of $(D+1)$-dimensional unit vectors in a 2D plane.

We then note that a further projection from 2D to 1D would result in the familiar distribution for the dot product of two random unit vectors in $D+1$ dimensions:
\begin{eqnarray}
L &\equiv& \sin(\psi)\mathbf{\hat{s}}\cdot \mathbf{\hat{n}}\\
P(L) & = & \int_{\sin^{-1}(L)}^{\pi/2}  P(\mathbf{\hat{s}}\cdot \mathbf{\hat{n}}=L/\sin{\psi})\frac{d\psi}{\sin\psi} \nonumber \\
&\propto & (1-L^2)^{(D-2)/2} \label{e:projection},
\end{eqnarray}
where $\psi$ is the angle from the $\mathbf{\hat{x}}$ axis, so that $P(\psi)\propto 1$, and the $\frac{1}{\sin\psi}$ term in Eq.\ \ref{e:projection} is equal to $\frac{d\mathbf{\hat{s}}\cdot \mathbf{\hat{n}}}{dL}$.

We can then show that $P(\mathbf{\hat{s}}\cdot \mathbf{\hat{n}}) \propto \mathbf{\hat{s}}\cdot \mathbf{\hat{n}} (1-(\mathbf{\hat{s}}\cdot \mathbf{\hat{n}})^2)^{(D-3)/2}$ (i.e., Lambertian emission) is the solution that satisfies Eq.\ \ref{e:projection}.  We find that:
\begin{eqnarray}
    &&\int_{\sin^{-1}(L)}^{\pi/2}  \frac{L}{\sin\psi}\left(1-\frac{L^2}{\sin^2\psi}\right)^{(D-3)/2}\frac{d\psi}{\sin\psi} \\
    &=&\int_{L}^{1} \left(1-u^2\right)^{(D-3)/2}\frac{du}{\sqrt{1-\frac{L^2}{u^2}}},
\end{eqnarray}
using $u = L/\sin\psi$.  This evaluates to:
\begin{eqnarray}
&& (1 - L^2)^{(D-3)/2} \sqrt{u^2 -L^2} \nonumber \\
    &&\times\quad \left.{}_2F_1\left(\frac{1}{2}, \frac{3 - D}{2}; \frac{3}{2}; \frac{L^2 - u^2}{L^2-1}\right)\right|_{u=L}^{u=1} \label{e:hyper1}\\
    &=&(1 - L^2)^{(D-2)/2} {}_2F_1\left(\frac{1}{2}, \frac{3 - D}{2}; \frac{3}{2}; 1\right) \label{e:hyper2}\\
    &\propto& (1-L^2)^{(D-2)/2},
\end{eqnarray}
as desired, noting that ${}_2F_1(a,b;c;0)=1$ for arbitrary $(a,b,c)$ in Eq.\ \ref{e:hyper1}, and that ${}_2F_1\left(\frac{1}{2}, \frac{3 - D}{2}; \frac{3}{2}; 1\right)$ in Eq.\ \ref{e:hyper2} is a constant of integration that only depends on $D$.

\section{Ray tracing algorithm using velocities instead of weights}

\label{a:velocity}

We can transform the ray-tracing formula (which changes the direction of propagation but not the magnitude) back into a velocity-dependent sampler, with the speed $|\mv|$ equal to a function $S(\mx)$ (corresponding to an effective weight of $W=S^{-1}$).  Notably, we cannot generally write down a potential $U(\mathbf{x})$ to describe the dynamics of this transformation.  For example, with standard ray tracing, the ray speed does not change, regardless of the path.

The change in $v$ given by:
\begin{equation}
    \frac{d|\mv|}{dt} = |\mv|\frac{d|\mv|}{ds} = S(\mx)^2\cos\phi \left|\nabla_\mx \ln(S(\mx))\right|,
\end{equation}
where $\phi$ is the angle between the speed gradient and $\mv$ (for $D>1$; otherwise, $\phi=0$).  For $D>1$, $\phi$ changes throughout a given timestep, with:
\begin{equation}
    \phi(t) = \phi_i + \theta(t) - \theta_i = 2\arctan\left(\tan\left(\frac{\theta_i}{2}\right)\exp\left(-S(\mx)\Delta t|\nabla_\mx \ln(n(\mx))|\right)\right)+\phi_i-\theta_i,
\end{equation}
so we find using Mathematica that:
\begin{equation}
    |\mv_f| = |\mv_i| + S(\mx)\cdot\frac{ F(\mx,S(\mx))}{F(\mx,n(\mx))}\left[ \cos(\phi_i-\theta_i)\left(\ln\left(\frac{1+\cos(\theta_i)}{1+\cos(\theta_f)}\right)+\Delta t\cdot F(\mx,n(\mx))\right)+\sin(\phi_i-\theta_i)(\theta_f-\theta_i)\right]
\end{equation}
with $\theta_i=0$ if $\nabla_\mx n(\mx) = 0$, and where the operator $F$ performs:
\begin{equation}
    F(\mx,G(\mx)) = S(\mx)|\nabla_\mx \ln(G(\mx))|.
\end{equation}
As noted previously in the paper, a velocity-dependent sampler may be more susceptible to path heating when stochastic gradients are used.

\section{Symplectic Integration}

\label{a:symplectic}

We note that published fully symplectic integration methods for ray-tracing exist \citep{Satoh03,Li16a,Ohno20,McKeon23}.  Most of these adopt the convention $\mathbf{p} \equiv n \frac{d\mx}{ds}$, with the Hamiltonian $\mathcal{H} = \frac{1}{2}(p^2 - n^2)$.  In these formulations, $\frac{d\mathcal{H}}{dp_i} = n\frac{dx_i}{ds}$, so authors often define a ``time'' as $dt\equiv \frac{1}{n}ds$, which means that the effective sample weighting density would scale as $n$.  An alternate symplectic integration scheme suggested by the master's thesis of Jorg Portegies\footnote{\url{https://pure.tue.nl/ws/portalfiles/portal/46968663/778374-1.pdf}} is to maintain the definition of $\mathbf{p} \equiv n \frac{d\mx}{ds}$, but to adopt $\mathcal{H} = |\mathbf{p}| - n$.  As a result, $\frac{d\mathcal{H}}{dp_i} = \frac{dx_i}{ds}$, so the sample weighting density is uniform per unit path length.

Fully symplectic integration has the advantage that it carries better guarantees about convergence scaling with timestep size.  However, both methods above (as well as symplectic methods for nonseparable Hamiltonians) carry the disadvantage of needing to account for ``off-shell'' values of the Hamiltonian to maintain reversibility.  Specifically, the paths of light rays satisfy $\mathcal{H}=0$ for both definitions above, but real integrators will deviate, arriving at a new location with $\mathcal{H}= \Delta\mathcal{H}\ne 0$.  As a result, the likelihood function $\mathcal{L}(\mx)$ needs to be lifted to include $\Delta \mathcal{H}$ (e.g., as $\mathcal{L}(\mx,\Delta \mathcal{H}) = \mathcal{L}(\mx)\mathcal{L}(\Delta\mathcal{H})$) to maintain reversibility.  Unfortunately, the dynamics of systems with $\mathcal{H}\ne 0$ are different (e.g., $\mathcal{H}<0$ implies a bound system that cannot escape to infinity), and so the Jacobian becomes messy for such cases.  While future investigation of symplectic integrators is certainly interesting, the approach we adopt in this work carries the advantages of 1) always remaining on-shell (so that there are no insurmountable potential barriers), 2) having a simple Jacobian, and 3) efficiently suppressing noise from stochastic gradients.

\section{Comparison to Hamiltonian Monte Carlo for Standard Distributions}

\begin{figure}
\capstart
\centering
\vspace{-5ex}
\includegraphics[width=0.6\columnwidth]{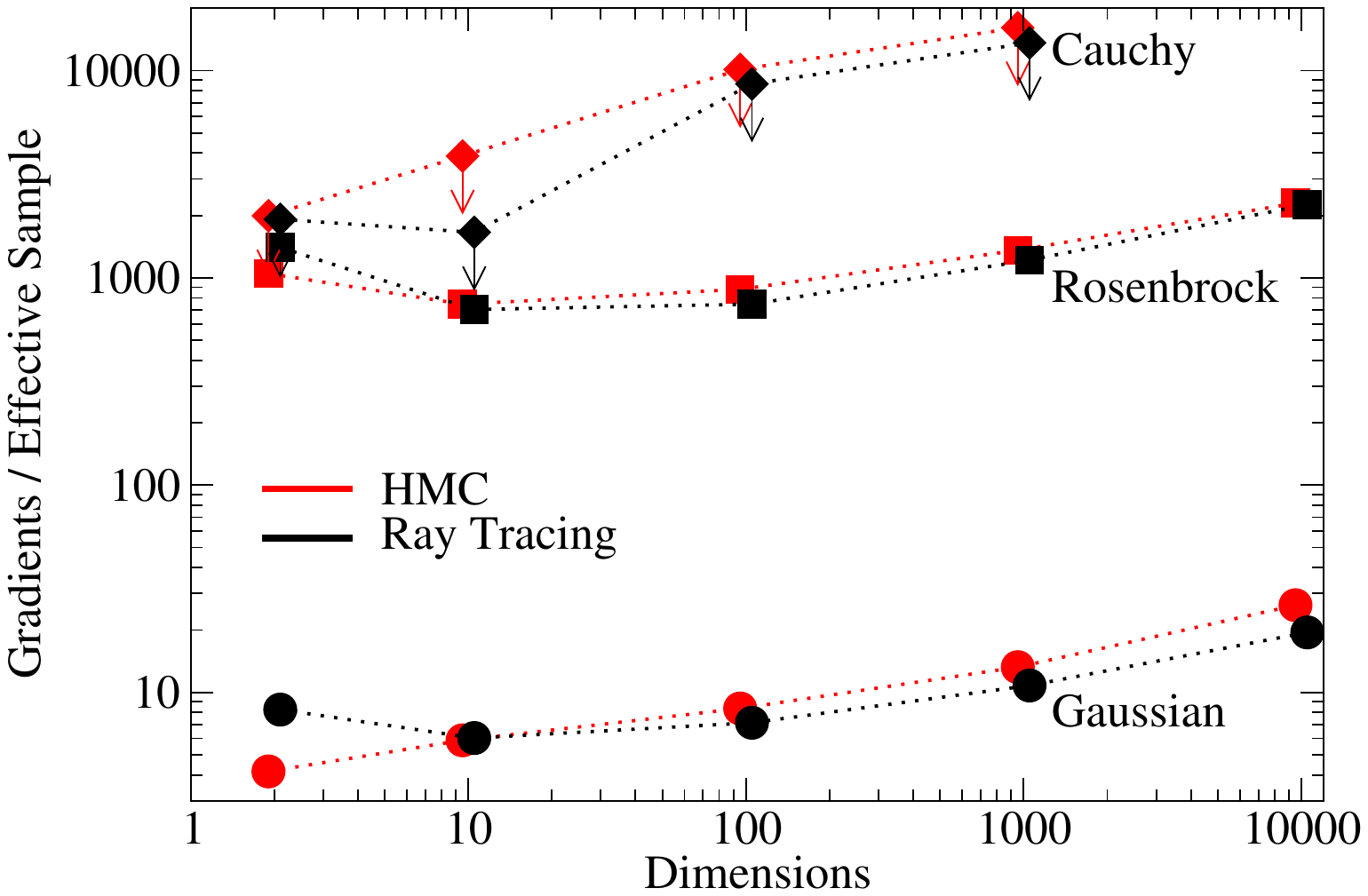}\\[-4ex]
    \caption{Comparison between sampling efficiencies (defined as gradients per effective sample) when perfect gradients are used, for ray tracing and HMC.  No strong performance advantages are observed.  The minimum dimension shown is 2, since ray tracing is not defined for $D=1$.  Upper limits are shown for the Cauchy distribution, since the sampling efficiency as a function of step size and trajectory length had many local minima, and it is not guaranteed that hand-tuning achieved the global minimum.}
    \vspace{3ex}
    \label{f:ess_hmc}
\end{figure}

\label{a:hmc}

In this section, we compare the hyperparameter tuning necessary for both ray tracing and HMC to achieve optimal sampling efficiency.   We use the same autocorrelation time definition as in Section \ref{s:applications}, and define the sampling efficiency as $E=1/(G\tau)$, where $G$ is the number of gradients per trajectory and $\tau$ is the autocorrelation time in units of trajectories per independent sample.  We hand-tune both step sizes and trajectory lengths for both HMC and ray tracing to achieve optimum effective sample sizes per gradient computed.  In particular, we felt that it would be useful for readers to know how to change existing HMC parameters to be able to use ray tracing; the short answer is that ray tracing seems to prefer similar (or slightly larger) step sizes, as well as somewhat longer trajectory lengths than HMC (see Table \ref{t:parameters}).  To make the comparison easier with existing reader codebases, we do not use continuous momentum refreshment here.

In the tests here, we do not find a strong advantage (i.e., more than a factor of a few) for either HMC or ray tracing in any test (summarized in Fig.\ \ref{f:ess_hmc}).  This stands in contrast to some past results---e.g., \cite{Robnik22} argued for stronger differences between HMC and microcanonical HMC (which should have performance very similar to ray tracing for $D>10$). 
We can interpret this most simply from the theoretical basis in Section \ref{s:hmc_connection}.  For high-dimensional spaces, HMC and ray tracing (as well as microcanonical HMC) all have very similar sample weighting.  As well, due to the narrow range of likelihood values in the typical set, there simply is not much freedom to explore regions outside of the typical set, and so none of the samplers enjoys a strong advantage.  For lower-dimensional spaces, it would seem that a ray tracing-like method may have an advantage (e.g., in Fig.\ \ref{f:overview}).  But in practice, using a large step size with HMC results in a similar effect as within-trajectory resampling for the momentum vector, giving it similar performance as ray tracing.  We were unable to contrive an example (even for multimodal distributions) that led to more than a factor of $\sim 2$ difference between the samplers' optimal tuning when a perfect gradient was used.  \modified{The performance improvement noted in past studies like \cite{Robnik22} is more extensively discussed in Section \ref{s:performance_claims}.}


 All tests in this section were performed on a Mac M1 Max laptop using the public \texttt{JAX} implementation of ray tracing and HMC, which employs DKD leapfrog integration.

\subsection{Independent Gaussian Distributions}

Many samplers do well on Gaussian distributions, so this test functions as an easily interpretable base efficiency for HMC and ray tracing.   For the Gaussian tests, we initialize a single walker on the typical set, perform burn-in for 4000 trajectories, and perform 4000 more trajectories for the autocorrelation time estimation step.  Both HMC and ray tracing have short autocorrelation times on these tests, so this provides 1000 or more independent samples by the time the chain is complete.

\begin{table}
\capstart
\centering
\begin{tabular}{lc|ccccc}
\hline
\hline
\multicolumn{2}{l|}{\textbf{Gaussian Distribution}} & 2D & 10D & 100D & 1000D & 10000D\\
\hline
Ray Tracing & Step Size & 0.71 & 0.71 & 0.52 & 0.26 & 0.17\\
& Path Length & $\pi/2$ & $\pi/2$ & $\pi/2$ & $\pi/2$ & $\pi/2$ \\
\hline
HMC & Step Size & 0.67 & 0.5 & 0.36 & 0.21 & 0.15\\
& Path Length & $\pi/2$ & $\pi/2$ & $\pi/2$ & $\pi/2$ & $\pi/2$ \\\hline
\hline
\multicolumn{2}{l|}{\textbf{Rosenbrock Distribution}} & 2D & 10D & 100D & 1000D & 10000D\\
\hline
Ray Tracing & Step Size & 0.01 & 0.015 & 0.013 & 0.0085 & 0.0045\\
& Path Length & 3.9  & 3.7 & 2.6 & 2.5 & 2.5 \\
\hline
HMC & Step Size & 0.012 & 0.017 & 0.01 & 0.007 & 0.0045\\
& Path Length & 3.4 & 3.2 & 2.5 & 2.5 & 2.4\\
\hline
\hline
\multicolumn{2}{l|}{\textbf{Cauchy Distribution}} & 2D & 10D & 100D & 1000D \\
\hline
Ray Tracing & Step Size & 0.6 & 0.37 & 0.3 & 0.24\\
& Path Length & 55 & 65 & 70 & 70\\
\hline
HMC & Step Size & 0.5 & 0.4 & 0.32 & 0.23\\
& Path Length & 55 & 55 & 60 & 70\\
\hline
\end{tabular}
\caption{Hand-tuned parameters for ray tracing and HMC to produce the sampling efficiencies shown in Fig.\ \ref{f:ess_hmc}.}
\label{t:parameters}
\end{table}

\subsection{Rosenbrock Distributions}

The multidimensional Rosenbrock distribution we use here is defined by:
\begin{equation}
    \log \mathcal{L}(\mathbf{x}) = \sum_{i=1}^{D/2}   \left[100(x_{2i-1}^{2}-x_{2i})^{2}+(x_{2i-1}-1)^{2}\right].\label{e:rosenbrock}
\end{equation}
This can be recognized as a sheared multivariate Gaussian distribution, from which exact samples can be drawn using:
\begin{eqnarray}
    x_{2i-1} & = & \mathcal{N}\left(1,0.5\right)\nonumber\\
    v & = & \mathcal{N}(0,0.005)\nonumber\\
    x_{2i} & = &v + x_{2i-1}^2,\label{e:rosen_sample} 
\end{eqnarray}
where $\mathcal{N}(\mu,\sigma^2)$ refers to a normal distribution with mean $\mu$ and variance $\sigma^2$.  The untransformed distribution (Eq.\ \ref{e:rosenbrock}) can be challenging to sample because it has a narrow parabolic channel along which probability is distributed, and so many steps are typically needed for gradient-based samplers to traverse its entire length.

We find that false ``convergence'' can occur with both HMC and ray tracing, particularly for large step sizes that do not allow exploration of narrow probability channels.  To measure true convergence, we perform 1000 trials of 1000 independent point samples each (using Eq.\ \ref{e:rosen_sample}), and calculate the 68$^\mathrm{th}$ percentile ranges for the sample mean and sample standard deviation along each dimension.  For each dimensionality $D$ that we show here, we require at least $90\%$ of the tests for the mean and standard deviation of each dimension to be within the 1000-independent-sample ranges.  For both ray tracing and HMC, we sample 10000 trajectories before estimating the autocorrelation times.

\subsection{Cauchy Distributions}

The multidimensional Cauchy distribution we use here is defined by:
\begin{equation}
    \ln \mathcal{L}(\mx) = -D\ln\pi -\sum_{i=1}^D \log(1+x_i^2)
\end{equation}
This can be challenging for samplers due to its wide probability tails, which can require very long trajectories to traverse.  The Cauchy distribution has no well-defined average (indeed, its sample mean is itself Cauchy-distributed), and has infinite variance.  Hence, the usual approach to test for convergence is to choose a different moment of the distribution that is well defined.  Here, we choose the distribution of log-likelihoods for individual dimensions, which are distributed as:
\begin{equation}
    P(\ln \mathcal{L}) = \left(\pi\exp(-\ln \mathcal{L})-\pi^2\right)^{-1/2},
\end{equation}
 defined for $\ln \mathcal{L}<-\ln\pi$.  This distribution has a well-defined average ($\sim -2.53102$) and standard deviation ($\sim 1.8138$).  As for the previous subsection, we compute the 68$^\mathrm{th}$ percentile ranges for the sample means and sample standard deviations of 1000 trials of 1000 independently-drawn points from the Cauchy distribution (which are obtained by taking the ratio of two independent variables both distributed as $\mathcal{N}(0,1)$).  As above, we consider a chain converged if at least $90\%$ of the tests for the mean and standard deviation of each dimension fall within the 1000-independent-sample ranges.  For both samplers, we sample 200000 trajectories before estimating the autocorrelation times.

\section{Further Generalizations}
\label{a:further_generalizations}
\modified{Here, we consider more general classes of samplers than those allowed by Snell's law.  For example, we could examine a more general conservation law, such as:}
\begin{equation}
    \mmodified{f(\mathcal{L}_1,\theta_1)\cdot \sin \theta_1 = f(\mathcal{L}_2,\theta_2)\cdot \sin \theta_2},
\end{equation}
\modified{or in differential form:}
\begin{equation}
    \mmodified{\frac{d\theta_1}{d\theta_2} = \frac{f(\mathcal{L}_2,\theta_2) \cos\theta_2 + \frac{df(\mathcal{L}_2,\theta_2) }{d\theta_2}\sin\theta_2}{f(\mathcal{L}_1,\theta_1) \cos\theta_1 + \frac{df(\mathcal{L}_1,\theta_1) }{d\theta_1}\sin\theta_1}.}
\end{equation}
\modified{The radiance ratio will be:}
\begin{equation}
    \mmodified{\frac{L_2}{L_1} = \frac{\cos\theta_1 d\Omega_1}{\cos\theta_2 d\Omega_2} = \frac{d\theta_1}{d\theta_2} \cdot \frac{\sin^{D-2}\theta_1\cos\theta_1}{\sin^{D-2}\theta_2\cos\theta_2} = \left(\frac{f(\mathcal{L}_2,\theta_2)}{f(\mathcal{L}_1,\theta_1)}\right)^{D-2}\frac{f(\mathcal{L}_2,\theta_2)  + \frac{df(\mathcal{L}_2,\theta_2) }{d\theta_2}\tan\theta_2}{f(\mathcal{L}_1,\theta_1) + \frac{df(\mathcal{L}_1,\theta_1) }{d\theta_1}\tan\theta_1}.}
\end{equation}
\modified{Hence, we can recover a fair constant-speed sampler ($L\propto\mathcal{L}$) under the condition:}
\begin{equation}
    \mmodified{\mathcal{L} = f(\mathcal{L},\theta)^{D-2}\left(f(\mathcal{L},\theta)  + \frac{df(\mathcal{L},\theta)}{d\theta}\tan\theta\right),}
\end{equation}
\modified{which has the general solution:}
\begin{equation}
    \mmodified{f(\mathcal{L},\theta) = \left(\mathcal{L}+g(\mathcal{L})\csc^{D-1}\theta\right)^{\frac{1}{D-1}},}
\end{equation}
\modified{with $g(\mathcal{L})$ an arbitrary function of $\mathcal{L}$.}  
\modified{The appropriate update for $\theta$ becomes:}
\begin{equation}
\mmodified{\frac{d\theta}{ds} = -|\nabla_\mx \ln f|\cdot\left(\sin \theta+\frac{g(\mathcal{L})}{\mathcal{L}}\csc^{D-2}\theta\right).}\label{e:tumbler}
\end{equation}
\modified{Intuitively, this broad family of fair samplers can be thought of as ``tumblers'' in the sense that the angle $\theta$ is driven toward a direction parallel to the gradient by the $\sin\theta$ term in Eq.\ \ref{e:tumbler}, but the other terms give the angle an extra rotation rate, so that the angle $\theta$ can ``tumble through'' the gradient direction (or avoid the gradient direction, if the sign is opposite), both of which prevent the sampler from tracking too close to the gradient direction.  In principle, this could lead to further noise suppression at the cost of potentially longer burn-in times, but we leave a test of this hypothesis to future work.}

\modified{As another option, we could attempt to find constant-speed samplers that have an oversampling rate that changes with the path length, e.g., for some desired $\frac{d}{ds}\ln(\mathcal{R}(s))$.  Formally, this implies:}
\begin{equation}
\mmodified{\frac{d \ln L}{ds} = \cos(\theta)|\nabla_\mx \ln \mathcal{L}(\mx)| + \frac{d\ln \mathcal{R}}{ds}.}\label{e:volume_change}
\end{equation}
\modified{Intuitively, an increased oversampling rate acts like a stronger likelihood gradient, so the sampler will be drawn towards higher-likelihood regions than it would otherwise, and a lower oversampling rate would have the opposite effect.  From the definition of the radiance (Section \ref{s:overview}), and assuming that we only change $\theta$ (as opposed to $\psi_{1\ldots{D-2}}$), we have:}
\begin{equation}
    \mmodified{\frac{d\ln L}{ds} = (2-D)\cot \theta \frac{d\theta}{ds}+\tan\theta\frac{d\theta}{ds}-\frac{d}{ds}\ln d\theta}.\label{e:differential_jacobian}
\end{equation}
\modified{Equating these two formulas lets us solve for an update with an arbitrary oversampling rate.  For example, we could find solutions of the form:}
\begin{equation}
    \mmodified{\frac{d\theta}{ds} = -h(\theta)\frac{\cos(\theta)}{D-1}|\nabla_\mx \ln \mathcal{L}(\mx)|}.
\end{equation}
\modified{Plugging this equation into Eqs.\ \ref{e:volume_change} and \ref{e:differential_jacobian} results in the following differential equation for $h(\theta)$:}\footnote{\modified{Of note, some care is required in dealing with the term $\frac{d}{ds}\ln d\theta$.  The  interpretation as $\frac{d}{ds} \frac{d\theta}{ds}$ is complicated by the fact that the path taken (i.e., the meaning of $s$) is $\theta$-dependent.  A more transparent approach is to define $x_0$ as the direction of the likelihood gradient, and then to write $\frac{d}{ds} = \cos\theta \frac{d}{dx_0}$, leading to $\frac{d}{ds}\ln d\theta = \cos \theta \frac{\partial}{\partial\theta}\left[\frac{d\theta}{dx_0}\right] = \cos\theta \frac{\partial}{\partial \theta}\left[\sec \theta \frac{d\theta}{ds}\right]$}}
\begin{equation}
    \mmodified{\frac{dh(\theta)}{d\theta} =  D-1 - h(\theta)\cdot\left[(D-2)\cot\theta -\tan\theta\right] + \sec\theta \cdot \left[\frac{D-1}{|\nabla_x \ln \mathcal{L}(\mx)|}\cdot\frac{d\ln \mathcal{R}}{ds}\right]}.
\end{equation}
\modified{This can be solved for any desired $\frac{d\ln \mathcal{R}}{ds}$.  For example, if $\frac{d\ln \mathcal{R}}{ds}$ is independent of $\theta$, then one simple solution in 2D is:}
\begin{equation}
    \mmodified{\frac{d\theta_{2D}}{ds} = -\sin(\theta)|\nabla_x \ln \mathcal{L}(\mx)| - \theta\cdot\frac{d\ln \mathcal{R}}{ds}.}
\end{equation}
\modified{Of course, the form of $\frac{d}{ds}\ln(\mathcal{R}(s))$ must also be reversible.  A simple choice is to adopt $\frac{d}{ds}\ln(\mathcal{R}(s)) = -A\cdot \mathrm{sgn}(\sin(s))$, where $\mathrm{sgn}(x)$ is the sign function ($+1$ if $x$ is positive, $-1$ if $x$ is negative, and $0$ otherwise).  This is reversible if the path stops at $s\in2\pi\mathbb{Z}$, as this choice leads to $\frac{d}{d(-s)}\ln(\mathcal{R}(2\pi\mathbb{Z}-s)) = -\frac{d}{ds}\ln(\mathcal{R}(-s))=\frac{d}{ds}\ln(\mathcal{R}(s))$.  Following the intuition above, this form for $\frac{d}{ds}\ln(\mathcal{R}(s))$ will push the sampler into lower-likelihood regions before returning back to the original typical set, functioning as a reversible ``likelihood arc'' sampler that can more efficiently explore multi-modal distributions, remains constant-speed (and so is resilient to gradient noise), and is straightforward to implement.  Future work (Behroozi et al., in prep.) will examine whether this and related methods offer better performance for exploring neural networks.}

\section{Neural Network Details}

\subsection{UniverseMachine Stellar Mass Estimates with a 1433-parameter Multi-Layer Perceptron}

\label{a:um1560}

To enable comparison between ray tracing and HMC, we use a relatively simple network architecture (i.e., fully-connected network layers).  We found that converging geometries (i.e., fewer neurons in each successive layer) and flat geometries (similar neuron count in each successive layer) tended to produce higher fractions of stuck dimensions---i.e., predictions that had long autocorrelation times.  We also found that smoother activation functions tended to produce fewer stuck dimensions.  We hence use a fan geometry (each successive layer having more neurons) with scaled exponential linear unit (SELU; \citealt{SELU}) activation functions, with a final fully-connected linear layer producing the single output prediction.  Specifically, after the input layer, the successive hidden layers have 8, 16, 24, and 32 neurons, followed by the final linear layer, leading to a total of 1433 trainable parameters.

 For the input data, we use $\sim$ 1 million halos subsampled from the \textsc{UniverseMachine} DR1 catalog (\citealt{BWHC19}, based on the Bolshoi-Planck simulation; \citealt{Klypin14,RP16b}), with an 80\%/20\% training/validation split.  As inputs, we use single-snapshot dark matter halo properties (i.e., the scale factor, the halo mass, the maximum circular velocity, and two random seeds), and we train networks to match the corresponding galaxy stellar mass according to the \textsc{UniverseMachine}.  Because the \textsc{UniverseMachine} relies on the entire halo growth history to generate a stellar mass, it is impossible for the networks to exactly fit the results given the input variables, and so typical best root mean squared errors achievable with the Adam optimizer \citep{Kingma14} are $\sim 0.35$ dex.  Inputs and outputs were all normalized to have zero mean and unit variance.  For all trials, we scale the initial Kaiming \citep{He15} starting positions by a factor of 3 (for a typical  norm of $|\mx_0|\sim 12$) to explore a greater diversity of parameter space.  Initial momenta (and refreshes) are drawn from random unit Gaussians (for a typical momentum norm of $|\mv_0|\sim \sqrt{1560} \sim 39.5$).
 
 For both HMC and ray tracing, we scale the loss (mean-squared error) by a factor of 5000, and use mini-batch sizes of 20000, so each training epoch has about $400$ steps, and the typical $\sigma_{\ln \mathcal{L}}$ is about $10$.
We tested a range of timesteps to see where the trajectories tended to fail, settling on $\Delta t=0.0002$ for both ray tracing and HMC.  For both HMC and ray tracing, we adopt a stochastic momentum refresh fraction of $f=5\Delta t$, and run 10 walkers for 1000 epochs each.   For HMC, we also do a pre-burn-in with a momentum refresh fraction of $f=50\Delta t$ for 5 epochs (allowing the excess kinetic energy gained from starting at an initially very unlikely position to decay).  We fail trajectories that have $\Delta E$ (HMC) or $\Delta \ln L/\mathcal{L}$ (ray tracing) of more than $5\sigma_{\ln \mathcal{L}}\sim 50$, following Section \ref{s:stochastic}.  For comparison with existing practice, we also run Adam fitting \citep{Kingma14} for 50 epochs with a learning rate of 0.02 (notably, the standard learning rate of $0.001$ did not result in convergence),  for 1000 different starting positions.  Sampling and fitting were performed on an NVIDIA RTX 5090 using \texttt{PyTorch}.  Since the network was not complex, throughput on this GPU was only $\sim 2-3\times$ faster than running on an Apple M1 Max laptop.

\subsection{Metallicity Estimates with ResNet-34}

\label{a:resnet}

In our simplified version of the \cite{Wu19} study, we use 1/10 of the available SDSS galaxy images for our main analysis, for a total of $N_\mathrm{train}=$11,664 images in the training set and 1,296 images in the validation set, as this allowed us to have multiple independent training sets drawn from the same underlying distribution.  As in the \cite{Wu19} study, we replace the last layer of ResNet-34 with a fully connected linear layer with a single output to perform regression to the renormalized spectroscopic metallicity (i.e., the metallicity shifted and rescaled to have zero mean and unit variance).  We find that scaling the MSE loss by a factor of $4\times 10^{4}$ (corresponding to $D_\mathrm{eff}\sim 1.4 \times 10^4 \sim N_\mathrm{train}$, or $0.06$\% of the full parameter space) allows indefinite sampling without overfitting.  We use a timestep size of $\Delta t = 2\times 10^{-6}$, corresponding to a spatial step of $\Delta s \sim 0.01$, or an angular step of initially $\Delta \phi \sim 10^{-4}$.    For the main run, we use Kaiming initialization for a single MCMC walker, running for $5\times 10^5$ epochs with a mini-batch size of 32.  This mini-batch size results in a total GPU memory usage of $\sim 2$GB, i.e., small enough to run on any iPhone released in the past 7 years.  To mimic the approach in the \cite{Wu19} study, we also use Adam to train the last layer of the ResNet-34 for 2 epochs, and then allow training of all layers simultaneously for 8 epochs; this is repeated $1000$ times to estimate the variance of fitting results.  We also run a second walker for 5000 epochs with the same training data and a different initial seed, as well as a third walker with an independent (i.e., no overlap) training set of 11,664 images and a different initial seed.  Sampling and fitting were performed initially on an NVIDIA RTX 3060 and later on an NVIDIA RTX 5090 using \texttt{PyTorch}.  The RTX 3060 averaged 1400 epochs/day (5$\times 10^5$ steps/day), and the RTX 5090 averaged 5600 epochs/day (2$\times 10^6$ steps/day).

\subsection{Exploring GPT-2 Posteriors}

\label{a:gpt2}

To explore GPT-2 posteriors, we use the \texttt{nanoGPT} implementation,\footnote{\url{https://github.com/karpathy/nanoGPT}} which we modified 1) to use the ray tracing algorithm for parameter updates, and 2) to lower GPU memory usage by $\sim$10\% with better garbage collection.  We used a mini-batch size of 5 (with 1024 tokens in each sample), this being the largest mini-batch size that could fit on our 32GB RTX 5090.  For training data, we used a 100 billion-token subset of the \texttt{fineweb-edu} dataset \citep{Penedo24}, taken from the public \texttt{llm.c} implementation, due to the original WebText training set used by OpenAI being private.  We initialized the model parameters from public 1.5B-parameter GPT-2 model weights that had been trained using the same \texttt{fineweb-edu} dataset.\footnote{\url{https://huggingface.co/karpathy/gpt2_1558M_final3_hf}, chosen because the author's later version with a longer training time quickly explodes with further training, per \url{https://huggingface.co/karpathy/gpt2_1558M_final4_hf}.} 

To perform MCMC sampling, we used shuffled training set data, but to estimate the loss evolution over training epochs for posterior autocorrelation tests, we used a fixed random sample of 400$\times$5120 tokens each from the training and validation sets.  This ensured that most computations were spent in parameter updates as opposed to computing the loss function precisely.  With the RTX 5090, we could complete $\sim$120000 mini-batches per day, covering about $600$M input tokens per day (163 days per epoch).  We used a timestep size of $\Delta t=5\times 10^{-6}$, corresponding to a parameter space step of $\Delta s=0.2$, or an angular step of $\Delta \phi = 2\times 10^{-4}$.

\section{\modified{Convergence Using the Vehtari Split-$\hat{R}$ Estimator}}

\begin{figure}
\capstart
\vspace{-10ex}
\hspace{-8ex}
    \includegraphics[width=0.55\columnwidth]{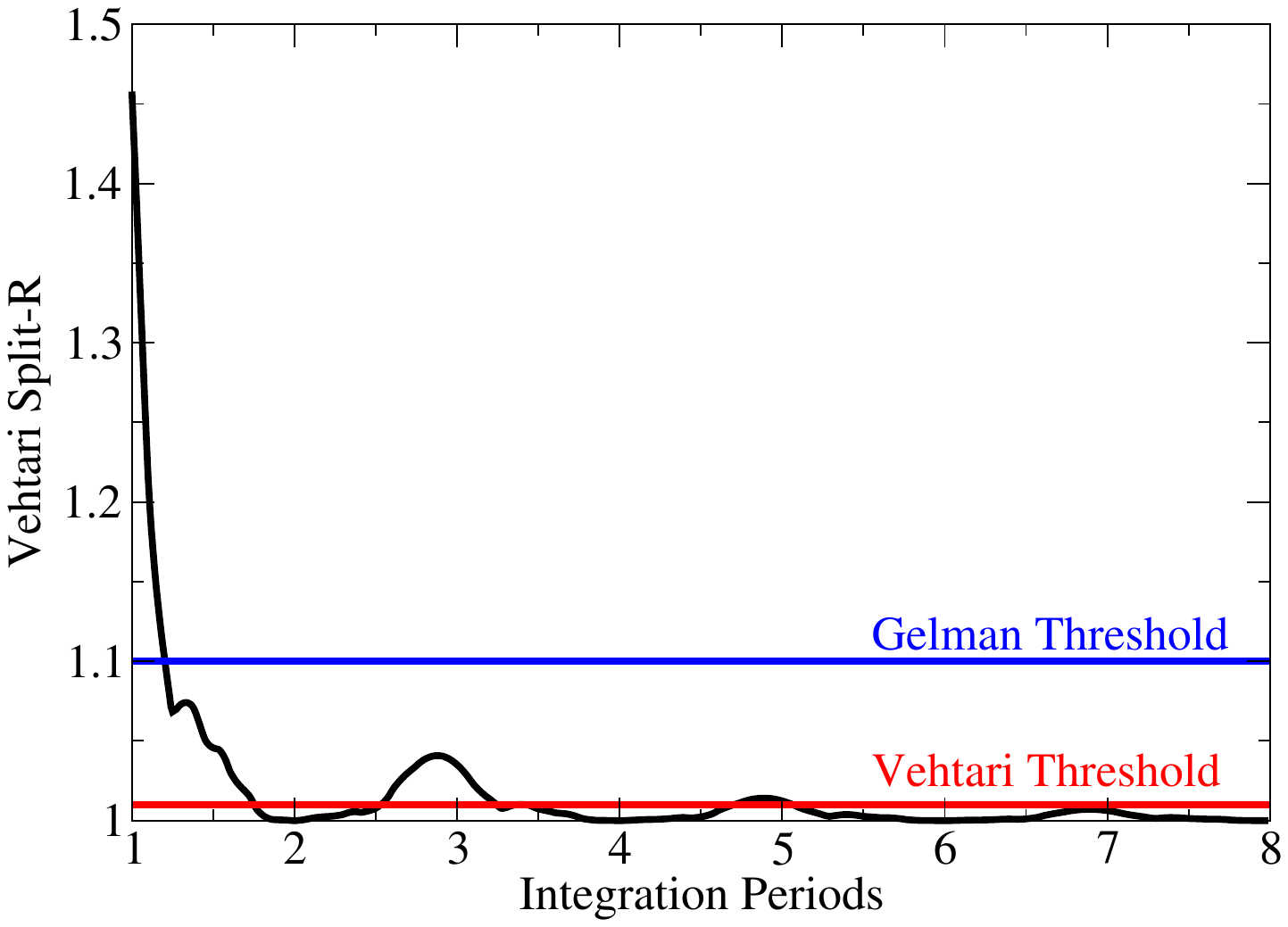}\hspace{-5ex}
     \includegraphics[width=0.55\columnwidth]{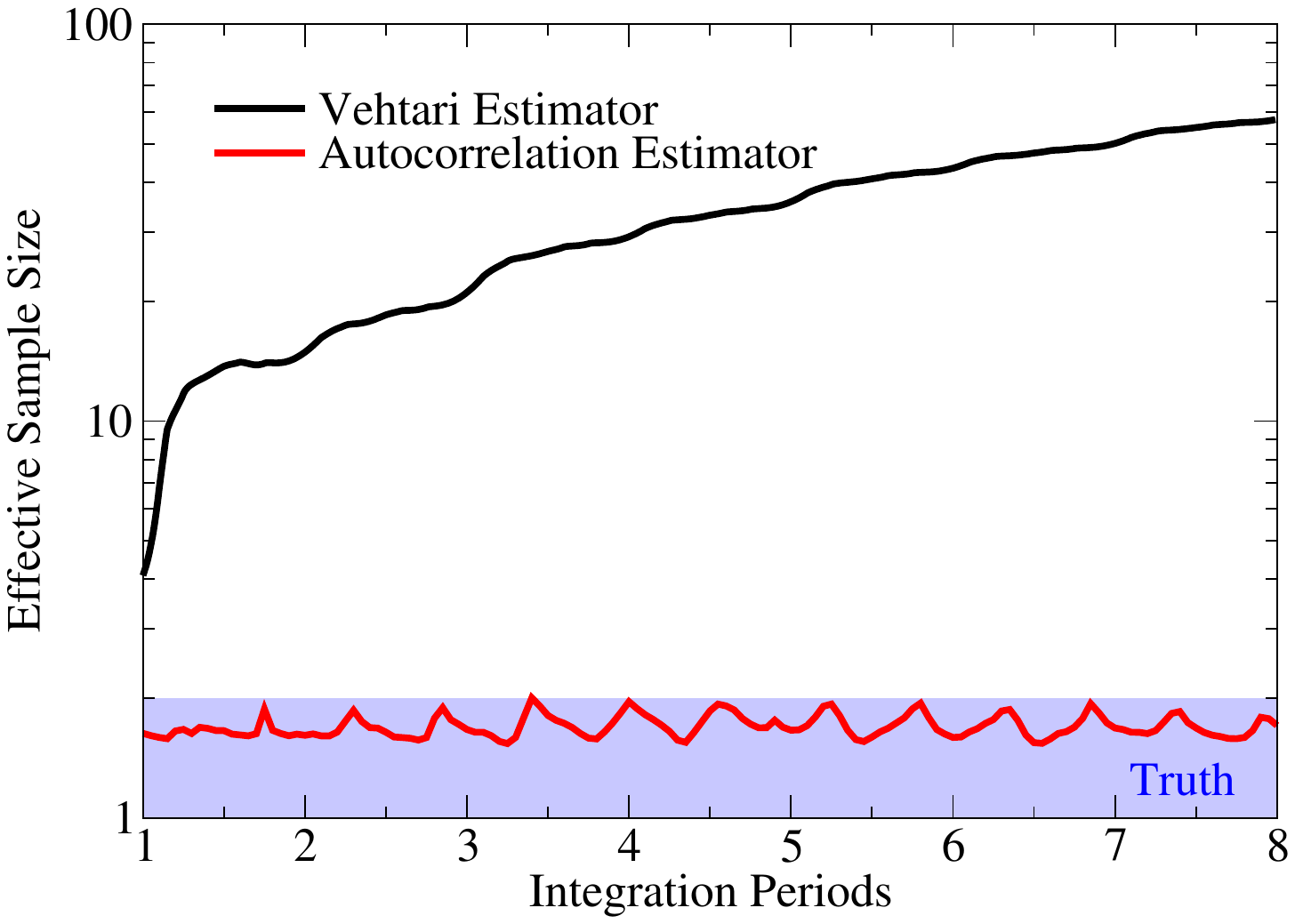}\\[-6ex]
     \caption{\modified{\textbf{Left} panel: Split-$\hat{R}$ convergence diagnostic for the \citet{Vehtari21} estimator for two HMC chains in a unit 2D Gaussian distribution, started out-of-phase at identical distances from the origin and integrated for many periods without momentum refreshment.  Because the moments of both chains converge to identical values (and because any subsample of either chain has identical moments), the \citet{Vehtari21} estimator asymptotes to perfect convergence.  \textbf{Right} panel: effective sample size (ESS) estimates for the same chains, using the ESS estimator in \citet{Vehtari21} and the autocorrelation estimator in this paper.  The \citet{Vehtari21} estimator (along with all other moment comparison estimators) shows a false increase with integration period, whereas the autocorrelation estimator correctly recognizes that distant samples are still strongly correlated.  The actual effective sample size for the chains is between 1 and 2, depending on the summary statistic: the phases are independent, but the radii are identical.}}
     \label{f:vehtari}
     \capstart

    \hspace{-8ex}\includegraphics[width=0.55\columnwidth]{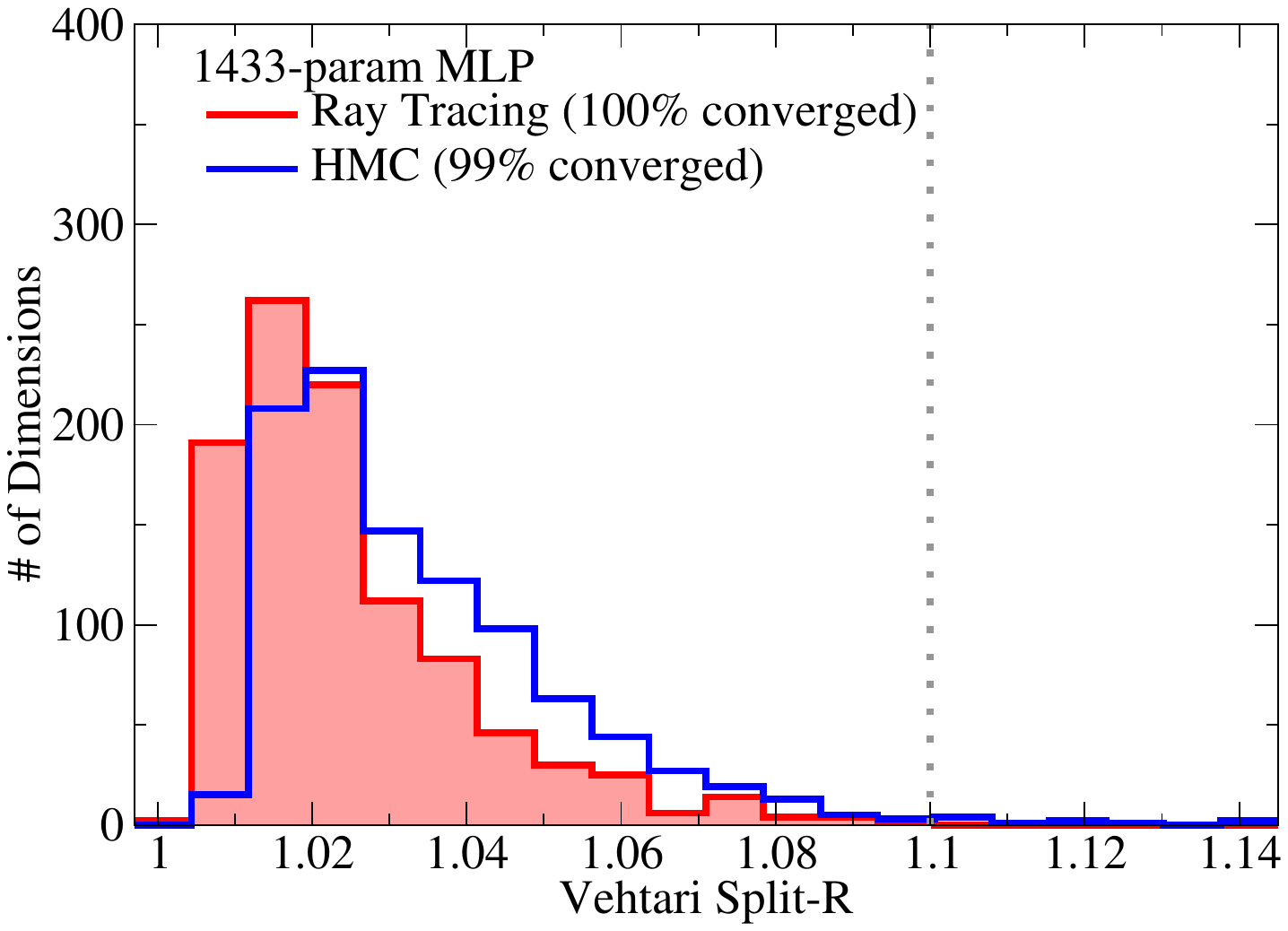}\hspace{-5ex}\includegraphics[width=0.55\columnwidth]{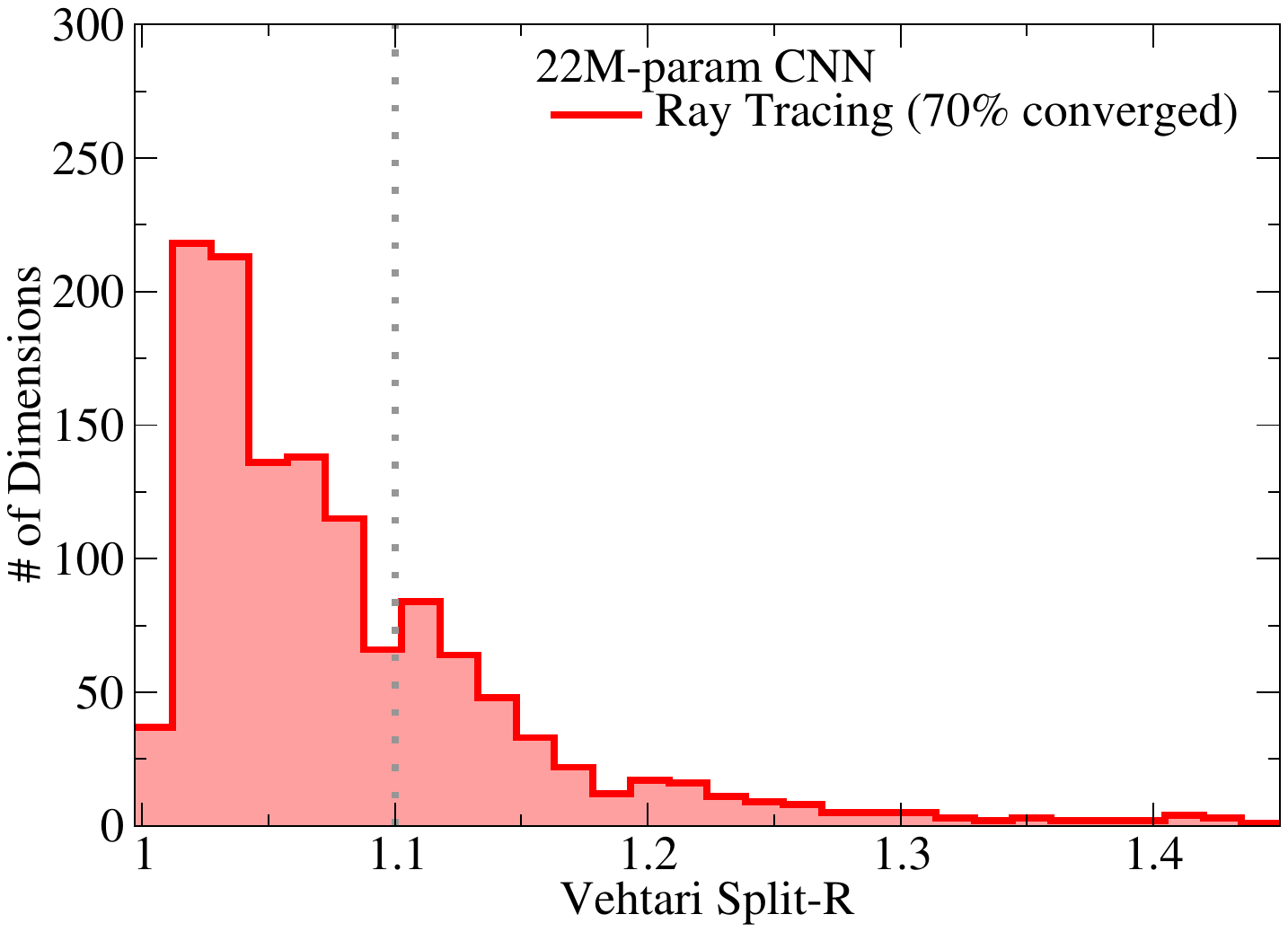}\\[-6ex]
    \caption{\modified{\textbf{Left} panel: Split-$\hat{R}$ convergence diagnostic for the \citet{Vehtari21} estimator for the ray tracing and HMC chains in Section \ref{s:um}.  Ray tracing and HMC have largely comparable performance, similar to the autocorrelation estimates in Fig.\ \ref{f:acor}.  \textbf{Right} panel: Split-$\hat{R}$ convergence diagnostic for the \citet{Vehtari21} estimator for the ray tracing chains in Section \ref{s:resnet}.}}
    \label{f:vehtari_cnn}

\capstart
    \vspace{-6ex}
    \hspace{-8ex}\includegraphics[width=0.55\columnwidth]{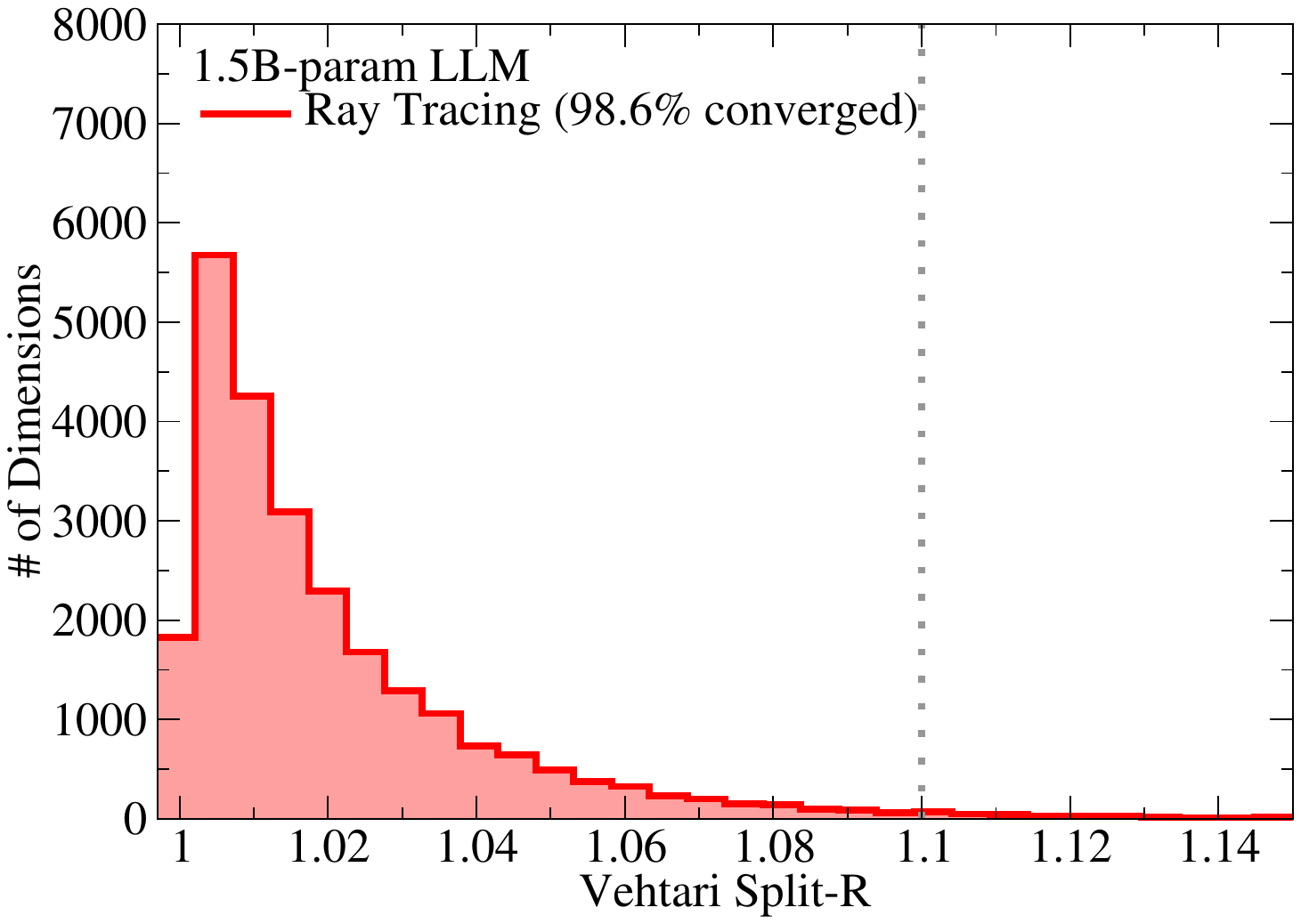}\hspace{-5ex}\includegraphics[width=0.55\columnwidth]{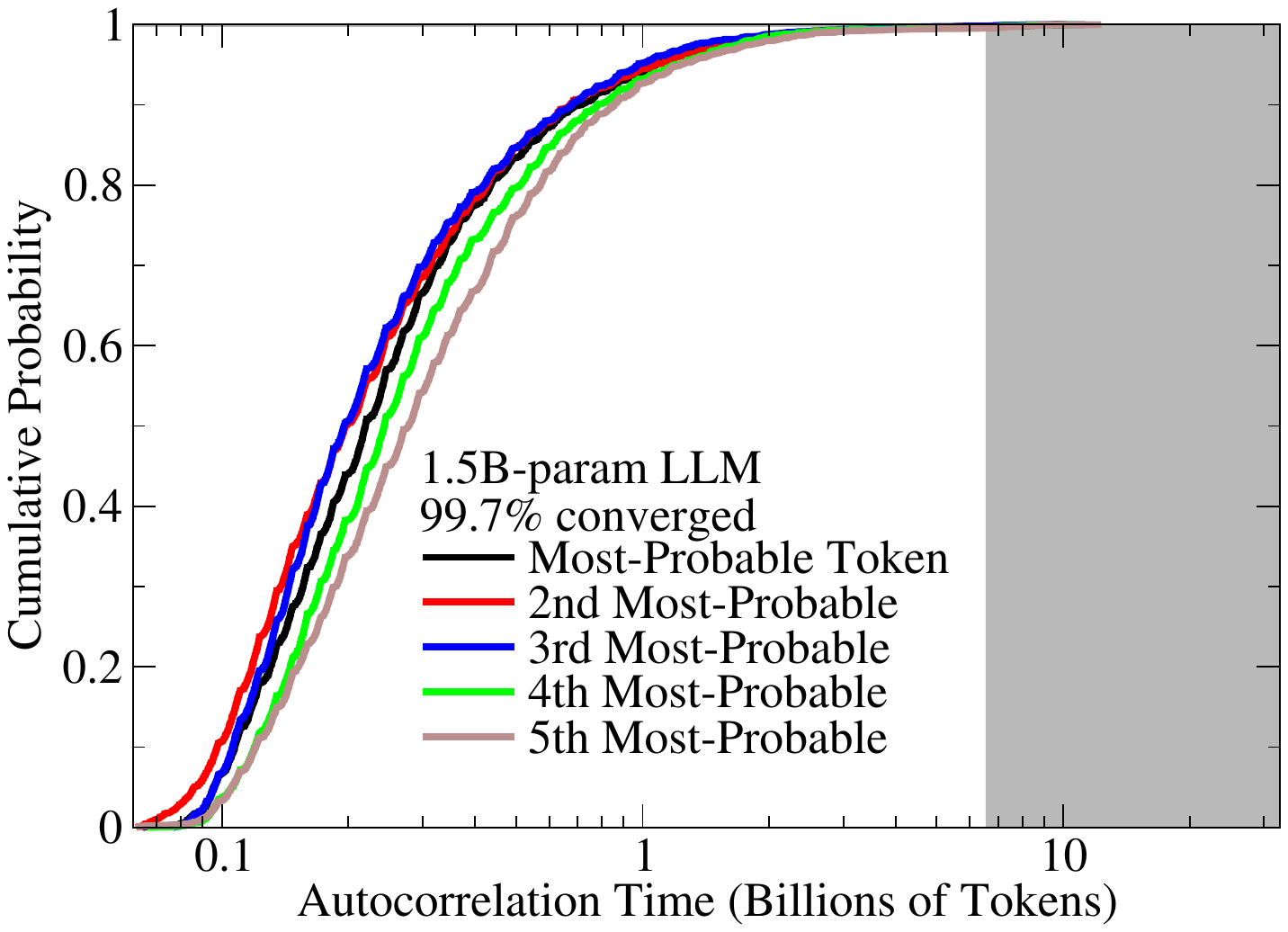} 
    \vspace{-6ex}
    \caption{\modified{\textbf{Left} panel: Standard autocorrelation time diagnostic for the two-chain version of the large language model in Section \ref{s:1B}.  The second chain saved snapshots at 20$\times$ larger intervals, so the minimum autocorrelation time resolution is $\sim$100 million tokens, pushing the minimum of the distribution towards higher values as compared to Fig.\ \ref{f:gpt2_autocorrelation}. \textbf{Right} panel: convergence diagnostic for the \citet{Vehtari21} estimator for the ray tracing chains in Section \ref{s:1B}.}}
\label{f:vehtari_llm}    
\end{figure}

\label{a:convergence}

\modified{Convergence tests used in the literature are often one of two types: 1) autocorrelation estimators (like that used in the main body of this paper), and 2) comparison of moments across and/or within chains (e.g., \citealt{Gelman92} and the more modern update with folding and variance sensitivity in \citealt{Vehtari21}).  Each type of estimator has strengths and weaknesses. } 

\modified{We do not use moment comparisons in the main paper, as they can be insensitive to trajectory over-integration, which necessitates a very strict convergence criterion ($|\hat{R}-1|<0.01$ in \citealt{Vehtari21}) to avoid false positives.  For example, we show in Fig.\ \ref{f:vehtari} the \cite{Vehtari21} $\hat{R}$ for two HMC chains that are started from identical distances from the center of a unit Gaussian distribution, and integrated indefinitely without momentum refreshment.  The longer the integration time, the more identical are all the moments of both chains; further, any chain subsamples will again have identical moments.  In contrast, the autocorrelation estimator used in this paper recognizes that samples are correlated across arbitrary time intervals, leading to a very low estimated sample size, and implying a lack of convergence.}

\modified{Nonetheless, for reference, we show the \citet{Vehtari21} split-$\hat{R}$ estimates for the three neural networks that were explored in this paper.  Figs.\ \ref{f:vehtari_cnn} and \ref{f:vehtari_llm} show that convergence is essentially identical for the \citet{Vehtari21} split-$\hat{R}$ estimator using the \cite{Gelman92} threshold and the adopted autocorrelation time estimator in this paper.  For reference, \cite{Izmailov21} also used the \cite{Gelman92} threshold with a split-$\hat{R}$ convergence estimator, so the convergence estimate in Section \ref{s:resnet} is a reasonable comparison.  We also show the two-chain autocorrelation time convergence estimate for the large language model application in Section \ref{s:1B} in Fig.\ \ref{f:vehtari_llm}.  For technical reasons, the number of samples saved for the second chain was $20\times$ fewer than the first chain, resulting in a lower autocorrelation time resolution; however, the two-chain autocorrelation time estimate had an essentially identical convergence fraction ($99.7\%$ vs. $99.8\%$) as the higher resolution single-chain run, and so the autocorrelation figure presented in the main text is from the single-chain run.}

\section{\modified{Sensitivity of $D_\mathrm{eff}$ to network parameter count}}

\label{a:sensitivity}

\begin{figure}
\vspace{-6ex}
\hspace{-8ex}
\capstart

    \hspace{-8ex}\includegraphics[width=0.55\columnwidth]{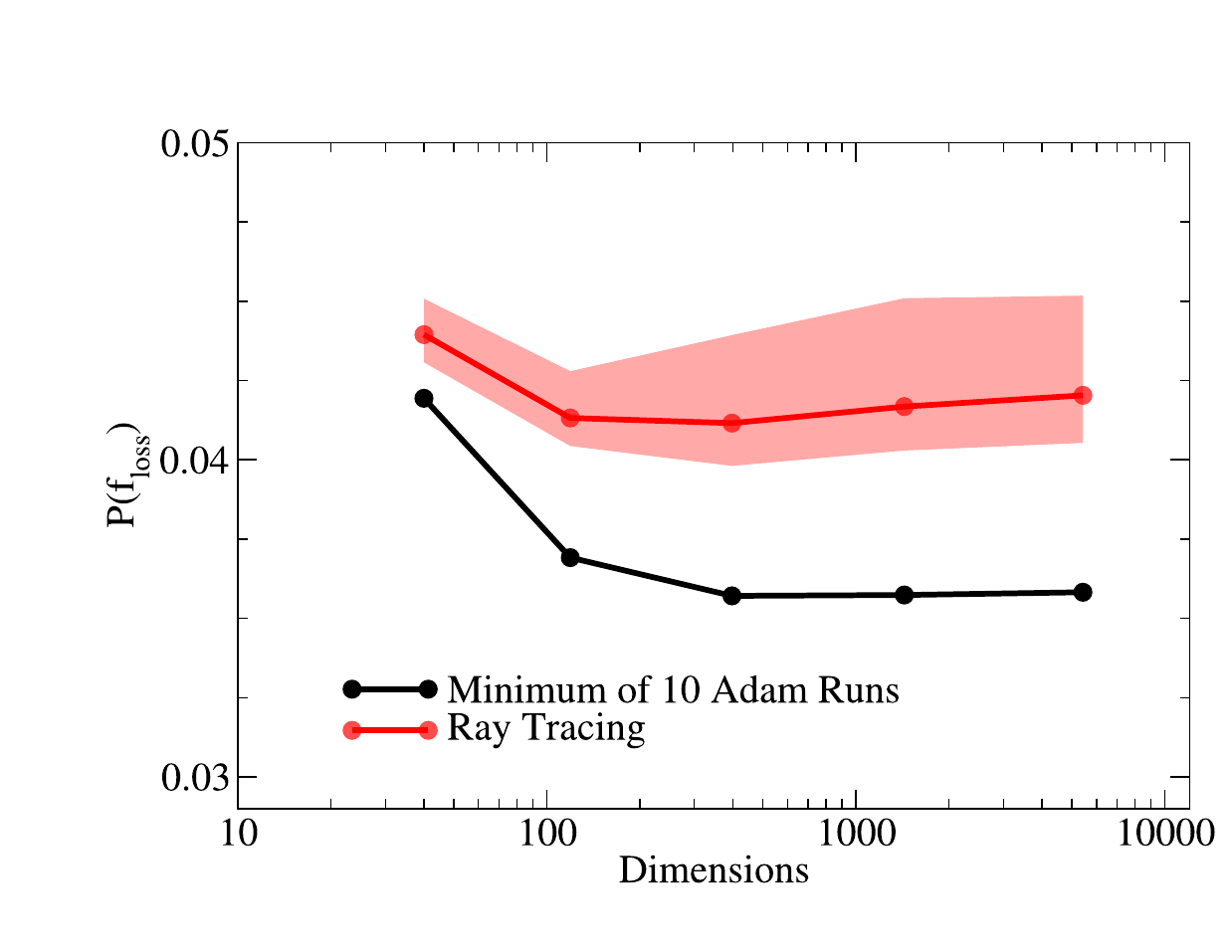}\hspace{-5ex}
     \includegraphics[width=0.55\columnwidth]{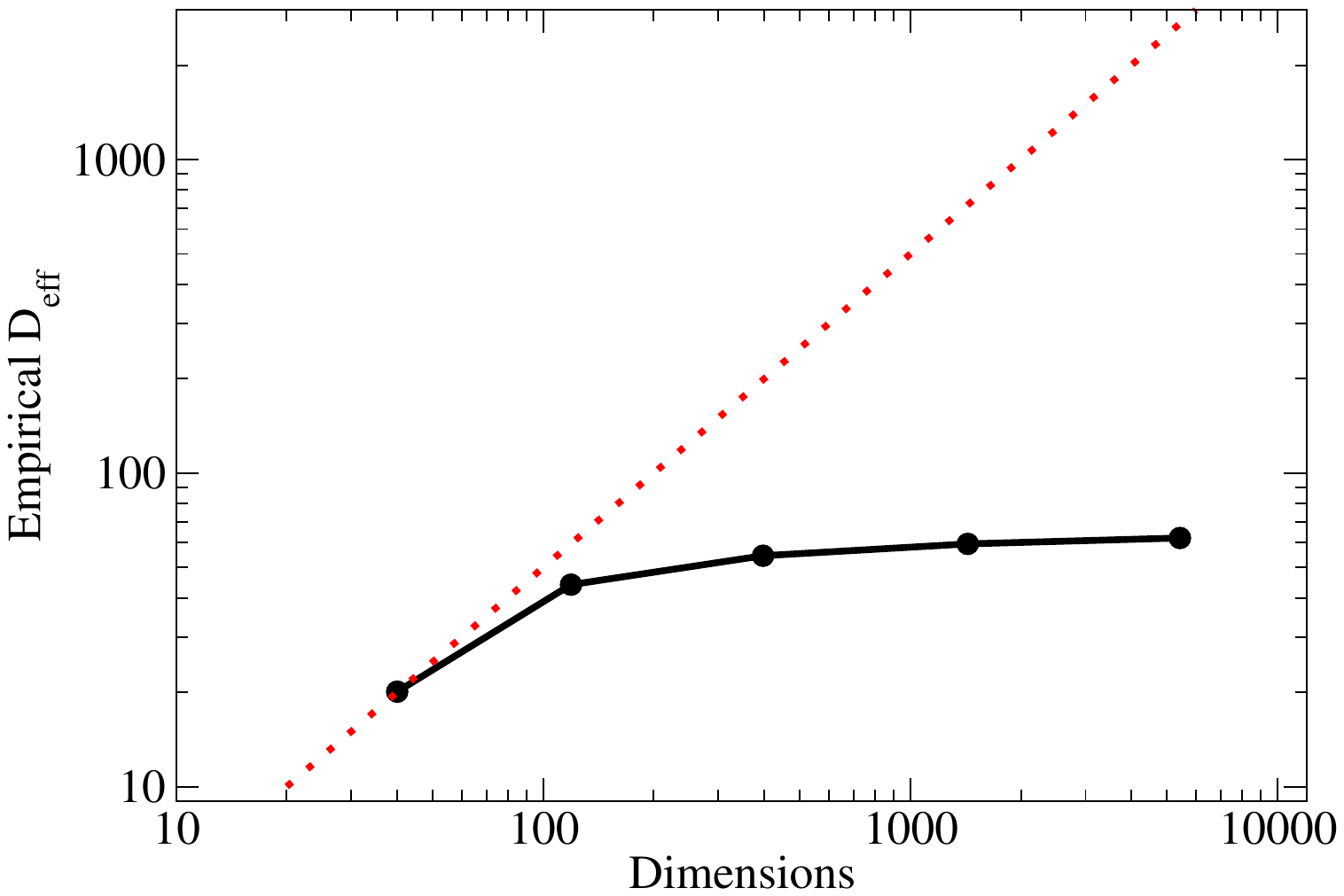}\\[-6ex]
     \caption{\modified{\textbf{Left} panel: stability of the loss function (as applied to the validation set) with respect to parameter count for the problem in Section \ref{s:um}.  Since the problem is relatively simple, adding more parameters to the network does not improve performance.  The black line shows the minimum of 10 Adam runs with 100 epochs each, and the red line with error bars shows the median and 68$^\mathrm{th}$ percentiles of the distribution of $P(f_\mathrm{loss})$ for a fixed likelihood function (i.e., $\ln \mathcal{L} = -5000 f_\mathrm{loss}$ independent of network parameter count).    \textbf{Right} panel: the black line shows implied effective dimensionality of the manifold (per Eq.\ \ref{e:approx_likelihood}), which stabilizes around 60.  The actual number of parameters in the underlying \textsc{UniverseMachine} model used to generate the dataset is similar, at around 50 (see text).  For comparison, the red dotted line shows what would occur if the effective dimensionality scaled with the parameter count, which would be expected if the network performance were limited by parameter count instead of by problem complexity.}}
     \label{f:sensitivity}
\end{figure}

\modified{Sections \ref{s:1B} and \ref{s:resnet} provided examples where the effective number of dimensions $D_\mathrm{eff}$ was limited by the network parameter count and the training set size, respectively; we hypothesized in Section $\ref{s:um}$ that the problem complexity would also set a natural upper limit on $D_\mathrm{eff}$.  To test this hypothesis, we performed a sensitivity analysis of the loss function for the problem in Section \ref{s:um},  varying the network parameter count while holding the likelihood function fixed (i.e., $\ln \mathcal{L}(\mx) = -5000 f_\mathrm{loss}(\mx)$.  We kept the shape and activation function of the network the same as in Section \ref{a:um1560} and scaled the number of neurons in each layer; that is, we had 4 hidden layers with $1x$, $2x$, $3x$, and $4x$ neurons, with $x=1,2,4,8,$ and $16$.  Since the number of parameters is roughly proportional to the square of the number of neurons, the total parameter counts ranged from 40 to 5425.  We run tests with both the Adam optimizer (10 instances run for 100 epochs each) and ray tracing (two chains run for 1500 epochs each) for all network parameter counts.}

\modified{The left panel of Fig.\ \ref{f:sensitivity} shows how the network loss distribution changes with parameter count.    For the lowest parameter count (40), the network does not have enough flexibility to match the underlying manifold (at least with standard optimization algorithms).  However, at larger parameter counts, the loss distribution appears to stabilize for both Adam and ray tracing.  The implied effective dimensionality from Eq.\ \ref{e:approx_likelihood} is shown in the right panel of Fig.\ \ref{f:sensitivity}, which stabilizes at a value around 60.}

\modified{A few lessons emerge from these tests.  First, even though the implied effective dimensionality may be less than the network parameter count, the network geometry may be such that it is not possible to use all the network's parameters effectively.  The likelihood of this occurring is expected to be less if the effective dimensionality is an order of magnitude below the network parameter count, but the only way to know for sure is to do a sensitivity test with an increased number of parameters.}

\modified{Second, the effective dimensionality does seem to be stable for simple problems, where the underlying manifold is comfortably within the function space of the network employed.  In this particular case, the empirical effective dimension count ($\sim$60) is similar to the number of parameters used by the \textsc{UniverseMachine} to generate the underlying dataset ($\sim 52$, counting 44 free \textsc{UniverseMachine} parameters, 2 fixed \textsc{UniverseMachine} parameters, 5 cosmology parameters, and the choice of $\Delta_{vir}$ for halo definition).  Future studies will reveal whether this accord is a coincidence or a useful method for estimating the true dimensionality of the underlying problem manifold.}

\end{document}